\documentclass[12pt]{iopart}
\usepackage{iopams}

\usepackage{xr}
\usepackage[english]{babel}
\usepackage[latin2]{inputenc}
\usepackage{t1enc}
\expandafter\let\csname equation*\endcsname\relax
\expandafter\let\csname endequation*\endcsname\relax
\usepackage{amsmath}
\usepackage{amssymb}
\usepackage[mathscr]{euscript}
\usepackage{dsfont}
\usepackage{bbm}
\usepackage{tabularx}
\usepackage{graphicx}
\usepackage{enumerate}
\usepackage{mathtools}
\usepackage[thmmarks]{ntheorem-hyper}
\usepackage{tikz-cd}
\usepackage{accents}

\usepackage[normalem]{ulem}
\usepackage{color}
\definecolor{blue}{rgb}{0,0,1}
\definecolor{red}{rgb}{1,0,0}

\newlength{\mathindentorig}
\theoremheaderfont{\bf}\theorembodyfont{\it}\theoremsymbol{\fbox{}}\newtheorem{Thm}{Theorem}
\theoremheaderfont{\bf}\theorembodyfont{\sl}\theoremsymbol{\setlength{\unitlength}{1mm}\begin{picture}(2.5,2.5)\linethickness{2.5mm}\put(0,1.25){\line(1,0){2.5}}\end{picture}}\theoremstyle{plain}\newtheorem{Prf}{Proof}
\theoremheaderfont{\bf}\theorembodyfont{\it}\theoremsymbol{}\newtheorem{Def}[Thm]{Definition}
\theoremheaderfont{\bf}\theorembodyfont{\sl}\theoremsymbol{}\newtheorem{Rem}[Thm]{Remark}
\theoremheaderfont{\bf}\theorembodyfont{\sl}\theoremsymbol{}
\theoremheaderfont{\bf}\theorembodyfont{\it}\theoremsymbol{}
\theoremheaderfont{\bf}\theorembodyfont{\sl}\theoremsymbol{}\newtheorem{Exa}[Thm]{Example}
\theoremheaderfont{\bf}\theorembodyfont{\it}\theoremsymbol{\fbox{}}\newtheorem{Col}[Thm]{Corollary}

\newcommand{\I}{\mathrm{i}}
\newcommand{\R}{\mathbb{R}}
\newcommand{\N}{\mathbb{N}}

\newcommand{\C}{\mathbb{C}}
\newcommand{\de}{\mathrm{d}}
\newcommand{\Ker}{\mathop{\mathrm{Ker}}\nolimits}
\newcommand{\Ran}{\mathop{\mathrm{Ran}}\nolimits}

\newcommand{\RR}{\mathcal{R}}
\newcommand{\F}{\mathbb{F}}
\newcommand{\Fx}{\mathbb{F}^{\times}}
\newcommand{\FT}{\mathbb{F}_{\!{}_{T}}}
\newcommand{\FxT}{\mathbb{F}_{\!{}_{T}}^{\times}}
\newcommand{\T}{\mathcal{T}}
\newcommand{\Ta}{\mathcal{T}_{a}}
\newcommand{\Th}{\mathcal{T}_{h}}
\newcommand{\E}{\mathbf{E}}
\newcommand{\MDS}{\mathbf{M}}
\newcommand{\MDShat}{\hat{\mathbf{M}}}

\newcommand{\dpsi}{\delta\!\psi}
\newcommand{\dpsiT}{\delta\!\psi_{\!{}_{T}}}
\newcommand{\varphiT}{\varphi_{\!{}_{T}}}
\newcommand{\dJ}{\delta\!J}

\newcommand{\KK}{\mathtt{K}}
\newcommand{\1}{\mathds{1}}
\newcommand{\iotabig}{\text{\Large$\iota$}}
\newcommand{\M}{\mathcal{M}}
\newcommand{\K}{\mathcal{K}}
\newcommand{\supp}{\mathop{\mathrm{supp}}\nolimits}
\newcommand{\timesW}{\mathbin{{\times}_{\!\!{}_{W}}}}
\newcommand{\EE}{\mathcal{E}}
\newcommand{\DD}{\mathcal{D}}
\newcommand{\LL}{\mathrm{L}}
\newcommand{\Rad}{\mathrm{Rad}}
\newcommand{\DV}{D\!V}
\newcommand{\Dv}{D\!v}
\newcommand{\dv}{\delta\!v}
\newcommand{\dC}{\delta\!C}
\newcommand{\defin}[1]{\textit{\textbf{#1}}}
\newcommand{\HH}{\mathcal{H}}
\newcommand{\otimesHS}{\mathbin{{\otimes}\!_{{}_{\mathrm{HS}}}}}

\begin{document}

\title[On generally covariant mathematical formulation of Feynman integral in Lorentz signature]{On generally covariant mathematical formulation of Feynman integral in Lorentz signature}

\author{Andr\'as L\'aszl\'o}
\address{Wigner Research Centre for Physics, Budapest}
\ead{laszlo.andras@wigner.hu}

\begin{abstract}
It is widely accepted that the Feynman integral is one of the most promising 
methodologies for defining a generally covariant formulation of nonperturbative 
interacting quantum field theories (QFTs) without a fixed prearranged causal background. 
Recent literature suggests that if the spacetime metric is not fixed, e.g.\ because 
it is to be quantized along with the other fields, one may not be able 
to avoid considering the Feynman integral in the original Lorentz signature, without Wick rotation. 
Several mathematical phenomena are known, however, which are at some point showstoppers 
to a mathematically sound definition of Feynman integral in Lorentz signature. The 
Feynman integral formulation, however, is known to have a differential reformulation, 
called to be the master Dyson--Schwinger (MDS) equation for the field correlators. In 
this paper it is shown that a particular presentation of the MDS equation can be cast 
into a mathematically rigorously defined form: the involved function spaces and operators 
can be strictly defined and their properties can be established. Therefore, MDS equation 
can serve as a substitute for the Feynman integral, in a mathematically sound formulation 
of constructive QFT, in arbitrary signature, without a fixed background causal structure. 
It is also shown that even in such a generally covariant setting, 
there is a canonical way to define the Wilsonian regularization of the MDS equation. The 
main result of the paper is a necessary and sufficient condition for the regularized MDS 
solution space to be nonempty, for conformally invariant Lagrangians. This 
theorem also provides an iterative approximation algorithm for obtaining regularized 
MDS solutions, and is guaranteed to be convergent whenever the solution space is nonempty. 
The algorithm could eventually serve as a method for putting Lorentz signature QFTs 
onto lattice, in the original metric signature.
\end{abstract}


\maketitle

\section{Introduction}
\label{secIntroduction}

By now, a lot is known about the mathematically sound formulation of 
interacting quantum field theory (QFT), using perturbation theory \cite{Henneaux1994}. 
However, still until now, there is no widely accepted concise mathematical 
formulation known for nonperturbative interacting QFT. 
Strictly speaking, as of now, it is only conjectured that eventually one could well-define 
an interacting QFT model in a nonperturbative manner, in a constructive way, 
e.g.\ as specified by a Lagrangian. 
A well known promising attempt for the nonperturbative approach is the 
algebraic quantum field theory (AQFT) \cite{Todorov1990,Fewster2020}. 
AQFT is known to capture several important qualitative aspects of the 
QFT formalism in physics, such as the spin-statistics theorem, 
but there are no known concrete AQFT constructions in the compexity 
of e.g.\ a 3+1 dimensional full quantum electrodynamics. Concrete AQFT models, as of now, 
are only known for free particles in arbitrary dimensions, 
or for simple systems, such as discrete Ising models in 1+1 dimensional 
and discrete spacetimes, or for particular simple systems in 
spacetime dimensions typically lower than 3+1. 
There are also recent advances of perturbative AQFT on causal sets, in 
which framework concrete interacting models are constructed by now 
\cite{Fewster2019}, assuming a finite system of causal sets. 
Due to the difficulties of nonperturbative formulation, 
the perturbative rigorous formulation of 
constructive QFT (pQFT) was seriously considered by a number of authors 
\cite{Hollands2002,Dutsch2001,Brunetti2003,Dabrowski2014,Dutsch2019}. In particular, \cite{Hollands2008} proves the perturbative renormalizability of 
Yang--Mills interactions over globally hyperbolic spacetimes. 
Moreover, a generally covariant framework was already developed \cite{Costello2011}. 
However, it is generally thought that the only promising framework, which could be 
capable of formalizing \emph{nonperturbative} 
interacting generally covariant QFT models in the continuum limit, 
is likely to be the Feynman integral formulation \cite{Feynman2010}. 

A lot is known about Feynman intergrals \cite{Glimm1987,Velhinho2017}, but in Lorentz signature, 
without taking a Wick rotation, it is seems to be still not a completely 
understood mathematical construction, although the modern literature seems 
to tighten the noose on the measure theoretically well defined 
Feynman integral \cite{Albeverio2008,Gill2008}. 
Other authors \cite{Montaldi2017} argue, that Feynman integral should not be, strictly speaking, 
understood in the measure theoretical sense, i.e.\ in the sense of infinitesimal 
summation, but in a more generalized sense. That kind of picture is indeed 
supported by the fact that e.g.\ 
for a fermionic system, the Feynman integral is defined 
as a Berezin integral, which indeed has little link with integration 
in terms of infinitesimal summation. 
To complicate the picture, recent literature suggests \cite{Hollands2008,Feldbrugge2017a,Feldbrugge2017b,Baldazzi2019} that in order to apply 
Feynman integral formalism to a generally covariant setting, in which case the 
a background spacetime metric is not fixed, the applicability of the usual Wick 
rotation from Lorentzian to Euclidean signature can be problematic.

The above issues with the Lorentz signature Feynman integral formulation 
can be circuimvented using the well known differential reformulation of Feynman integral formalism, 
called to be the master Dyson--Schwinger (MDS) equations for the field correlators 
(see e.g.\ \cite{Weigand2014} for a didactic review). From the usually presented 
form of the MDS equation in the QFT literature, it is not immediately evident 
that the function spaces and operators involved in the MDS equation are well defined, 
and are not merely symbolical summaries of heuristic QFT protocols. 
In this paper, however, it is argued that with the right choice of variables, 
these objects can be made mathematically well defined, and as such, the MDS equation 
can be used to substitute the Feynman integral for a mathematically sound definition 
of constructive nonperturbative generally covariant QFT. 
It will be also shown, that in these variables the Wilsonian regularized version of the MDS equation 
can also be canonically defined in a generally covariant setting, 
which is not yet described in the literature. 
The main result of the paper is a theorem about a necessary and sufficient 
condition for the regularized MDS equation to have nonempty solution space, 
for theories with classically conformally invariant Lagrangians. 
The pertinent theorem is constructive in the sense that it provides a 
(probably slowly converging) iterative algorithm for approximating MDS solutions, 
which is guaranteed to be convergent whenever the solution space is nonempty. 
This method can eventually be also employed for doing lattice QFT-like calculations 
in arbitrary signatures, in particular, in the original Lorentz signature.

The structure of the paper is as follows. In Section~\ref{secFeyn} the 
heuristic form of the MDS equation is recalled, as derived from the heuristic 
Feynman integral formulation in usual QFT. The rest of the paper intends to keep mathematical rigor. 
In Section~\ref{secBasicQFT} the function spaces and operators needed to define 
the (unregularized) MDS operator are presented. In Section~\ref{secWeakReg}, these are generalized 
in the distributional sense, and the Wilsonian regularized version of the MDS 
equation is invoked and justified. Section~\ref{secExist} is dedicated for the main theorem 
of the paper about a necesary and sufficient condition for the existence 
of solutions of the regularized MDS equation, for conformally 
invariant Lagrangians. 
\ref{secEL} was added in order to pin down the precise continuity 
properties of a typical Euler--Lagrange functional in a standard classical field theory, 
which is key in the construction. 
\ref{secWR} was added for completeness, in order to define the Wilsonian renormalizability 
in a generally covariant way, given the notion of Wilsonian regularization. 
The paper heavily relies on the theory of non-normable 
topological vector spaces (TVS), and therefore a supplementary material \cite{Laszlo2022s} 
is provided for a recollection of important and sometimes counterintuitive theorems on these, for 
readers not specialized in the theory of TVS.

\section{Feynman integral and the heuristic form of the MDS equation}
\label{secFeyn}

We briefly recall the justification of the MDS equation in the Feynman 
integral formulation of QFT. Let $F$ denote the space of all (that is, off-shell) 
smooth classical field configurations. As expanded in \ref{secEL}, 
in most models it is safe to assume that $F$ is a topological affine space, 
such that its subordinate vector space $\F$, the space of smooth field variations, 
carryies a nuclear Fr\'echet topology. The affineness of $F$ is necessary in order to naturally 
accomodate gauge fields. 
$\F^{*}$ will denote the topological dual of $\F$, understood with the standard strong dual topology. 
In the Feynman integral formulation of QFT, it is postulated that the evaluation 
method for Feynman type (i.e., causally ordered) quantum vacuum expectation value 
of observables in a (not necessarily unique) vacuum state $\rho$ is the 
following. Given a fixed reference field 
$\psi_{0}\in F$ and test functionals $J_{1},\dots,J_{n}\in\F^{*}$, the causally ordered 
quantum vacuum expectation value of the polynomial observable 
$(J_{1}|\cdot{-}\psi_{0})\cdots(J_{n}|\cdot{-}\psi_{0}):\,F\rightarrow\R$ is 
declared to be
\begin{eqnarray}
 \int\limits_{\psi\in F} (J_{1}|\psi{-}\psi_{0})\cdots(J_{n}|\psi{-}\psi_{0})\;\; \e^{\I \frac{1}{\hbar}S(\psi)} \;\de\rho(\psi) \;\Bigg/\; \int\limits_{\psi\in F} \e^{\I \frac{1}{\hbar}S(\psi)} \;\de\rho(\psi),
\label{eqExpHeur}
\end{eqnarray}
where the symbol $\de\rho(\cdot)$ denotes the hypothetical Feynman measure 
corresponding to a vacuum state $\rho$, $(\cdot|\cdot)$ denotes 
the duality pairing form between $\F^{*}$ and $\F$, whereas $S:\,F\rightarrow\R$ 
is the action functional of the underlying classical field theory. In the heuristic calculations, 
$\de\rho(\cdot)$ is handled as if it were a Lebesgue measure on $F$, and as if 
$\e^{\I \frac{1}{\hbar}S(\cdot)}\,\de\rho(\cdot)$ were a finite measure, 
having finite moments and analytic Fourier transform. 
A sign change $\hbar\mapsto{-}\hbar$ would correspond to a reversal in the causal ordering, 
if there were any \emph{a priori} causal structure over the spacetime manifold 
(which in fact, is not needed to be assumed at this point). 
The hypotethical \emph{partition function} condenses all these information about 
the state $\rho$, and would be a mapping
\begin{eqnarray}
 Z_{\hbar,\psi_{0}}:\quad \F^{*}\rightarrow\C,\quad J\mapsto Z_{\hbar,\psi_{0}}(J) := \int\limits_{\psi\in F} \e^{\I\,\left(J\vert\psi{-}\psi_{0}\right)}\; \e^{\I \frac{1}{\hbar}S(\psi)} \;\de\rho(\psi),
\label{eqFeyn}
\end{eqnarray}
i.e.\ the formal Fourier transform of the hypothetical measure $\e^{\I \frac{1}{\hbar}S(\cdot)}\,\de\rho(\cdot)$. 
The collection of \emph{$n$-field correlators}
\begin{eqnarray}
 G_{\hbar,\psi_{0}}^{(n)} & := & \left.\left((-\I)^{n}\,\frac{1}{Z_{\hbar,\psi_{0}}(J)}\; D^{(n)}Z_{\hbar,\psi_{0}}(J)\right)\right\vert_{J=0}
\label{eqCorr}
\end{eqnarray}
is an other means to rephrase these information about the state $\rho$, 
and also can be used to evaluate the quantum expectation values Eq.(\ref{eqExpHeur}) 
by simple duality pairing, like 
$\big(J_{1}{\otimes}{\dots}{\otimes}J_{n}\,\big\vert\,G_{\hbar,\psi_{0}}^{(n)}\big)$. 
Here $D^{(n)}Z_{\hbar,\psi_{0}}$ is assumed to behave like the $n$-th Fr\'echet derivative of the partition function 
$J\mapsto Z_{\hbar,\psi_{0}}(J)$, implicitly assuming that $Z_{\hbar,\psi_{0}}$ is $n$-times continuously 
Fr\'echet differentiable (and for fermion fields, this differentiation is assumed to be a graded differentiation). 
Since the partition function would be a map 
$Z_{\hbar,\psi_{0}}:\F^{*}\rightarrow\C$, the collection of field correlators 
$G_{\hbar,\psi_{0}}:=\big(G_{\hbar,\psi_{0}}^{(0)},G_{\hbar,\psi_{0}}^{(1)},{\dots},G_{\hbar,\psi_{0}}^{(n)},{\dots}\big)$ 
would sit in 
$\T(\F):=\mathop{\bigoplus}\limits_{n\in\N_{0}}\mathop{\otimes}\limits^{n}\F$, 
i.e.\ in the tensor algebra of $\F$, or more precisely in a graded-symmetrized subspace of $\T(\F)$.

Let $E(\psi):=D_{F}S(\psi)$ denote the Euler--Lagrange functional, i.e.\ the 
derivative of the action functional $S$, evaluated at the classical field configuration 
$\psi\in F$. It would be a map 
$E:\,F\times\F\rightarrow\R,\,(\psi,\dpsi)\mapsto \left(E(\psi)\big\vert\dpsi\right):=\left(D_{F}S(\psi)\big\vert\dpsi\right)$, 
being linear in its second variable, since it is a derivative. 
In the usual QFT protocol it is assumed that the EL functional 
$E$ is multipolynomial, and thus so is the real valued map 
$\psi\mapsto \left(E(\psi)\,\big\vert\,\dpsi\right)$ for 
any fixed field variation $\dpsi\in\F$. Let 
$\E((-\I)D_{\F^{*}}+\psi_{0})$ be the multipolynomial differential 
operator defined by the polynomial coefficients of the Euler-Lagrange 
functional $E$. 
Applying the usual rules of formal Fourier transform, a function $Z:\F^{*}\rightarrow\C$ is 
of the form Eq.(\ref{eqFeyn}), up to a complex multiplyer, if and only if 
it satisfies the master Dyson--Schwinger (MDS) equation
\begin{eqnarray}
 \left.\Big(\; \E((-\I)D_{\F^{*}}+\psi_{0})\; Z \;\Big)\right\vert_{J} & = & -\hbar\,J\; Z(J)
 \qquad\qquad\qquad (\forall J\in\F^{*}),
\label{eqMDS8}
\end{eqnarray}
see e.g.\ \cite{Weigand2014} for a didactic derivation. 
The operational meaning of this usual presentation of the MDS equation might not 
seem immediately evident. However, expressing $Z_{\hbar,\psi_{0}}$ via its formal 
Taylor series, encoded by the collection of field correlators $G_{\hbar,\psi_{0}}\in\T(\F)$, 
the MDS equation Eq.(\ref{eqMDS8}) is seen to be equivalent to
\begin{eqnarray}
 \mathrm{we\;search\;for}\; G\in\T(\F) \;\mathrm{such\;that}: \cr
 \Big. G^{(0)} = 1 \quad\mathrm{and}\quad 
 \iotabig_{(\E_{\psi_{0}}\vert\dpsi)}\,G = \I\,\hbar\,L_{\dpsi}\,G \qquad (\forall\,\dpsi\in\F).
\label{eqAMDS3}
\end{eqnarray}
The symbols of this equation mean the following. 
$L_{\dpsi}$ denotes the left-multiplication operator 
in the tensor algebra $\T(\F)$ by the one-vector $\dpsi\in\F$. The symbol $\iotabig_{p}$ 
denotes the left-insertion operator by some element $p$ from the topological 
dual space of $\T(\F)$. The map $E_{\psi_{0}}:\,\F\rightarrow\F^{*}$ is 
defined via $E_{\psi_{0}}:=E\circ(\mathrm{I}_{\F}+\psi_{0})$ from the 
original Euler--Lagrange functional $E:\,F\rightarrow\F^{*}$, i.e.\ it is 
the Euler--Lagrange functional with respect to a fixed reference 
field $\psi_{0}\in F$, re-expressed on the space of field variations $\F$. 
Since it was assumed to be multipolynomial, it can eventually be regarded as a linear map
$E_{\psi_{0}}:\,\T(\F)\rightarrow\F^{*}$. 
As such, it may be identified with an element 
$\E_{\psi_{0}}\in\big(\T(\F)\big)^{*}\otimes\F^{*}$, and correspondingly 
$(E_{\psi_{0}}\vert\dpsi)$ with $(\E_{\psi_{0}}\vert\dpsi)\in\big(\T(\F)\big)^{*}\,$ ($\forall\,\dpsi\in\F$), 
which then has the corresponding left-insertion operator 
$\iotabig_{(\E_{\psi_{0}}\vert\dpsi)}$ acting over $\T(\F)$. 
The spaces and operators involved in Eq.(\ref{eqAMDS3}) would be 
perfectly meaningful if the space of fields $F$ were finite dimensional, 
and could be used as a substitute for Feynman integral formulation 
Eq.(\ref{eqExpHeur}), regardless of e.g.\ a metric signature or other auxiliary information on the 
details of the underlying classical theory described by the Euler--Lagrange functional $E$. 
In Section~\ref{secBasicQFT} it shall be shown that the 
pertinent objects can be made well-defined even when $F$ is 
indeed the infinite dimensional space of smooth off-shell field configurations in a realistic field theory. 
The Eq.(\ref{eqAMDS3}) presentation of the MDS equation does not seem to be described in the literature.

In QFT, it is also necessary to consider the Wilsonian regularized version of the 
Feynman integral. Wilsonian regularization 
means performing the Feynman integral Eq.(\ref{eqExpHeur}) on a 
subspace of off-shell fields with their high frequency modes suppressed. 
In a generally covariant setting the meaning of this might not seem immediately 
evident, but Wilsonian regularized Feynman type expectation value of the observable 
$(J_{1}|\cdot{-}\psi_{0})\cdots(J_{n}|\cdot{-}\psi_{0}):\,F\rightarrow\R$ can be postulated as
\begin{eqnarray}
  \int\limits_{\dpsi\in\RR[\F]} (J_{1}|\dpsi)\cdots(J_{n}|\dpsi)\;\; \de\RR_{*}\mu_{\psi_{0}}(\dpsi) \;\Bigg/\; \int\limits_{\dpsi\in\RR[\F]} 1 \;\de\RR_{*}\mu_{\psi_{0}}(\dpsi)
\label{eqExpHeurW}
\end{eqnarray}
with $\psi_{0}\in F$ and $J_{1},\dots,J_{n}\in\F^{*}$ as previously, 
where $\RR:\,\F\rightarrow\F$ is some continuous 
linear operator, $\RR[\F]\subset\F$ denotes the image of $\F$ by $\RR$, 
the symbol $\mu_{\psi_{0}}$ stands for the pushforward of the hypothetical finite measure 
$\e^{\I \frac{1}{\hbar}S(\cdot)}\,\de\rho(\cdot)$ on $F$ via the map $F\rightarrow\F,\,\psi\mapsto(\psi{-}\psi_{0})$, 
and $\RR_{*}\mu_{\psi_{0}}$ stands for the pushforward of the measure $\mu_{\psi_{0}}$ on $\F$ to $\RR[\F]$ by $\RR$. 
The map $\RR$ can be called a \emph{regulator}, and typically it is a 
convolution operator by some test function in case of theories over an affine 
spacetime (can be generalized for arbitrary spacetimes as well), 
and Eq.(\ref{eqExpHeurW}) means 
nothing but the natural pushforward Feynman integration on the subspace $\RR[\F]\subset\F$, given that 
the original Feynman integration Eq.(\ref{eqExpHeur}) on $F$ was meaningful. The map $\RR$ implements 
the high frequency damping. 
Using the fundamental formula of integral substitution, one infers that the Wilsonian regularized MDS equation 
on the field correlators reads
\begin{eqnarray}
 \Big. G^{(0)} = 1 \quad\mathrm{and}\quad 
 \iotabig_{(\E_{\psi_{0}}\vert\dpsi)}\,G = \I\,\hbar\,L_{\RR\,\dpsi}\,G \qquad (\forall\,\dpsi\in\F)
\label{eqAMDS3W}
\end{eqnarray}
in the analogy of Eq.(\ref{eqAMDS3}), where again $L_{\RR\,\dpsi}$ is the 
left-multiplication in $\T(\F)$ by the one-vector $\RR\,\dpsi\in\F$.
As shall be expanded in Section~\ref{secWeakReg}, the pertinent objects can be made well-defined 
similarly to that of the unregularized MDS equation. 
The Wilsonian regularized MDS equation Eq.(\ref{eqAMDS3W}) does not seem to be described in the literature.

From this point on, we drop the heuristic arguments, and all the statements and 
formulas are intended to be mathematically rigorous. The aim 
of this paper is to show that the objects involved in Eq.(\ref{eqAMDS3}) and 
Eq.(\ref{eqAMDS3W}) are mathematically well defined, and to establish the 
fundamental properties of the solution spaces of the pertinent equations.

\section{Mathematically rigorous definition of the unregularized MDS operator}
\label{secBasicQFT}

As detailed in \ref{secEL}, in a generic classical field theory, it is safe 
to assume that the space of off-shell fields $F$ is the affine space of 
smooth sections of a real finite dimensional affine bundle over a real 
finite dimensional smooth base manifold. The space of field variations 
$\F$ are comprised of differences of elements in $F$, and as such it is the 
vector space of smooth sections of the real finite dimensional 
vector bundle subordinate to our affine bundle, understood with the standard $\EE$ smooth function topology, 
which is known to be nuclear Fr\'echet. 
Within $\F$, there is the space of test field variations $\FT$, 
comprised of compactly supported smooth sections, with the standard $\DD$ test 
function topology. For the sake of genericity, in this section we avoid 
using the knowledge that $F$, $\F$ and $\FT$ are these concrete spaces, 
they will be considered abstract spaces instead. 
The symbol ${}^{*}$ shall denote strong topological dual. 
See \cite{Laszlo2022s} and the Appendix of \cite{Costello2011} 
for a condensed summary on the theory of topological vector spaces.

\begin{Def}
Let $F$ be a real affine space, with a subordinate real topological vector 
space $\F$. Let the topology on $\F$ be nuclear Fr\'echet (in short, NF space, 
see also \cite{Laszlo2022s}-Remark\ref{S-remNuclear}, the $\EE$ smooth function space is the archetype of an NF space). 
We call $F$ the \defin{space of classical field configurations} and the subordinate vector 
space $\F$ the \defin{space of classical field variations}. Let 
$\FT\subset\F$ be some subspace of $\F$, endowed with 
a topology not weaker than $\F$. Let $\FT$ be 
either nuclear Fr\'echet or the strict inductive limit of a countable system 
of nuclear Fr\'echet spaces with closed adjacent images (in short, LNF space, 
see also \cite{Laszlo2022s}-Remark\ref{S-remNuclear}, the $\DD$ test function space is the acrhetype of an LNF space). 
Then, we call $\FT$ the \defin{space of test field variations}.
\label{defF}
\end{Def}

As detailed in \ref{secEL}, in a generic concrete classical field theory, the Euler--Lagrange 
functional is the derivative of the action functional, with its linear variable 
restricted to the space of test field variations, so that the Euler--Lagrange functional 
becomes an everywhere defined map. 
It is also shown to be a jointly sequentially continuous map in its two variables. 
This justifies the following abstract definition.

\begin{Def}
Let $E:\,F\times\FT\rightarrow\R,\,(\psi,\dpsiT)\mapsto E(\psi,\dpsiT)$ be a jointly sequentially continuous map which is linear 
in its second variable. Then, $E$ will be called a \defin{classical Euler--Lagrange (EL) functional}. 
(By means of \ref{secEL}~Theorem\ref{thmEcont}(ii,iii), then $E$ is also separately continuous in 
its two variables, and when viewed as a map 
$E:\,F\rightarrow\FT^{*},\,\psi\mapsto E(\psi)$, it is continuous.) 
Given a $\dpsiT\in\FT$, when the second argument of $E$ is evaluated, 
it will be denoted by $(E\,\vert\,\dpsiT):\,F\rightarrow\R$, which is then a continuous map. 
When that map is evaluated at some $\psi\in F$, we denote it by 
$\big(E(\psi)\,\big\vert\,\dpsiT\big)\in\R$. We call the equation
\begin{eqnarray}
 \Big. \mathrm{we\;search\;for}\; \psi\in F \;\mathrm{such\;that:}\qquad
 \forall\;\dpsiT\in\FT:\quad \big(E(\psi)\,\big\vert\,\dpsiT\big) = 0
\label{eqELc}
\end{eqnarray}
the \defin{classical Euler--Lagrange (EL) equation}.
When $E$ is viewed as a map ${E:\,F\rightarrow\FT^{*}}$, 
given any fixed field $\psi_{0}\in F$, we use the notation 
$E_{\psi_{0}}:=E\circ(\mathrm{I}_{\F}+\psi_{0})$, which will then be a 
continuous map $E_{\psi_{0}}:\,\F\rightarrow\FT^{*}$, 
and $\psi_{0}$ will be called a \defin{reference field}. 
(By construction, for all $\psi\in F$ the identity 
$E(\psi)=E_{\psi_{0}}(\psi{-}\psi_{0})$ holds.)
\label{defEL}
\end{Def}

In order to define the MDS operator, we will need to invoke the notion of a topologized tensor algebra made out of $\F$. 
For that, recall the below facts.

\begin{Rem}
In this remark block let $\mathbb{U}$ denote a 
nuclear Fr\'echet (NF) or strong dual of a nuclear Fr\'echet (DNF) space. 
(See also \cite{Laszlo2022s}-Remark\ref{S-remNuclear}.)
\begin{enumerate}[(i)]
 \item \label{remFpropmult} For all $n\in\N_{0}$, the completed topological tensor product $\mathop{\otimes}\limits^{n}\mathbb{U}$ 
is meaningful (e.g.\ understood with the projective tensor product topology), 
and is NF or DNF, respectively. Moreover, in the analogy of 
finite dimensional vector spaces, the pertinent tensor product can be 
implemented via the multiplicative realization. That is, it is topologically 
isomorphic to the space of the jointly continuous $n$-fold multilinear forms 
on the strong dual space of $\mathbb{U}$. (See also \cite{Laszlo2022s}-Remark\ref{S-remNuclear}.)
 \item With the same assumptions, one has that for all $n\in\N_{0}$, the identity 
$\big(\mathop{\otimes}\limits^{n}\mathbb{U}\big)^{*}\equiv\mathop{\otimes}\limits^{n}\mathbb{U}^{*}$ 
holds. (See also \cite{Laszlo2022s}-Remark\ref{S-remNuclear}.)
 \item Given a countable system of NF or a countable system of DNF spaces, 
their cartesian product can be equipped with a vector space structure and 
with the product (also called Tychonoff or initial or projective) topology. 
This is the weakest topology such that the canonical projections of the cartesian product are continuous. 
With this, it will become an NF or DNF space, respectively. 
(See also \cite{Laszlo2022s}-Remark\ref{S-remCartesian}, \cite{Laszlo2022s}-Remark\ref{S-remNuclear}.) 
Therefore, the Tychonoff tensor algebra 
$\T(\mathbb{U}):=\mathop{\bigoplus}\limits_{n=0}^{\infty}\mathop{\otimes}\limits^{n}\mathbb{U}$ 
is meaningful and is NF or DNF, respectively. (The symbol $\mathop{\bigoplus}\limits_{n=0}^{\infty}:=\mathop{\bigtimes}\limits_{n=0}^{\infty}$ 
as set operation, but we use rather $\bigoplus$ for vector spaces.)
 \item Given a countable system of NF or a countable system of DNF spaces, 
in their cartesian product vector space there is the subspace of the elements 
with all zero except for finite entries, which subspace is called the algebraic direct sum space. 
This can be equipped with 
the locally convex direct sum (also called final or injective) topology. 
This is the strongest topology such that the canonical injections of the cartesian product are continuous. 
With this, it will become an NF or DNF space, respectively (see also \cite{Laszlo2022s}-Remark\ref{S-remCartesian}, \cite{Laszlo2022s}-Remark\ref{S-remNuclear}). 
Therefore, the algebraic tensor algebra 
$\Ta(\mathbb{U}):=\mathop{\oplus}\limits_{n=0}^{\infty}\mathop{\otimes}\limits^{n}\mathbb{U}$ 
with the locally convex direct sum topology is meaningful and is NF or DNF, respectively.
 \item One has that 
$\big(\T(\mathbb{U})\big)^{*}\equiv\T_{a}(\mathbb{U}^{*})$ and 
$\big(\T_{a}(\mathbb{U})\big)^{*}\equiv\T(\mathbb{U}^{*})$. (See also \cite{Laszlo2022s}-Remark\ref{S-remCartesian}.)
 \item The Tychonoff tensor algebra 
has a jointly continuous bilinear map 
${{\otimes}:\,{\T(\mathbb{U})\times\T(\mathbb{U})}\rightarrow\T(\mathbb{U})}$, the 
tensor algebra multiplication, with a unit element $\1:=(1,0,0{\dots})\in\T(\mathbb{U})$ 
(consequences of \cite{Laszlo2022s}-Remark\ref{S-remJoint}). The subspaces of $k$-tensors provide a grading of $\T(\mathbb{U})$. 
Quite trivially, the left multiplication operator for all $u\in\T(\mathbb{U})$ is a 
continuous linear map $L_{u}:\,\T(\mathbb{U})\rightarrow\T(\mathbb{U})$.
 \item Similarly, the algebraic tensor algebra 
has a jointly continuous bilinear map 
$\Ta(\mathbb{U})\times\Ta(\mathbb{U})\rightarrow\Ta(\mathbb{U})$, the 
tensor algebra multiplication, with a corresponding unit element 
(consequences of \cite{Laszlo2022s}-Remark\ref{S-remJoint}). 
The subspaces of $k$-tensors provide a grading of $\Ta(\mathbb{U})$. 
Quite trivially, the left multiplication operator is a 
continuous linear map $\Ta(\mathbb{U})\rightarrow\Ta(\mathbb{U})$.
 \item Since $\T(\mathbb{U})$ and 
$\Ta(\mathbb{U}^{*})$ are strong duals to each-other, and both of these are 
graded unital associative algebras with jointly continuous multiplications, 
by transposing the algebra multiplication and unit from the duals, one infers 
that both $\T(\mathbb{U})$ and $\Ta(\mathbb{U}^{*})$ are bialgebras, with corresponding 
coproduct and counit. The counit of $\T(\mathbb{U})$ is $b:\,\T(\mathbb{U})\rightarrow\R,\,G:=(G^{(0)},G^{(1)},{\dots})\mapsto b\,G:=G^{(0)}$, 
i.e.\ extraction of the scalar component, the symbol ``$b$'' standing for ``base'' or ``bottom form''.
 \item \label{remFpropInsertion} Due to the bialgebra nature of $\T(\mathbb{U})$, 
i.e.\ due to the existence of a continuous coproduct on $\T(\mathbb{U})$, for all 
$p\in\Ta(\mathbb{U}^{*})$ the corresponding left insertion operator 
$\iotabig_{p}:\,\T(\mathbb{U})\rightarrow\T(\mathbb{U})$ is meaningful, and 
is a continuous linear operator. 
More concretely, the left insertion operator $\iotabig_{p^{(n)}}$ with a 
$p^{(n)}\in\mathop{\otimes}\limits^{n}\mathbb{U}^{*}$ ($n\in\N_{0}$) exists, 
because for all $m\in\N_{0}$ ($m\geq n$) the tensor product $\mathop{\otimes}\limits^{m}\mathbb{U}$ 
can be identified with the space of 
$\mathop{\otimes}\limits^{n}\mathbb{U}^{*}\;\times\;\mathop{\otimes}\limits^{m{-}n}\mathbb{U}^{*}\rightarrow\R$ 
jointly continuous bilinear forms, as stated in (\ref{remFpropmult}). 
Similarly, the left insertion operators make sense in $\Ta(\mathbb{U})$, 
and is a continuous linear operator. 
(For the sake of distinction in terminology, we call merely the operators $\iotabig_{p^{(1)}}$ with 
$p^{(1)}\in\mathbb{U}^{*}\equiv\mathop{\otimes}\limits^{1}\mathbb{U}^{*}\subset\Ta(\mathbb{U}^{*})$ 
as insertion operators, whereas for generic $p^{(n)}\in\mathop{\otimes}\limits^{n}\mathbb{U}^{*}\subset\Ta(\mathbb{U}^{*})$ ($n\in\N_{0}$) 
or more generally for $p\in\Ta(\mathbb{U}^{*})$, we call the corresponding $\iotabig_{p^{(n)}}$ or 
$\iotabig_{p}$ as multipolynomial insertion operator.) 
For all $p\in\Ta(\mathbb{U}^{*})$, one has the identity $p=b\,\iotabig_{p}$. 
For the left insertion operator, we use the 
normalization convention such that for all $G^{(n)}\in\mathop{\otimes}\limits^{n}\mathbb{U}$ 
and $u\in\mathbb{U}$ and $p\in\mathbb{U}^{*}$ one has
$\iotabig_{p}\,L_{u}\;G^{(n)}=(n+1)\,(p|u)\,G^{(n)}$.
 \item A historical note: over an affine (Minkowski) spacetime, one can 
define the space of rapidly decreasing (Schwartz) functions $\mathcal{S}$, 
which is an NF space. The tensor algebra $\Ta(\mathcal{S})$ is referred to as 
Borchers--Uhlmann (BU) algebra (original papers: \cite{Borchers1962,Uhlmann1962}, 
and including a short review: \cite{Yngvason1973}). The Wightman functionals 
in QFT are understood to be in the space 
$\big(\Ta(\mathcal{S})\big)^{*}\equiv\T(\mathcal{S}^{*})$.
 \item By construction, the $\Ta$ topology is strongest tensor algebra 
topology, whereas $\T$ is the weakest. It is possible to define a natural 
topological tensor algebra which is in between the $\T$ and $\Ta$, in terms of topology strength. 
It will be motivated and introduced later, in Section~\ref{secExist}, and will be key 
to the presented construction, if one wishes to quantize analytic EL functionals, 
and not only polynomial ones.
\end{enumerate}
\label{remFprop}
\end{Rem}

\begin{Def}
Assume that the space of field variations as real nuclear Fr\'echet spaces 
has a direct sum splitting $\F=\F_{r}\oplus\F_{c}$, called 
\defin{the real-complex splitting}, 
where both $\F_{r}$ and $\F_{c}$ are closed (and therefore nuclear Fr\'echet), 
and $\F_{c}$ has a complex structure (i.e.\ it can be regarded as a complex 
nuclear Fr\'echet space). Denote by 
$\F_{r\C}:=\F_{r}\otimes\C$ the complexification of $\F_{r}$. 
Then, we use the notation $\F_{(\C)}:=\F_{r\C}\oplus\F_{c}$, and call it the 
\defin{space of field variations with complex structure}. 
(We assume that also $\FT\subset\F$ respects this splitting.)
\label{defRC}
\end{Def}

The above definition is necessary, because in field theory, certain fields 
(like variations of Dirac fields) sit in an inherently complex vector space, whereas other 
fields (like variations of gauge fields) sit in an inherently real vector space, and QFT 
assumes that the sectors not being inherently complex are complexified. 
In the most simple case, one has merely $\F_{(\C)}=\F$ if $\F$ was complex, or $\F_{(\C)}=\F\otimes\C$ if $\F$ was real.

\begin{Def}
Let the vector space of field variations admit a real-complex splitting 
$\F={\F_{r}\oplus\F_{c}}$, as in Definition\ref{defRC}.
Furthermore, assume a direct sum structure
$\F=\mathop{\oplus}\limits_{i=1}^{f}\F_{i}$, such that for each 
$i=1,{\dots},f$ the subspace $\F_{i}$ is either entirely within $\F_{r}$ or in 
$\F_{c}$ and are closed (thus, also nuclear Fr\'echet), and let there be integers $s_{i}\in\{0,1\}$ associated 
to each subspace $\F_{i}$ ($i=1,{\dots},f$). Then, the subspaces 
$\F_{1},{\dots},\F_{f}$ are called the \defin{flavor sectors}, 
and their associated integers $s_{1},{\dots},s_{f}$ are called 
\defin{bosonic or fermionic labels}. 
(We assume that also $\FT\subset\F$ respects this splitting.)
\label{defLabels}
\end{Def}

In the most simple case, there is only one single flavor sector, globally 
endowed with a bosonic or fermionic label. For invoking the MDS equation, 
we will need the graded-symmetrized subspace of $\T(\F_{(\C)})$, 
according to the bosonic and fermionic labels. In order to establish that algebra, 
the following remark is useful.

\begin{Rem}
Whenever $\F$ is split as $\F=\mathop{\oplus}\limits_{i=1}^{f}\F_{i}$ into 
flavor sectors with bosonic / fermionic labels $s_{i}$ ($i=1,{\dots},f$), as 
in Definition\ref{defLabels}, then for all $n\in\N_{0}$ one may introduce 
a continuous linear representation $U_{\pi}$ of a permutation group element $\pi\in\Pi_{n}$ on 
the space $\mathop{\otimes}\limits^{n}\F_{(\C)}$ as follows (see also \cite{Dubin1989b}~Chapter4). 
Take an element $x_{1}{\otimes}{\dots}{\otimes}x_{n}\in\mathop{\otimes}\limits^{n}\F_{(\C)}$, 
where each factor $x_{i}$ ($i=1,{\dots},n$) resides in some $\F_{(\C)j}$ ($j=1,{\dots},f$). 
Then, set
\begin{eqnarray*}
U_{\pi}(x_{1}{\otimes}{\dots}{\otimes}x_{n}) & := & (-1)^{s_{1}\,\sigma_{1}(\pi)+{\dots}+s_{f}\,\sigma_{f}(\pi)}\; x_{\pi(1)}{\otimes}{\dots}{\otimes}x_{\pi(n)},
\end{eqnarray*}
where $\sigma_{i}(\pi)\in\{0,1\}$ ($i=1,{\dots},f$) is the parity of the perumation 
$\pi$ within each index block. 
The map $U_{\pi}$ can then be linearly 
extended in $\mathop{\otimes}\limits^{n}\F_{(\C)}$. Due to the NF property of 
the involved spaces, the topology defining seminorms on $\mathop{\otimes}\limits^{n}\F_{(\C)}$ 
may be taken to be such that $U_{\pi}$ are continuous (\cite{Dubin1989b}~Chapter4), 
therefore can uniquely be extended as acting as a continuous linear map 
$U_{\pi}:\,\mathop{\otimes}\limits^{n}\F_{(\C)}\rightarrow\mathop{\otimes}\limits^{n}\F_{(\C)}$, 
thus defining the signed permutation operator on the entire space $\mathop{\otimes}\limits^{n}\F_{(\C)}$. 
Therefore, on each space $\mathop{\otimes}\limits^{n}\F_{(\C)}$ the continuous linear projection operator
\begin{eqnarray*}
P_{n} := \frac{1}{n!}\;\sum_{\pi\in\Pi_{n}} U_{\pi}
\end{eqnarray*}
can be defined.
The family of operators $P_{n}$ ($n\in\N_{0}$) on the spaces 
$\mathop{\otimes}\limits^{n}\F_{(\C)}$ can be joined as a single grading preserving continuous linear projection operator 
${P:\,\T(\F_{(\C)})\rightarrow\T(\F_{(\C)})}$. This signed symmetrizer projection 
operator $P$ has the following properties against the tensor algebra multiplication:
\begin{eqnarray*}
 P(x\,y) \;=\; P(P(x)\,y) \;=\; P(x\,P(y)) \;=\; P(P(x)\,P(y)) \quad(\forall\,x,y\in\T(\F_{(\C)})).
\end{eqnarray*}
Therefore, the closed subspace $\Ker(P)$ is a two-sided ideal in $\T(\F_{(\C)})$. 
(The presented approach was inspired by \cite{Dubin1989b}~Chapter4.)
\label{remDubin}
\end{Rem}

Using the fact that the closed subspace of an NF space 
is also NF and that the factor space of an NF space 
with a closed subspace is also NF (see also \cite{Laszlo2022s}-Remark\ref{S-remNuclear}), 
the following definition is meaningful.

\begin{Def}
Let the space of field variations $\F$ admit flavor sectors 
$\F_{i}$ and bosonic / fermionic labels $s_{i}$ ($i=1,{\dots},f$), 
as in Definition\ref{defLabels}, and corresponding signed symmetrization 
projector $P$ as in Remark\ref{remDubin}.
Then the factor algebra $A(\F_{(\C)}):=\T(\F_{(\C)})/\Ker(P)$ is called 
the \defin{field algebra}. Clearly, it is a unital associative 
algebra, and a nuclear Fr\'echet (NF) topological vector space, 
with jointly continuous algebra multiplication 
$\bullet:\,A(\F_{(\C)})\times A(\F_{(\C)})\rightarrow A(\F_{(\C)})$. 
The topological transpose $P^{*}$ of $P$ allows the analogous construction 
in the strong dual of $\T(\F_{(\C)})$, which makes it also a unital associative 
algebra with jointly continuous algebra multiplication, and therefore $A(\F_{(\C)})$ 
retains the bialgebra structure from $\T(\F_{(\C)})$.

Since the complementing projection operator $I{-}P$ to $P$ is also continuous, 
as topological vector spaces one may naturally identify $A(\F_{(\C)})$ 
with the closed subspace $\Ker(I{-}P)=\Ran(P)\subset\T(\F_{(\C)})$. 
Using this linear topological identification, the algebraic product 
$\bullet$ may be pushed forward from $A(\F_{(\C)})$ to the subspace 
$\Ran(P)\subset\T(\F_{(\C)})$. That is, as usual, the algebra $A(\F_{(\C)})$ 
may be regarded as a closed subspace of $\T(\F_{(\C)})$. On that space the 
product $\bullet$ can be traced back to the tensor algebra product 
$\otimes$, with the identity: for all $x\in\mathop{\otimes}\limits^{m}\F_{(\C)}$ 
and $y\in\mathop{\otimes}\limits^{n}\F_{(\C)}$, one 
has $x\bullet y=\frac{(m+n)!}{m!\,n!}\,P(x\otimes y)$. 
The unit element, the counit map, as well as the insertion operator by a one-form $p^{(1)}\in\F_{(\C)}^{*}$ 
coincides to the one defined on $\T(\F_{(\C)})$. The strong dual of 
$A(\F_{(\C)})$ may be identified with the corresponding subspace of $\Ta(\F_{(\C)}^{*})$. 
Whenever not confusing, we will suppress the multiplication symbol $\bullet$.
\label{defFieldA}
\end{Def}

The above definition was necessary, because in QFT the Feynman type field correlators 
are graded-symmetrized, i.e.\ they sit rather in $A(\F_{(\C)})$ than in 
$\T(\F_{(\C)})$. (In the most simple case one has that 
$A(\F_{(\C)})$ is $\bigvee(\F_{(\C)})$ or $\bigwedge(\F_{(\C)})$.) 
As expanded above, the left multiplication operator (given some $\dpsi\in\F$) is the same as $L_{\dpsi}$ in 
$\T(\F_{(\C)})$, with a subsequent graded-symmetrization 
and combinatorial normalization. It shall be denoted 
by the same symbol $L_{\dpsi}$, when not confusing. 
According to the chosen normalization conventions, in the algebra 
$A(\F_{(\C)})\subset\T(\F_{(\C)})$, the counit map $b$ 
and the left-insertion operator $\iotabig_{p}$ by a one-form 
$p\in\F_{(\C)}^{*}$ literally coincide with the corresponding operators in $\T(\F_{(\C)})$. 
Due to the graded-symmetrization,
one has that for all $G\in A(\F_{(\C)})$, and for all $\dpsi\in\F_{(\C)}$, $\dJ\in\F_{(\C)}^{*}$ 
from the same fermionic sector 
$\left(\iotabig_{\dJ}\,L_{\dpsi}+L_{\dpsi}\,\iotabig_{\dJ}\right)\,G = (\dJ|\dpsi)\,G$ holds, whereas 
$\left(\iotabig_{\dJ}\,L_{\dpsi}-L_{\dpsi}\,\iotabig_{\dJ}\right)\,G = (\dJ|\dpsi)\,G$ holds otherwise.

Until Section~\ref{secExist}, for the sake of simplicity we assume that the EL 
functional $E:\,F\rightarrow\FT^{*}$ is multipolynomial, which is defined 
as follows.

\begin{Def}
Let $E:\,F\times\FT\rightarrow\R$ be an EL functional as in Definition\ref{defEL}. 
We say that the EL functional $E$ is \defin{multipolynomial}, 
whenever there exists a reference field $\psi_{0}\in F$, such that 
there exists an element 
$\E_{\psi_{0}}\in\big(A(\F_{(\C)})\big)^{*}\otimes\FT^{*}\subset\Ta(\F_{(\C)}^{*})\otimes\FT^{*}$, 
for which
\setlength{\mathindent}{0.0\mathindentorig}
\begin{eqnarray*}
\forall\,\psi\in F,\;\dpsiT\in\FT:\quad \big(E_{\psi_{0}}(\psi{-}\psi_{0})\,\big\vert\,\dpsiT\big)=\Big(\E_{\psi_{0}}\,\Big\vert\,\big(1,\,\mathop{\otimes}\limits^{1}(\psi{-}\psi_{0}),\,\mathop{\otimes}\limits^{2}(\psi{-}\psi_{0}),\,\dots\big)\otimes\dpsiT\Big)
\end{eqnarray*}
\setlength{\mathindent}{\mathindentorig}$\!\!$
holds. 
(Note that then for all $\psi_{0}\in F$ there exists the corresponding element $\E_{\psi_{0}}$.)
When an element $\dpsiT\in\FT$ is contracted with $\E_{\psi_{0}}$ in its 
last tensorial entry, we will use the notation 
$(\E_{\psi_{0}}\vert\dpsiT)$ to denote the corresponding element of 
$\big(A(\F_{(\C)})\big)^{*}\subset\Ta(\F_{(\C)}^{*})$. 
\label{defMultip}
\end{Def}

Given $(\E_{\psi_{0}}\vert\dpsiT)$ as above, 
it has a corresponding multipolynomial insertion operator over the 
tensor algebra $\T(\F_{(\C)})$, as stated in Remark\ref{remFprop}(\ref{remFpropInsertion}). 
We shall denote that by the symbol $\iotabig_{(\E_{\psi_{0}}\vert\dpsiT)}$.

\begin{Def}
Let $\hbar$ be a fixed real number. 
Let $F$, $\F$, $\FT$ as in Definition\ref{defF}. 
Let $E:\,F\times\FT\rightarrow\R$ as in Definition\ref{defEL}, and assume that it is 
multipolynomial as in Definition\ref{defMultip}. 
Let $A(\F_{(\C)})$ be the field algebra as in Definition\ref{defFieldA}. 
Then, for some fixed reference field $\psi_{0}\in F$ and fixed test field $\dpsiT\in\FT$ the operator
\setlength{\mathindent}{0.0\mathindentorig}
\begin{eqnarray}
 \MDS_{\hbar,\psi_{0},\dpsiT}:\quad A(\F_{(\C)})\rightarrow A(\F_{(\C)}) ,\quad G\mapsto \MDS_{\hbar,\psi_{0},\dpsiT}\,G \,:=\, \Big(\,\iotabig_{(\E_{\psi_{0}}\vert\dpsiT)} \;-\; \I\,\hbar\,L_{\dpsiT}\,\Big)\,G
\label{eqMDSop}
\end{eqnarray}
\setlength{\mathindent}{\mathindentorig}$\!\!$
is called the \defin{unregularized master Dyson--Schwinger (MDS) operator}.
We call the below equation the \defin{unregularized master Dyson--Schwinger (MDS) equation}:
\begin{eqnarray}
 \Big. \mathrm{we\;search\;for}\; (\psi_{0},G_{\psi_{0}})\in F\times A(\F_{(\C)}), \;\mathrm{such\;that:} \cr
 \Big. b\,G_{\psi_{0}} = 1, \qquad\mathrm{and}\qquad 
 \forall\;\dpsiT\in\FT:\quad \MDS_{\hbar,\psi_{0},\dpsiT}\,G_{\psi_{0}} = 0.
\label{eqMDS}
\end{eqnarray}
\label{defMDS}
\end{Def}

The MDS formulation of QFT can be though of as a construction, where the 
objects of interest are elements of $F\times A(\F_{(\C)})$, and 
the selection equation for the physically realized 
such elements is the MDS equation. In Section~\ref{secWeakReg} it 
shall be shown that some finetuning (regularization) to this idea is needed, 
as is well known in the QFT literature.

\begin{Def}
Any continuous map $O:\,F\rightarrow\R$ is called an \defin{observable}, similarly as 
in a classical field theory. Given a fixed $\psi_{0}\in F$, we 
use the notation $O_{\psi_{0}}:=O\circ(\mathrm{I}_{\F}+\psi_{0})$, 
which is then a continuous map $O_{\psi_{0}}:\,\F\rightarrow\R$, and one has 
$O(\psi)=O_{\psi_{0}}(\psi{-}\psi_{0})$ for all $\psi\in F$ and observable $O$. 
An observable $O:\,F\rightarrow\R$ is called \defin{multipolynomial observable}, 
whenever for some reference fields $\psi_{0}\in F$, there exists an element 
$\mathbf{O}_{\psi_{0}}\in\Ta(\F^{*})$, such that for all $\psi\in F$, one has 
$O_{\psi_{0}}(\psi{-}\psi_{0})=\Big(\mathbf{O}_{\psi_{0}}\,\Big\vert\,\big(1,\,\mathop{\otimes}\limits^{1}(\psi{-}\psi_{0}),\,\mathop{\otimes}\limits^{2}(\psi{-}\psi_{0}),\,\dots\big)\Big)$. 
(If it holds, it then holds for any $\psi_{0}\in F$.)
\label{defObs}
\end{Def}

\begin{Def}
Given a solution $(\psi_{0},G_{\psi_{0}})\in F\times A(\F_{(\C)})$ of 
the MDS equation, the \defin{(Feynman type) quantum expectation value} of the multipolynomial 
observable $O:\,F\rightarrow\R$ at the solution $(\psi_{0},G_{\psi_{0}})$ is
$\mu_{{}_{(\psi_{0},G_{\psi_{0}})}}(O) := \left(\mathbf{O}_{\psi_{0}}\,\big\vert\,G_{\psi_{0}}\right)$.
\label{defQExp}
\end{Def}

We note that the above construction can be extended also to non-polynomial 
but analytic EL functionals and observables as well. 
For that, however, a stronger topology is needed on the tensor algebra of 
$\F$, which we will address later in Section~\ref{secExist}.

\begin{Exa}
For a scalar $\varphi^{4}$ model over a fixed Minkowski spacetime $\M$, the MDS operator reads as follows. 
Let $\mathrm{v}$ be the affine constant maximal form over $\M$ 
(corresponding to the Lebesgue measure). Denote by $\Box$ the Minkowski wave operator. 
Set $F:=\F:=C^{\infty}(\M,\R)$ and $\FT:=C^{\infty}_{c}(\M,\R)$. 
Then, the EL functional is
$E:\,F\times\FT\rightarrow\R,\,(\psi,\,\dpsiT)\mapsto\int_{\M}\dpsiT\,\Box\psi\,\mathrm{v}+\int_{\M}\dpsiT\,\psi^{3}\,\mathrm{v}$. 
For any fixed test field ${\dpsiT\in\FT}$ the corresponding MDS operator can be expressed as
\setlength{\mathindent}{0.0\mathindentorig}
\begin{eqnarray}
\big(\MDS_{\hbar,\psi_{0},\dpsiT}\,G\big)^{(n)}(x_{1},{\dots},x_{n})=\cr
 \quad \int_{y\in\M}\dpsiT(y)\,\Box_{y}G^{(n+1)}(y,x_{1},{\dots},x_{n})\,\mathrm{v}(y)+\int_{y\in\M}\dpsiT(y)\,G^{(n+3)}(y,y,y,x_{1},{\dots},x_{n})\,\mathrm{v}(y) \cr
 \qquad \Big.-\I\,\hbar\,n\,\tfrac{1}{n!}\textstyle\sum_{\pi\in\Pi_{n}}\dpsiT(x_{\pi(1)})\,G^{(n-1)}(x_{\pi(2)},{\dots},x_{\pi(n)})
\label{eqPhi4}
\end{eqnarray}
\setlength{\mathindent}{\mathindentorig}$\!\!$
at the reference field $\psi_{0}=0$ (for all $G\in A(\F_{(\C)})\,{=}\bigvee(\F\otimes\C)$ and $n\in\N_{0}$ and $x_{1},{\dots},x_{n}\in\M$, where 
$\Pi_{n}$ denotes the set of permutations of the symbols $1,{\dots},n$).
\label{exaPhi4}
\end{Exa}

\begin{Exa}
For a pure Yang--Mills model (possibly non-abelian, i.e.\ self-interacting) over a fixed spacetime 
$(\M,g_{ab})$, the MDS operator reads as follows (Penrose abstract indices ${}^{abc\dots}$ and 
${}_{abc\dots}$ are used for tangent tensors and their duals, respectively). 
Let $\mathrm{v}$ be the canonical volume form associated to the 
spacetime metric $g_{ab}$. Let $F$ denote the affine space of covariant 
derivation operators over some vector bundle $V(\M)$ with some given structure 
group $\mathcal{G}$ (internal or gauge group). Then, any two covariant 
derivations $\nabla,\nabla'\in F$ has a difference tensor (Yang--Mills potential) 
$A:=\nabla'-\nabla$ residing in the space of smooth sections of $T^{*}(\M){\otimes}V(\M){\otimes}V^{*}(\M)$, 
denoted by $\F$. Let $\FT$ denote the space compactly supported sections from $\F$. 
Then, the EL functional is 
$E:\,F\times\FT\rightarrow\R,\,(\nabla,\,A_{T})\mapsto\int_{\M}A_{T}{}_{d}\cdot\left(-\tilde{\nabla}_{e}\,\mathrm{v}\,g^{ec}g^{db}P(\nabla)_{cb}\right)$, 
the symbol $P(\nabla)_{ab}$ denoting the curvature tensor of $\nabla$ and 
$\cdot$ denoting the pointwise trace form on the sections of $V(\M){\otimes}V^{*}(\M)$, 
whereas $\tilde{\nabla}$ denoting an extension of $\nabla$ to 
the mixed tensor algebra of $V(\M)$ and $T(\M)$ with an arbitrary torsion-free 
covariant derivation on $T(\M)$. (The pertinent differential operator expression 
involving $\tilde{\nabla}$ is known to be uniquely defined, see also Remark~\ref{remDiv} in \ref{secEL}.) 
Specially, fix a covariant derivation $\nabla\in F$ as a reference field, then 
the EL functional with respect to this reference field $\nabla$ reads as 
$E_{\nabla}:\,\F\times\FT\rightarrow\R,\,(A,\,A_{T})\mapsto\int_{\M}A_{T}{}_{d}\cdot\left(-\tilde{\nabla}_{e}\,\mathrm{v}\,g^{ec}g^{db}\,(\tilde{\nabla}_{c}A_{b}-\tilde{\nabla}_{b}A_{c}+[A_{c},A_{b}])\right)$. 
Given a test field variation $A_{T}\in\FT$, the corresponding MDS operator is
\setlength{\mathindent}{0.0\mathindentorig}
\begin{eqnarray}
\big(\MDS_{\hbar,\nabla,A_{T}}\,G\big)^{(n)}(x_{1},{\dots},x_{n})_{a_{1}{\dots}a_{n}}{}^{\alpha_{1}{\dots}\alpha_{n}}=\cr
 \quad \int_{y\in\M} \Big( -A_{T}(y)_{d}{}^{\gamma}\,K(y)_{\gamma\beta}\,\tilde{\nabla}^{y}_{e} \,\mathrm{v}(y)\,g(y)^{ec} \,g(y)^{db}\,\tilde{\nabla}^{y}_{c} \,G^{(n+1)}(y,x_{1},{\dots},x_{n})_{ba_{1}{\dots}a_{n}}{}^{\beta\alpha_{1}{\dots}\alpha_{n}} \cr
 \Big.\qquad\quad\;\; + A_{T}(y)_{d}{}^{\gamma}\,K(y)_{\gamma\beta}\,\tilde{\nabla}^{y}_{e} \,\mathrm{v}(y)\,g(y)^{ec} \,g(y)^{db}\,\tilde{\nabla}^{y}_{b} \,G^{(n+1)}(y,x_{1},{\dots},x_{n})_{ca_{1}{\dots}a_{n}}{}^{\beta\alpha_{1}{\dots}\alpha_{n}} \cr
 \Big.\quad - A_{T}(y)_{d}{}^{\gamma}\,K(y)_{\gamma\beta}\,\tilde{\nabla}^{y}_{e} \,\mathrm{v}(y)\,g(y)^{ec} \,g(y)^{db} \,G^{(n+2)}(y,y,x_{1},{\dots},x_{n})_{[cb]a_{1}{\dots}a_{n}}{}^{\delta\,\varepsilon\,\alpha_{1}{\dots}\alpha_{n}}\,C(y)_{\delta\,\varepsilon}^{\beta} \Big) \cr
 \quad \Big.-\I\,\hbar\,n\,\tfrac{1}{n!}\textstyle\sum_{\pi\in\Pi_{n}}A_{T}(x_{\pi(1)})_{a_{\pi(1)}}{}^{\alpha_{\pi(1)}}\,G^{(n-1)}(x_{\pi(2)},{\dots},x_{\pi(n)})_{a_{\pi(2)}{\dots}a_{\pi(n)}}{}^{\alpha_{\pi(2)}{\dots}\alpha_{\pi(n)}}
\label{eqYM}
\end{eqnarray}
\setlength{\mathindent}{\mathindentorig}$\!\!$
for all $G\in A(\F_{(\C)})\,{=}\bigvee(\F\otimes\C)$ and $n\in\N_{0}$ and $x_{1},{\dots},x_{n}\in\M$, where 
$\Pi_{n}$ denotes the set of permutations of the symbols $1,{\dots},n$, 
and the Penrose abstract indices ${}^{\alpha\,\beta\,\gamma\,\delta\,\varepsilon\dots}$ were used 
for the Lie algebra of the structure group $\mathcal{G}$, with $K_{\alpha\beta}$ 
denoting the index notation of the trace form, and $C_{\alpha\beta}^{\gamma}$ 
denoting the index notation of the commutator.
\end{Exa}

\section{The weak (distributional) and the Wilsonian regularized MDS operator}
\label{secWeakReg}

\begin{Def}
Let $F,\,\F,\,\FT,\,A(\F_{(\C)}),\,E$ be as in Definition\ref{defMDS}.
We call the EL functional $E:\,F\times\FT\rightarrow\R$ to be 
\defin{free} or \defin{non-interacting}, whenever the corresponding 
continuous map $E:\,F\rightarrow\FT^{*}$ is affine. 
We call the Euler--Lagrangian functional \defin{interacting} otherwise.
(Note that by construction, for a free EL functional, given any reference field $\psi_{0}\in F$, 
the map $E_{\psi_{0}}(\cdot)-E_{\psi_{0}}(0):\,\F\rightarrow\FT^{*}$ is linear. 
If in addition, $\psi_{0}$ were an EL solution, then $E_{\psi_{0}}(\cdot):\,\F\rightarrow\FT^{*}$ is linear.)
\label{defFree}
\end{Def}

\begin{Rem}
It is seen that if $(\psi_{0},\,G_{\psi_{0}})\in F\times A(\F_{(\C)})$ were a solution to the 
unregularized MDS equation Eq.(\ref{eqMDS}), and the reference field $\psi_{0}\in F$ is 
an EL solution, and $E$ is non-interacting, then 
$\iotabig_{(\E_{\psi_{0}}\vert\dpsiT)}\,G_{\psi_{0}}^{(2)}=\I\,\hbar\,\dpsiT$ 
holds for all test fields $\dpsiT\in\FT$.
\label{remFree}
\end{Rem}

\begin{Col}
Let the EL functional $E$ be the one of the free wave or Klein--Gordon equation 
over Minkowski spacetime. In that case, the solution space of the unregularized 
MDS equation Eq.(\ref{eqMDS}) is empty, whenever $\hbar\neq 0$.
\end{Col}

The above is rather evident by means of Remark\ref{remFree}: 
the correlator $G_{\psi_{0}}^{(2)}$ would need to be proportional 
to a fundamental solution (Green's functional), which does not sit 
in the space of smooth correlators $\F{\otimes}\F$, but is at best a distribution. 
It is thus tempting to extend the definition of the MDS equation 
in the weak (distributional) sense, so that free theories can have MDS solutions. 
In order to define the distributional sense fields, 
one needs to use the information that the EL functional 
$E:\,F\times\FT\rightarrow\R$ is actually that of a concrete classical field theory. 
Namely, that $F$ is the space of smooth sections of an affine bundle, 
$\F$ is the space of smooth sections of its subordinate vector bundle, and 
$\FT$ is the space of compactly supported smooth sections of that vector bundle.

\begin{Rem}
In order to define the weak MDS operator, we will need to substitute 
$A(\F_{(\C)})$ with its distributional version, which is expanded below.
\begin{enumerate}[(i)]
 \item Assume that $F$ is the space of smooth sections of an affine bundle over 
the base manifold $\M$, with subordinate vector bundle $U(\M)$, whose smooth 
sections span the space $\F$. 
Take the densitised dual of that vector bundle, ${U\!^{\times}\!(\M):=U^{*}(\M)\otimes\mathop{\wedge}\limits^{\dim(\M)}T^{*}(\M)}$, 
and denote the space of its smooth sections by $\F^{\times}$ correspondingly. 
Take the $n$-fold external tensor product bundle ${U\!^{\times}\!(\M)\boxtimes\dots\boxtimes U\!^{\times}\!(\M)}$ of that, which will then be 
a vector bundle over the $n$-fold cartesian product ${\M\times\dots\times\M}$ as base manifold. 
The space of smooth sections of this vector bundle shall be denoted by 
$\Fx_{n}$, which has its natural $\EE$ topology which is nuclear Fr\'echet (NF), 
and is topologically isomorphic to $\mathop{\otimes}\limits^{n}\F^{\times}$ 
by means of Schwartz kernel theorem. 
It has the subspace of compactly supported sections, 
denoted by $\FxT{}_{n}$ and is a dense subspace within $\Fx_{n}$ in the $\EE$ topology. 
The space $\FxT{}_{n}$ with its natural $\DD$ topology becomes a 
countable strict inductive limit of nuclear Fr\'echet spaces with 
closed adjacent images (LNF space) whenever the base manifold $\M$ is 
noncompact, and is nuclear Fr\'echet (NF) if $\M$ is compact 
(see \cite{Laszlo2022s}-Remark\ref{S-remNuclear}). The strong dual of the space 
$\FxT{}_{n}$ is denoted by $(\FxT{}_{n}\big)^{*}$ with 
its natural $\DD^{*}$ topology. It is a DLNF space when $\M$ is noncompact, 
and $DNF$ when $\M$ is compact. One has that 
$\mathop{\otimes}\limits^{n}\F\subset(\FxT{}_{n}\big)^{*}$, i.e.\ the latter 
space can be regarded as the space of distributional $n$-field correlators.
 \item In the above construction we avoided using completed topological 
tensor product $\mathop{\otimes}\limits^{n}\FxT$, as that space is topologically 
not isomorphic to $\FxT{}_{n}$ whenever we are in the realm of LNF spaces, 
i.e.\ when $\M$ is noncompact (although they are isomorphic as linear spaces, 
the latter has a stronger topology). 
This slight complication is mentioned in more details in \cite{Laszlo2022s}-Remark\ref{S-remJoint}(\ref{S-remJointDD}). 
The pertinent issue is absent, whenever $\M$ is compact 
($\FxT{}_{n}\equiv\mathop{\otimes}\limits^{n}\FxT$ topologically, in that case, i.e.\ one does not need to distinguish them on compact manifolds).
 \item One can form the algebraic tensor algebra 
$\Ta(\FxT)$, defined as the algebraic direct sum $\mathop{\oplus}\limits_{n=0}^{\infty}\FxT{}_{n}$ 
equipped with the locally convex direct sum topology. Its topology will be 
LNF whenever $\M$ is noncompact, and NF if $\M$ is compact. 
$\Ta(\FxT)$ forms a unital associative algebra, 
with (at least) separately continuous multiplication.
 \item The tensor algebra of distributional field variations $\T\big((\FxT)^{*}\big)$ is defined to be the 
space $\big(\Ta(\FxT)\big)^{*}$. It is topologically isomorphic 
to $\mathop{\bigoplus}\limits_{n=0}^{\infty}(\FxT{}_{n})^{*}$, by means of 
\cite{Laszlo2022s}-Remark\ref{S-remCartesian}. It is a DLNF space when $\M$ is noncompact, and 
DNF space if $\M$ is compact. It is also a unital associative algebra, with an 
(at least) separately continuous algebra multiplication.
 \item The distributional graded-symmetrized field algebra 
$A\big((\FxT{}_{(\C)})^{*}\big)\subset\T\big((\FxT{}_{(\C)})^{*}\big)$ can be defined in the analogy of Definition\ref{defFieldA}. 
Clearly, the smooth field algebra $A(\F_{(\C)})$, is dense in 
the distributional sense field algebra $A\big((\FxT{}_{(\C)})^{*}\big)$.
\end{enumerate}
\label{remFTprop}
\end{Rem}

\begin{Rem}
The MDS operator of a non-interacting EL functional can be naturally extended in the distributional sense, as follows.
\begin{enumerate}[(i)]
 \item A continuous linear operator $A:\,\F\rightarrow\F$ is said to possess 
a \defin{formal transpose}, if there exsists a continuous linear 
operator $A^{t}:\,\FxT\rightarrow\FxT$, such that for all 
$\dpsi\in\F$ and $p_{{}_{T}}\in\FxT$ one has that
$\int_{\M} (A\,\dpsi)\,p_{{}_{T}}=\int_{\M} \dpsi\,(A^{t}\,p_{{}_{T}})$, 
with $\M$ being the underlying manifold. 
The topological transpose ${\big(A^{t}\big)^{*}:\,\big(\FxT\big)^{*}\rightarrow\big(\FxT\big)^{*}}$ of the 
formal transpose operator is called the \defin{distributional extension} of $A$.
 \item The notion of formal transpose can be generalized to operators 
$A:\,\T(\F)\rightarrow\T(\F)$, being of the type $A^{t}:\,\Ta(\FxT)\rightarrow\Ta(\FxT)$, and 
$\big(A^{t}\big)^{*}:\,\big(\Ta(\FxT)\big)^{*}\rightarrow\big(\Ta(\FxT)\big)^{*}$ being the distributional extension of $A$.
 \item One may note that for all $\dpsiT\in\FT$ and 
$G\in A(\F_{(\C)})$ and $p\in\Ta(\FxT)$ one has that 
$(p\,\vert\,L_{\dpsiT}G)=(\iotabig_{\dpsiT}p\,\vert\,G)$. 
Moreover, the linear map $\iotabig_{\dpsiT}:\Ta(\FxT)\rightarrow\Ta(\FxT)$ is 
continuous. Therefore, $\iotabig_{\dpsiT}$ is the formal transpose of $L_{\dpsiT}$. 
Consequently, the operator $L_{\dpsiT}$ admits a distributional extension 
$A((\FxT{}_{(\C)})^{*})\rightarrow A((\FxT{}_{(\C)})^{*})$, being the 
topological transpose of $\iotabig_{\dpsiT}$.
 \item Whenever $E$ is the EL functional of a non-interacting 
classical field theory, and $\psi_{0}\in F$ is fixed, then for each $\dpsiT\in\FT$ there exists a 
unique element $\pi_{{}_{T}}\in\FxT$, such that 
$(E_{\psi_{0}}(\dpsi)\vert\dpsiT) - (E(\psi_{0})\vert\dpsiT)=\int_{\M}\pi_{{}_{T}}\,\dpsi$ for 
all $\dpsi\in\F$ (see also \ref{secEL}). Therefore, one has that 
$\iotabig_{(\E_{\psi_{0}}\vert\dpsiT)-(E(\psi_{0})\vert\dpsiT)}=\iotabig_{\int_{\M}\pi_{{}_{T}}\,(\cdot)}$. Because of 
that, for any $G\in A(\F_{(\C)})$ and $p\in\T_{a}(\FxT)$ one has the identity 
$\big(p\,\big\vert\,\iotabig_{(\E_{\psi_{0}}\vert\dpsiT)-(E(\psi_{0})\vert\dpsiT)}\,G\big)=\big(L_{\pi_{{}_{T}}}\,p\,\big\vert\,G\big)$, 
i.e.\ the formal transpose of $\iotabig_{(\E_{\psi_{0}}\vert\dpsiT)}$ exists, being 
the continuous linear map $L_{\pi_{{}_{T}}}+(E(\psi_{0})\vert\dpsiT)\,I:\,\Ta(\FxT)\rightarrow\Ta(\FxT)$. 
Consequently, the operator $\iotabig_{(\E_{\psi_{0}}\vert\dpsiT)}$ admits a 
distributional extension $A((\FxT{}_{(\C)})^{*})\rightarrow A((\FxT{}_{(\C)})^{*})$.
 \item The above construction clearly fails for interacting classical field theories, 
since then the formal transpose of $\iotabig_{(\E_{\psi_{0}}\vert\dpsiT)}$ as a continuous 
linear map $\Ta(\FxT)\rightarrow\Ta(\FxT)$ cannot be defined. See e.g.\ the interaction 
term in Eq.(\ref{eqPhi4}) as an example.
 \item \label{remDistrvi} Let $E:\,F\times\FT\rightarrow\R$ be the EL functional 
of a classical field theory, and $J\in\FT^{*}$, then we call an element 
$\KK_{J}\in F$ a \defin{solution with a source $J$} whenever 
$\forall\dpsiT\in\FT:\;(E(\KK_{J})\,\vert\,\dpsiT)=(J\vert\dpsiT)$ holds. 
Specially, one may consider only $J\in\FxT\subset\Fx\subset\FT^{*}$. If
$\KK:\,\FxT\rightarrow F$ is a continous map, such that for all 
$J\in\FxT$ the field $\KK(J)\in F$ is a solution with a source $J$, then 
$\KK$ is called a \defin{fundamental solution}. (It may or may not exist, 
and if exists, it is typically not unique.)
\end{enumerate}
\label{remDistr}
\end{Rem}

\begin{Def}
Let $F,\,\F,\,\FT,\,A(\F_{(\C)}),\,E,\,\hbar$ be as in Definition\ref{defMDS}, and let 
$E:\,F\times\FT\rightarrow\R$ the EL functional of a non-interacting classical 
field theory as in the Definition\ref{defFree}. 
Fix a reference field $\psi_{0}\in F$. 
Then, by means of Remark\ref{remDistr}, for all $\dpsiT\in\FT$, the MDS 
operator can be extended as a continuous linear operator
${\MDS_{\hbar,\psi_{0},\dpsiT}:\,A((\FxT{}_{(\C)})^{*})\rightarrow A((\FxT{}_{(\C)})^{*})}$, 
called to be the \defin{weak or distributional master Dyson--Schwinger (MDS) operator}.
We call the equation
\begin{eqnarray}
 \Big. \mathrm{we\;search\;for}\; G_{\psi_{0}}\in A((\FxT{}_{(\C)})^{*}), \;\mathrm{\;such\;that:} \cr
 \Big. b\,G_{\psi_{0}} = 1, \qquad\mathrm{and}\qquad 
 \forall\;\dpsiT\in\FT:\quad \MDS_{\hbar,\psi_{0},\dpsiT}\,G_{\psi_{0}} = 0.
\label{eqMDSw}
\end{eqnarray}
the \defin{weak or distributional master Dyson--Schwinger (MDS) equation}.
\label{defMDSw}
\end{Def}

\begin{Rem}
With the above notations, assume that the EL equation admits a fundamental solution 
$\KK:\,\FxT\rightarrow F,\,J\mapsto\KK(J)$ as in Remark\ref{remDistr}(\ref{remDistrvi}). 
Then, $E_{\psi_{0}}:\,\F\rightarrow\FT^{*}$ also has a 
corresponding fundamental solution $\KK_{\psi_{0}}:\,\FxT\rightarrow\F,\,J\mapsto\KK_{\psi_{0}}(J):=\KK(J){-}\psi_{0}$. 
Let $\psi_{0}$ be an EL solution, in which case $E_{\psi_{0}}:\,\F\rightarrow\FT^{*}$ becomes linear, and 
assume that $\KK_{\psi_{0}}$ can be chosen to be linear. 
Such a linear fundamental solution $\KK_{\psi_{0}}:\,\FxT\rightarrow\F$ can be naturally considered as an element 
$\KK_{\psi_{0}}^{(2)}\in\mathcal{L}(\FxT,\F)\subset(\FxT{}_{2})^{*}$. 
Assume moreover, that $\KK_{\psi_{0}}^{(2)}$ can be chosen to be invariant to the permutation symmetry of the field algebra. 
(E.g.\ for a wave or Klein--Gordon equation over Minkowski spacetime, the Feynman propagator would be such.)
\begin{enumerate}[(i)]
 \item Given these conditions, one may define the element 
$K_{\psi_{0}}:=(0,\,0,\,\I\,\hbar\,\KK_{\psi_{0}}^{(2)},\,0,\,0,{\dots})\in A\big((\FxT{}_{(\C)})^{*}\big)$, 
called to be the connected correlator, and one can 
take the ansatz $G_{\psi_{0}}:=\exp(K_{\psi_{0}})\in A\big((\FxT{}_{(\C)})^{*}\big)$. 
Then, $(\psi_{0},G_{\psi_{0}})\in F\times A\big((\FxT{}_{(\C)})^{*}\big)$ 
solves the weak (distributional sense) MDS equation Eq.(\ref{eqMDSw}).
 \item For the bosonic case, the above statement is seen trivially, by the fact that for all 
$\dpsiT\in\FT$ the insertion operator $\iotabig_{(\E_{\psi_{0}}\vert\dpsiT)}$ is 
an algebra derivation, and the field algebra is commutative, so one can use 
the formula for the derivative of exponential. If $\F$ has fermionic flavor 
sectors as well, then one can still trace the problem back to derivations 
acting on exponential: whenever $\dpsiT\in\FT$ 
resides in a single flavor sector, then for all $\dpsiT'$ from the same 
flavor sector, the linear map $L_{\dpsiT'}\,\iotabig_{(\E_{\psi_{0}}\vert\dpsiT)}$ 
is also an algebra derivation.
 \item Rather evidently, the above do not necessarily exhaust all the possible solutions. 
Typically, a fundamental solution $\KK_{\psi_{0}}$ satisfying the above is 
not unique. Moreover, one may add any term $\delta\!K_{\psi_{0}}$ to 
$K_{\psi_{0}}$ satisfying $b\,\delta\!K_{\psi_{0}}=0$ and 
$\forall\dpsiT\in\FT:\;\iotabig_{(\E_{\psi_{0}}\vert\dpsiT)}\,\delta\!K_{\psi_{0}}=0$, 
in which case $\exp(K_{\psi_{0}}+\delta\!K_{\psi_{0}})$ will still solve the weak MDS equation. 
In usual QFTs, these ambiquities are removed by further invariance requirements on $G_{\psi_{0}}$, 
which are not dealt with in the present paper.
 \item The existence of the assumed type of fundamental solution 
is guaranteed for any EL functional over an affine base manifold, 
whenever $E_{\psi_{0}}$ corresponds to a linear PDE with a multipolynomial 
differential operator, having constant coefficients. 
This is ensured by the celebrated Malgrange--Ehrenpreis theorem 
(\cite{Malgrange1956}~ChapitreI.1~Theor\'eme1 and \cite{Ehrenpreis1954}~Chapter6~Theorem10).
\end{enumerate}
\label{remAmb}
\end{Rem}

\begin{Col}
Let the EL functional $E$ be the one of a non-interacting classical field 
theory, which has a fundamental solution as in Remark\ref{remAmb}.
In that case, the solution space of the weak (distributional) MDS 
equation Eq.(\ref{eqMDSw}) is \defin{not} empty.
\label{corMDSw}
\end{Col}

According to Remark\ref{remAmb}, given 
a reference field $\psi_{0}\in F$, the solutions of the weak MDS equation 
Eq.(\ref{eqMDSw}) are not unique for free EL functionals. 
In the usual QFT constructions, this ambiquity is removed by additional 
requirements, such as Poincar\'e invariance of the solutions. 
In the presented construction, however such auxiliary conditions are not imposed, 
since in a generally covariant setting, it is not evident that the 
vacuum state should be required to be unique or not.

\begin{Rem}
For interacting models, the followings can be stated.
\begin{enumerate}[(i)]
 \item \label{remResolS} On one hand, there is a negative result: for a generic non-interacting 
EL functional, the unregularized MDS equation Eq.(\ref{eqMDS}) has no solutions.
 \item \label{remResolW} On the other hand, there is a positive result: for a generic non-interacting 
EL functional of a classical field theory having appropriate fundamental solution, the weak MDS equation Eq.(\ref{eqMDSw}) does have 
solutions, just as is the common wisdom in heuristic QFT.
 \item \label{remResolPhi4} For interacting models, the weak MDS operator cannot be an everywhere defined continuous 
operator acting on the space of distributional correlators. For instance, in a $\varphi^{4}$ 
model over Minkowski spacetime, the interacting part of 
the MDS operator does not have a formal transpose, as seen in Eq.(\ref{eqPhi4}).
This phenomenon occurs because the diagonal evaluation map 
of smooth functions ${\big((x,y)\mapsto G(x,y)\big)\longmapsto\big(z\mapsto G(z,z)\big)}$ 
$x,y,z\in\M$ cannot be extended to the distributions, in general.
 \item In order to remedy the above problem, one is tempted to view the 
everywhere defined continuous bilinear operator $\MDS_{\hbar,\psi_{0}}:\,A(\F_{(\C)})\times\FT\rightarrow A(\F_{(\C)})$ 
as a densely defined bilinear operator 
$\MDShat_{\hbar,\psi_{0}}:\,A((\FxT{}_{(\C)})^{*})\times\FT\rightarrowtail A((\FxT{}_{(\C)})^{*})$, 
via the natural dense linear embedding $A(\F_{(\C)})\subset A((\FxT{}_{(\C)})^{*})$ 
of the function sense correlators to the distributional sense correlators. 
Then, one is tempted to take its maximally extended operator, 
understood by its sequential closure. That is, a distributonal correlator 
would be in the domain of the extended $\MDShat_{\hbar,\psi_{0}}$, whenever 
it admits a function sense approximating sequence converging to it in the 
distributional sense, such that the evaluated $\MDS_{\hbar,\psi_{0}}$ on 
the approximator sequence is convergent in the distributional sense. The 
operator would be closable, whenever any two such approximator sequence of 
the same domain element yielded the same result. 
This strategy is made impossible by the fact that for all interacting 
EL functionals one can show that the MDS operator is not sequentially closable. 
(This occurs because the above diagonal evaluation map is so-called maximally 
non-closable, see \cite{Laszlo2022s}-Remark\ref{S-remClosure} for more details.)
 \item The celebrated H\"ormander's criterion \cite{Hormander1990} on the 
wave front set gives a sufficient condition for diagonal evaluation of 
multivariate distributions, but that condition is not applicable for the 
present problem. (E.g.\ already the wave front set of a solution to the distributional 
MDS equation generated from the Minkowski wave or Klein--Gordon equation is known to fail 
H\"ormander's sufficiency criterion, see \cite{Bar2009} Chapter 4 and \cite{Brouder2014}.)
 \item One can prove that the solution space of the unregularized MDS equation 
(understood over the smooth correlators) is always empty, regardless of the structure 
of the underlying base manifold $\M$ and the interactions in the EL functional 
(we plan to detail the proof in a different paper).
\end{enumerate}
\label{remResol}
\end{Rem}

In summary, the problem is that for interacting models, only the function sense 
MDS operator is well defined, but its solution space is always empty. 
In the non-interacting case, the MDS operator can be extended in the distributional 
sense, and its solution space has the right properties. However, the distributional 
extension of the MDS operator cannot be achieved for interacting models. In order 
to overcome this difficulty, one needs the regularized MDS operator, introduced below.

\begin{Def}
Let $F,\,\F,\,\FT,\,A(\F_{(\C)}),\,E,\,\hbar$ be as in Definition\ref{defMDS}. 
Fix a continuous linear operator $\RR:\,\F\rightarrow\F$. 
Given these, we call the operator
\setlength{\mathindent}{0.0\mathindentorig}
\begin{eqnarray}
 \MDS_{\hbar,\psi_{0},\RR,\dpsiT}:\; A(\F_{(\C)})\rightarrow A(\F_{(\C)}) ,\; G\mapsto \MDS_{\hbar,\psi_{0},\RR,\dpsiT}\,G := \Big(\,\iotabig_{(\E_{\psi_{0}}\vert\dpsiT)} - \I\,\hbar\,L_{\RR\,\dpsiT}\,\Big)\,G
\label{eqMDSopr}
\end{eqnarray}
\setlength{\mathindent}{\mathindentorig}$\!\!$
the \defin{$\RR$-regularized master Dyson--Schwinger (MDS) operator}. Moreover, we call
\begin{eqnarray}
 \Big. \mathrm{we\;search\;for}\; (\psi_{0},G_{\psi_{0}})\in F\times A(\F_{(\C)}), \;\mathrm{such\;that:} \cr
 \Big. b\,G_{\psi_{0}} = 1, \quad\mathrm{and}\quad
 \forall\;\dpsiT\in\FT:\; \MDS_{\hbar,\psi_{0},\RR,\dpsiT}\,G_{\psi_{0}} = 0
\label{eqMDSr}
\end{eqnarray}
the \defin{$\RR$-regularized master Dyson--Schwinger (MDS) equation}.
\label{defMDSr}
\end{Def}

The above definition is motivated by the Wilsonian regularization, 
heuristically stated in Eq.(\ref{eqAMDS3W}). If the base manifold $\M$ were 
an affine space, in order to achieve a Wilsonian regularization (UV frequency damping), 
the regularizer operator $\RR$ should be chosen as the convolution operator 
by a test function on $\M$. It is not difficult to see that in such a setting, 
for the non-interacting case, the Wilsonian regularized MDS equation 
Eq.(\ref{eqMDSr}) does have solutions in the space of smooth field correlators. 
Thus, Definition\ref{defMDSr} is expected to make sense also for interacting 
theories, since there is no problem with the diagonal evaluation map 
on the space of multivariate smooth functions. In order to adapt 
this construction to generic, non-affine manifolds $\M$, we invoke a notion 
of generalized convolution on smooth manifolds, see also \cite{Hormander2007,Shubin2001,Radzikowski1996}.

\begin{Rem}
The following terminologies are standard in the theory of pseudodifferential operators \cite{Hormander2007,Shubin2001,Radzikowski1996}, 
and generalizes the notion of convolution to manifolds.
\begin{enumerate}[(i)]
 \item A continuous linear map $C:\,{(\Fx)^{*}{\rightarrow}\F}$ is called a \defin{smoothing operator}, 
their space is denoted by $\Psi^{-\infty}$ in the literature. By Schwartz kernel theorem, 
such an operator can be identified with an element $\kappa\in\F{\otimes}\Fx$, 
i.e.\ $\kappa$ is a smooth section of the vector bundle $U(\M)\boxtimes U^{{\times}}(\M)$ 
over the base $\M{\times}\M$, with $\F$ being the space of smooth sections of 
$U(\M)$. This is emphasized by writing $C_{\kappa}$ instead, where $\big(C_{\kappa}\,\dpsiT\big)(x)=\int_{y\in\M}\kappa(x,y)\,\dpsiT(y)$ for all 
$\dpsiT\in\FT\subset(\Fx)^{*}$ and for all $x\in\M$.
 \item A smoothing operator $C_{\kappa}$ is called \defin{properly supported}, whenever 
the canonical projections from $\supp(\kappa)\subset\M{\times}\M$ onto each factor $\M$ 
is proper, i.e.\ the inverse images of compact sets are compact. In other words, 
for all compact subsets $\K\subset\M$ the closure of the sets 
${\{(x,y)\in\M{\times}\M\,\vert\,x\in\K,\;\kappa(x,y)\neq 0\}}$ and 
${\{(x,y)\in\M{\times}\M\,\vert\,y\in\K,\;\kappa(x,y)\neq 0\}}$ are compact. 
In that case, the map $C_{\kappa}$ can act as continuous linear maps 
$\FT\rightarrow\FT$, $\F\rightarrow\F$, $(\Fx)^{*}\rightarrow(\Fx)^{*}$, $(\FxT)^{*}\rightarrow(\FxT)^{*}$.
\end{enumerate}
\label{remPropSup}
\end{Rem}

\begin{Def}
Let $C_{\kappa}$ be a properly supported smoothing operator as in Remark\ref{remPropSup}. 
If it preserves the flavor sectors, then $\kappa$ is called a \defin{mollifying kernel}.
\label{defkappa}
\end{Def}

\begin{Rem}
A special example can shed some light on the role of $\kappa$ in Definition\ref{defkappa}.
Let $\M$ be a finite dimensional real affine space (``Minkowski spacetime''), 
the subordinate finite dimensional real vector space denoted by $T$ (``tangent space''). 
Let the vector bundle of fields be trivial, and trivialized compatibly with 
the affine structure. In that case, the fields, i.e.\ the elements of $\F$ are 
simply smooth functions from $\M$ to a finite dimensional real vector space, 
in which the classical fields take their values. 
Let us denote the identity operator of that finite dimensional real vector space 
by $I$. 
Due to the affineness of $\M$, up to a positive multiplier there exists 
a unique positive volume form field $\mathrm{v}$ which is parallel against the 
affine parallel transport (this corresponds to the Lebesgue measure). 
Take a compactly supported $C^{\infty}$ real valued scalar field 
$\rho:\,T\rightarrow\R$. Then, the field 
$(x,y)\mapsto\kappa(x,y):=\rho(x{-}y)\,\mathrm{v}(y)\,I$ is called a convolution kernel, 
and defines a mollifying kernel. 
For any element $\dpsi\in\F$ one has then that 
$C_{\kappa}\,\dpsi=\rho\star\dpsi$, i.e.\ $C_{\kappa}$ is the convolution 
operator by $\rho$. Similarly, for any element $p\in\Fx$ one has that 
$C_{\kappa}^{t}\,p=\rho^{t}\star p$, where $\rho^{t}$ is the reflected $\rho$ 
(for all $v\in T$, $\rho^{t}(v):=\rho({-}v)$). 
Due to the compact support of $\rho$, the $\kappa$ is indeed properly 
supported. Moreover, by construction, it is flavor sector preserving.
\label{remrho}
\end{Rem}

With the notion of mollifying kernel, one can define the Wilsonian regularization 
(UV frequency cutoff) also over generic manifolds. Namely, in Definition\ref{defMDSr}, 
one sets $\RR=C_{\kappa}$ for some mollifying kernel $\kappa$. In that case, 
we use the abbreviation $\MDS_{\hbar,\psi_{0},\kappa,\dpsiT}$ for 
$\MDS_{\hbar,\psi_{0},\RR,\dpsiT}=\MDS_{\hbar,\psi_{0},C_{\kappa},\dpsiT}$. 
It is seen that the regularized MDS equation is the analogy of the 
unregularized MDS equation Eq.(\ref{eqMDS}), but with a smoothing appearing in it.

\begin{Rem}
The following observation is useful for constructing concrete solutions of the 
regularized MDS equation. 
Let $\RR:\,\F\rightarrow\F$ be a continuous linear operator (typically, a smoothing operator $C_{\kappa}$ in our example). 
Then, it can be uniquely extended as a continuous grading preserving algebra derivation $\RR:\,\T(\F)\rightarrow\T(\F)$ 
of the unital associative topological graded algebra $\T(\F)$ via requiring the 
annihilation of unity ($\RR\,\1=0$), the preservation of the space of $n$-tensors ($n\in\N_{0}$), 
the Leibniz rule over tensor product, and coincidence with $\RR$ on the one-vectors. 
If $\RR$ is also preserving flavor sectors of $\F$, then it can be restricted to $A(\F_{(\C)})\subset\T(\F_{(\C)})$ 
as an algebra derivation. 
Similarly, the topological transpose operator $\RR^{*}:\,\F^{*}\rightarrow\F^{*}$ 
extends as a continuous linear operator $\RR^{*}:\,\Ta(\F^{*})\rightarrow\Ta(\F^{*})$. 
Assume moreover, that the 
pertinent operator $\RR$ on $\F$ has a formal transpose $\RR^{t}:\,\FxT\rightarrow\FxT$. 
Then, for the same reason it extends uniquely to $\Ta(\FxT)$ in the above manner, 
and as $(\RR^{t})^{*}$ to $\T((\FxT)^{*})$, and thus also to 
$A((\FxT{}_{(\C)})^{*})\subset\T((\FxT{}_{(\C)})^{*})$, if $\RR$ was flavor sector preserving. 
The operator $(\RR^{t})^{*}$ will not be distinguised in notation from $\RR$, 
since the former is the distributional extension of the latter. Similarly, 
$\RR^{t}$ will in general be denoted by $\RR^{*}$, since the latter is the distributional 
extension of the former.
\label{remDeriv}
\end{Rem}

\begin{Rem}
Use the assumptions of Remark\ref{remAmb}, and let $\kappa$ be a mollifying kernel. 
In that case, the $\kappa$-regularized fundamental solution $\tfrac{1}{2}C_{\kappa}\,\KK_{\psi_{0}}^{(2)}$ resides in $\mathop{\otimes}\limits^{2}\F$, 
and it is compatible with the permutation symmetry of the field algebra $A(\F_{(\C)})$. 
Let the base manifold be affine, and let $\kappa$ be specifically a convolution kernel by 
a symmetric test function. 
Then, $\tfrac{1}{2}C_{\kappa}\,\KK_{\psi_{0}}^{(2)}$ satisfies 
$\forall\dpsiT\in\FT:\;\iotabig_{(\E_{\psi_{0}}|\dpsiT)}\,\tfrac{1}{2}C_{\kappa}\,\KK_{\psi_{0}}^{(2)}=L_{C_{\kappa}\dpsiT}\,\1$. 
Define the element 
$K_{\psi_{0},\kappa}:=(0,\,0,\,\I\,\hbar\,\tfrac{1}{2}C_{\kappa}\,\KK_{\psi_{0}}^{(2)},\,0,\,0,{\dots})\in A(\F_{(\C)})$, 
called to be the smoothed connected correlator. Define the smoothed correlator 
with the ansatz $G_{\psi_{0},\kappa}:=\exp(K_{\psi_{0},\kappa})\in A(\F_{(\C)})$. 
Then, $(\psi_{0},G_{\psi_{0},\kappa})\in F\times A(\F_{(\C)})$ 
solves the $\kappa$-regularized MDS equation Eq.(\ref{eqMDSr}). 
In order to see this, one merely needs to repeat the proof of Remark\ref{remAmb}. 
(The object $\tfrac{1}{2}C_{\kappa}\,\KK_{\psi_{0}}^{(2)}$ e.g.\ for a 
wave or Klein--Gordon model over Minkowski spacetime, would correspond to the smoothed Feynman propagator.)
\label{remExist}
\end{Rem}

It is seen that in the above definition the trick is that although the 
fundamental solution $\KK_{\psi_{0}}^{(2)}\in\mathcal{L}(\FxT,\F)\subset(\FxT{}_{2})^{*}$ is merely 
defined in the distributional sense, but its $\kappa$-regularized version 
$\tfrac{1}{2}C_{\kappa}\,\KK_{\psi_{0}}^{(2)}$ sits in the space of 
smooth field correlators $\F{\otimes}\F$. Therefore, free theories of such kind 
will have smooth solutions of the $\kappa$-regularized MDS equation, and 
one does not need to go to the realm of distributional sense 
MDS equation, which is not applicable to interacting models. 
The replacement of the unregularized MDS operator with the $\kappa$-regularized 
MDS operator is called \emph{regularization}. A further sanity check on 
the presented Wilsonian regularization scheme of the MDS equation 
is the fact that such an equation would always have formal perturbative solutions 
if the field algebra were $\T(\F_{(\C)})$, and the base manifold $\M$ were 
affine. This can be seen to be an immediate consequence of the Malgrange--Ehrenpreis 
surjectivity theorem (\cite{Malgrange1956}~ChapitreI.1~Theor\'eme1 and \cite{Ehrenpreis1954}~Chapter6~Theorem10), 
but will be expanded in a different paper.

Having a rigorous and generally covariant formulation of Wilsonian regularization at hand, 
it is natural to ask the question whether it is possible to formulate Wilsonian 
renormalization using that. The answer is affirmative, and is addressed in \ref{secWR}.

\section{An existence condition for regularized MDS solutions}
\label{secExist}

In this section, we present an existence condition for the solutions of the 
Wilsonian regularized MDS equation.

\begin{Rem}
The followings spell out some facts about the topology of $\F$ and $\FT$.
\begin{enumerate}[(i)]
 \item Assume that the base manifold $\M$ underlying a concrete classical field theory 
is compact (with or without boundary, and if with boundary, we assume the cone condition). This 
is a realistic assumption for conformally invariant models, as for those, the theory 
can be reformulated on the compact manifold with boundary, underlying the 
conformally compactified spacetime.
 \item With the assumption as above, rather obviously, the space $\FT$ shall 
become also an NF space, similarly to $\F$. That is, $\FT$ becomes metrizable 
with all of its benefits: its topology will be sequential, and separetely 
continuous multilinear maps from it will become jointly continuous. 
(Recall that if $\M$ is compact then, either $\FT=\F$, or $\FT$ may be chosen to 
be the closed subspace of $\F$ consisting of fields vanishing at $\partial\M$ 
together with all of their derivatives. Its natural topology will become an $\EE$ 
function topology instead of $\DD$ type.)
 \item \label{remCompactMCH} It is also an elementary fact that over a compact base manifold $\M$, 
the space $\F$ and $\FT$ will not only be nuclear Fr\'echet, but also 
will admit continuous norms instead of merely continuous seminorms. 
By means of \cite{Laszlo2022s}-Remark\ref{S-remCH}, then they become nuclear Fr\'echet spaces 
with a countable increasing system of topology defining Hilbertian norms. 
Since these norms are simply Sobolev norms, they are Gel'fand compatible 
(see \cite{Laszlo2022s}-Remark\ref{S-remCH} and \cite{Laszlo2022s}-Remark\ref{S-remSobolev}), therefore by means of 
\cite{Laszlo2022s}-Remark\ref{S-remCH} they become NF spaces with the countably Hilbert (CH) 
property. Recall that if $\HH$ is a CH type NF space, then there exists a countable family 
\begin{eqnarray*}
 H_{0} \supset H_{1} \supset {\dots} \supset H_{n} \supset {\dots}
\end{eqnarray*}
of topological vector spaces, such that $\HH=\mathop{\bigcap}\limits_{n\in\N_{0}}H_{n}$ 
and is dense in all of the spaces $H_{n}$ ($n\in\N_{0}$), moreover their 
topologies are gradually strictly strengthening
\begin{eqnarray*}
 \tau_{0}\vert_{{}_{\HH}} \subset \tau_{1}\vert_{{}_{\HH}} \subset {\dots} \subset \tau_{n}\vert_{{}_{\HH}} \subset {\dots},
\end{eqnarray*}
and all of their topologies are complete and generated by a Hilbertian 
scalar product (that is, for each $n\in\N_{0}$, the space $H_{n}$ can be taken to be a Hilbert space), 
and for all $n\in\N_{0}$ there exists an integer $m\geq1$ such that the inclusion maps $i_{n+m,n}:\,H_{n+m}\rightarrow H_{n}$ are nuclear. 
(Specially, the spaces $H_{0},\,H_{1},\,{\dots}$ may be chosen such that all adjacent inclusion maps are nuclear.) 
The respective topology generating Hilbertian norms thus form an increasing system
\begin{eqnarray*}
 \Vert\cdot\Vert_{0} \leq C_{0,1}\,\Vert\cdot\Vert_{1} \leq {\dots} \leq C_{n{-}1,n}\,\Vert\cdot\Vert_{n} \leq {\dots} \cr
 \qquad \Big. (\text{for some } C_{0,1},{\dots},C_{n-1,n},{\dots}\in\R^{+}).
\end{eqnarray*}
The corresponding Hilbertian scalar products shall be denoted by
\begin{eqnarray*}
 \left<\cdot,\cdot\right>_{0}, \; \left<\cdot,\cdot\right>_{1}, \; {\dots}, \; \left<\cdot,\cdot\right>_{n}, \; {\dots}.
\end{eqnarray*}
\end{enumerate}
\label{remCompactM}
\end{Rem}

The proposed existence theorem will hinge on the fact that on the field algebra 
$A(\F_{(\C)})$ it is possible to naturally define a reasonable topology, somewhat stronger 
than the Tychonoff topology, such that it preserves the NF property coming from 
$\F$, and if present, the eventual CH property of $\F$ as well.

\begin{Rem}
We recall some findings on topologies of the tensor algebra of $\F$ \cite{Dubin1989a}.
\begin{enumerate}[(i)]
 \item Dubin and Hennings in their work (\cite{Dubin1989a}~Chapter3.1) introduces the notion 
of tensor algebra topology of the following kind. Let $\F$ be a nuclear Fr\'echet space. 
Then, they define the vector space
$\T(\F,\lambda,\mathfrak{p})$, where $\lambda$ is some topological 
subspace of the space of $\N_{0}\rightarrow\C$ sequences (it is a so-called \emph{K\"othe echelon space}), 
and $\mathfrak{p}$ is a family of Hilbertian seminorms on $\F$ defining its NF topology 
(recall that multiple seminorm families $\mathfrak{p}$ can define the same topology on $\F$). 
As a vector space, it is defined as follows:
\begin{eqnarray*}
 \T(\F,\lambda,\mathfrak{p}) & := & \Big\{ G\in\mathop{\bigoplus}\limits_{n=0}^{\infty}\mathop{\otimes}\limits^{n}\F \;\Big\vert\; \forall \Vert{\cdot}\vert\in\mathfrak{p}:\, \big(\Vert G^{(n)}\vert^{{\otimes}n}\big)_{n\in\N_{0}}\in\lambda \Big\},
\end{eqnarray*}
where for all topology defining Hilbertian seminorms $\Vert{\cdot}\vert\in\mathfrak{p}$ on $\F$, 
the symbol $\Vert{\cdot}\vert^{{\otimes}n}$ denotes the $n$-fold cross norm 
over $\mathop{\otimes}\limits^{n}\F$ originating from $\Vert{\cdot}\vert$, 
which is then also a Hilbertian seminorm (\cite{Dubin1989a}~Chapter3.1). 
The locally convex vector topology on $\T(\F,\lambda,\mathfrak{p})$ is defined 
by the system of seminorms 
$\big(G\mapsto\sum_{n=0}^{\infty}\vert u_{n}\vert\;\Vert G^{(n)}\vert^{{\otimes}n}\big)_{u\in\lambda\!^{\times},\Vert{\cdot}\vert\in\mathfrak{p}}$, 
where $\lambda\!^{\times}$ denotes the so-called K\"othe dual (which is, under, 
mild conditions, the strong topological dual) of the sequence space $\lambda$.
 \item Notable K\"othe echelon spaces include (\cite{Dubin1989a}~Chapter2.4):
\setlength{\mathindent}{0.0\mathindentorig}
\begin{eqnarray*}
 \phi\,   & := & \{ (u_{n})_{n\in\N_{0}}\in\C^{\N_{0}} \,\vert\, \exists m\in\N_{0}:\, \forall n>m:\, u_{n}=0 \} \quad \text{(terminating sequences)}, \cr
 h\,      & := & \{ (u_{n})_{n\in\N_{0}}\in\C^{\N_{0}} \,\vert\, \forall m\in\N_{0}:\, \exists C_{m}\in\R^{+}:\, \forall n\in\N_{0}:\, 2^{n\,m}\vert u_{n}\vert\leq C_{m}\}, \cr
 h'       & := & \{ (u_{n})_{n\in\N_{0}}\in\C^{\N_{0}} \,\vert\, \exists c\in\N:\, \exists m_{c}\in\N_{0}:\, \forall n\in\N_{0}:\, \vert u_{n}\vert\leq c\,2^{n\,m_{c}}\}, \cr
 \omega\, & := & \{ (u_{n})_{n\in\N_{0}}\in\C^{\N_{0}} \,\vert\, \text{any }u \} \qquad\qquad\qquad\qquad\qquad \text{(all sequences)}.
\end{eqnarray*}
\setlength{\mathindent}{\mathindentorig}$\!\!$
These are understood with their so-called normal topologies (\cite{Dubin1989a}~Chapter2.1). 
The space $\omega$ is the space of all sequences with the natural Tychonoff topology, 
the space $\phi$ is the space of finitely terminating sequences with the 
natural locally convex direct sum topology, whereas the space $h$ 
is known to be topologically isomorphic to the space $H(\C)$ of entire 
complex functions (\cite{Dubin1989a}~p.978). All of them are Hausdorff locally convex 
topological vector spaces, and the pairs $(\phi,\omega)$, $(h,h')$, $(h',h)$, $(\omega,\phi)$ 
are strong dual to each-other, and K\"othe duals to each-other. 
The spaces $\phi$, $h$, $\omega$ are metrizable, and thus Fr\'echet. 
The spaces $\phi$, $h$, $\omega$ have the so-called ``h'' property, because of which they are nuclear (\cite{Dubin1989a}~Chapter2.5). 
Therefore, $\phi$, $h$, $\omega$ are NF, and $h'$ is DNF. 
Specially, the space $h$ also has the countably Hilbert property.
 \item It is shown in \cite{Dubin1989a}~Chapter3.3 that specially for the sequence spaces 
$\lambda=\phi$ or $\lambda=h$ or $\lambda=\omega$, the tensor algebra 
$\T(\F,\lambda,\mathfrak{p})$ is independent of the choice of the representant 
$\mathfrak{p}$ of the topology defining Hilbertian seminorms on $\F$, thus 
one may write merely $\T(\F,\lambda)$ instead. Moreover, they inherit the 
NF property of $\F$, making $\T(\F,\lambda)$ a unital associative algebra with jointly continuous multiplication, with NF topology.
 \item It is rather easy to see that $\T(\F,\phi)$ is simply the algebraic 
tensor algebra $\Ta(\F)$ with its natural locally convex direct sum topology, 
$\T(\F,\omega)$ is simply the Tychonoff tensor algebra $\T(\F)$ with its natural 
Tychonoff topology.
 \item For $\lambda$ of the above types, it is shown in 
\cite{Dubin1989a}~Chapter3.3 that the topology defining seminorms on $\T(\F,\lambda)$ 
may be chosen to be Hilbertian seminorms
\begin{eqnarray}
 G & \mapsto & \sqrt{\sum_{n=0}^{\infty}\vert u_{n}\vert^{2}\;(\Vert G^{(n)}\vert^{{\otimes}n})^{2}}  \qquad ({u\in\lambda\!^{\times}, \; \Vert{\cdot}\vert\in\mathfrak{p}}),
\label{eqTDubin}
\end{eqnarray}
where $\mathfrak{p}$ is a representant of a topology defining family of 
Hilbertian seminorms on $\F$. From this, it is explicitely seen that 
whenever $\F$ is an NF space admitting a continuous Hilbertian norm, 
then the NF space $\T(\F,h)$ also admits a continuous Hilbertian norm.
 \item The explicit form of a representant of a topology defining countable 
family of increasing Hilbertian seminorms on $\T(\F,h)$, encoding its NF topology, can be given by:
\begin{eqnarray}
 G & \mapsto & \sqrt{\sum_{n=0}^{\infty}2^{m\,n}\;(\Vert G^{(n)}\vert_{m}^{{\otimes}n})^{2}}  \qquad (m\in\N_{0}),
\label{eqTDubinCH}
\end{eqnarray}
where $\Vert{\cdot}\vert_{0},\,\Vert{\cdot}\vert_{1},\,{\dots}$ is a representant of 
a topology defining countable system of increasing Hilbertian seminorms on 
$\F$, defining its NF topology. (The formula is the consequence of 
\cite{Dubin1989a}~Proposition3.7, but is also explicitely used in \cite{Vogt2009}.) 
From the above formula it is seen that whenever $\F$ is of CH type, then 
$\T(\F,h)$ is also of CH type (see also \cite{Laszlo2022s}-Remark\ref{S-remCH}). We will use the abbreviation 
$\Th(\F):=\T(\F,h)$, and will call it the analytic tensor algebra of $\F$, 
since the topology defining sequence space $h$ is isomorphic to the space of 
entire functions $H(\C)$.
 \item \label{remTDubinb} From Eq.(\ref{eqTDubinCH}) it is trivially read off that the counit map 
$b:\,\Th(\F)\rightarrow\R,\,G\mapsto b\,G:=G^{(0)}$ is continuous. 
Therefore, the corresponding projection operator $\1\,b$ onto the scalar 
sector and its complement $I{-}\1\,b$ is also continuous. 
Moreover, the pertinent complementing projection operators $\1\,b$ and 
$I{-}\1\,b$ are orthogonal projections with respect to the representants of 
Hilbertian sesquilinear forms from Eq.(\ref{eqTDubinCH}), 
and $b\,G=\left<\1,G\right>_{m}$ holds for all $G\in\Th(\F)$ and all $m\in\N_{0}$.
 \item From the Eq.(\ref{eqTDubinCH}) form of the Hilbertian seminorms on 
$\Th(\F)$ it is seen that this representant family has the property that 
whenever $\F$ is CH type NF space, then if its representant Hilbertian norm 
family is chosen to be such that the adjacent norms are nuclear against each-other, 
the adjacent Hilbertian norms defined by Eq.(\ref{eqTDubinCH}) are also nuclear 
against each-other. 
Similarly, whenever the adjacent norms on $\F$ are Hilbert--Schmidt 
against each-other, then the adjacent norms Eq.(\ref{eqTDubinCH}) are also 
Hilbert--Schmidt against each-other.
\end{enumerate}
\label{remTDubin}
\end{Rem}

In order to state our existence condition for the solutions of the 
regularized MDS equation, we will need to reconsider the space of 
field correlators to be based on $\Th(\F)$ with the analytic topology, 
and not on $\T(\F)$ with the Tychonoff direct sum topology. The reason is 
that for the construction to work, we need the eventual CH property of $\F$ 
to be inherited by its tensor algebra. Therefore, from this point on, 
the field algebra $A(\F_{(\C)})$ will be defined to be the appropriately 
symmetrized subspace of $\Th(\F_{(\C)})$ instead of $\T(\F_{(\C)})$ 
(see again Remark\ref{remDubin} and Definition\ref{defFieldA} for the 
technical construction of the symmetrized algebra).

\begin{Rem}
It is worth to verify that the tensor algebra topology 
on the new field algebra $A(\F_{(\C)})$, inherited from $\Th(\F_{(\C)})$, is 
not overly strict. For instance, one would like a typical solution of the 
regularized MDS equation for a non-interacting theory to be not excluded from our 
new, smaller field algebra $A(\F_{(\C)})$. By recalling Remark\ref{remExist}, 
one sees from Eq.(\ref{eqTDubinCH}) that the pertinent existent solution of the regularized MDS equation 
for a non-interacting theory indeed resides in the new, stricter field algebra as well.
\end{Rem}

In order to state an existence condition, we shall assume, like in Remark\ref{remCompactM}, 
that the base manifold $\M$ under the concrete theory is compact (with or without boundary, and 
if with boundary, we assume the cone condition, so that Sobolev and Maurin theorems hold, see \cite{Laszlo2022s}-Remark\ref{S-remSobolev}). 
As stated before, this is a realistic assumption in a conformally invariant 
theory, in which case the theory can be re-defined over a compact manifold with 
boundary (the conformal compactification of the would-be-spacetime).

\begin{Rem}
Assume that the base manifold $\M$ of the model is compact and its boundary, 
if not empty, has the cone condition. Then, the followings hold.
\begin{enumerate}[(i)]
 \item With such assumption, $\F$ and $\FT$ 
become countably Hilbert type NF spaces, which is then inherited by 
$\Th(\F_{(\C)})$, and thus by the field algebra $A(\F_{(\C)})$. From now on, 
let us use the abbreviation $\HH:=A(\F_{(\C)})$.
 \item In its original definition, the regularized MDS operator was a separetely continuous bilinear 
map $\MDS_{\psi_{0}}:\,\HH\times\FT\rightarrow\HH,\,(G,\dpsiT)\mapsto\MDS_{\psi_{0},\dpsiT}G$ 
(see also the original definition Eq.(\ref{eqMDSopr}), we suppress $\hbar$ and the fixed mollifying kernel in the notation in this chapter). 
Due to the compactness assumption on $\M$, the space $\FT$ becomes also metrizable, 
therefore the map $\MDS_{\psi_{0}}$ becomes jointly continuous (\cite{Laszlo2022s}-Remark\ref{S-remJoint}). 
Therefore, the regularized MDS operator, 
may be also viewed as a continuous linear map $\MDS_{\psi_{0}}:\,\HH\otimes\FT\rightarrow\HH$.
 \item Due to our compactness assumption on the base manifold $\M$, both $\FT$ and $\HH$ 
became countably Hilbert NF spaces, which technically means that in both spaces 
as well as on their tensor product, the properties Remark\ref{remCompactM}(\ref{remCompactMCH}) hold. 
Denote an associated chain of Hilbert spaces subordinate to $\HH$ by $H_{0},\,H_{1},\,{\dots}$, 
their Hilbertian norms by $\Vert{\cdot}\Vert_{0},\,\Vert{\cdot}\Vert_{1},\,{\dots}$ and 
their Hilbertian scalar products by $\left<\cdot,\cdot\right>_{0},\,\left<\cdot,\cdot\right>_{1},\,{\dots}$. 
Similarly, for $\FT$ denote by $F_{0},\,F_{1},\,{\dots}$ an associated chain 
of Hilbert spaces, their corresponding Hilbertian norms by 
$\Vert{\cdot}\Vert^{F}_{0},\,\Vert{\cdot}\Vert^{F}_{1},\,{\dots}$, and 
their Hilbertian scalar products by 
$\left<\cdot,\cdot\right>^{F}_{0},\,\left<\cdot,\cdot\right>^{F}_{1},\,{\dots}$. 
The associated chain of Hilbert spaces subordinate to $\HH\otimes\FT$ can 
be taken to be the Hilbert--Schmidt tensor product of the spaces 
$H_{m}{\otimesHS}F_{n}$ ($m,n\in\N_{0}$), with their canonical Hilbertian 
cross-norms and crossed Hilbertian scalar products. 
(Eventually, a subfamily of this, with strictly growing norms may also be considered instead.)
 \item \label{remM2nucl} Because of the nuclearity of the spaces $\HH$ and $\FT$, each Hilbertian norm 
in the above chains will have a stronger norm in the chain for which the embedding 
map becomes Hilbert--Schmidt, and eventually becomes nuclear, for large enough 
norms in the chain. 
(This can also be seen less abstractly on our concrete spaces as a consequence 
of the Maurin embedding theorem \cite{Laszlo2022s}-Remark\ref{S-remSobolev}(\ref{S-remSobolevII}).)
 \item The continuity of the linear map 
$\MDS_{\psi_{0}}:\,\HH\otimes\FT\rightarrow\HH$ in terms of these 
Hilbert space chains means that
\setlength{\mathindent}{0.0\mathindentorig}
\begin{eqnarray}
 \Big. \forall k\in\N_{0}:\; \exists m_{k},n_{k}\in\N_{0}:\; \exists C_{k,m_{k},n_{k}}\in\R^{+}:\cr
 \Big. \forall (G,\dpsiT)\in G\times\FT:\quad \Vert\MDS_{\psi_{0}}(G\otimes\dpsiT)\Vert_{k} \leq C_{k,m_{k},n_{k}}\, \Vert G\Vert_{m_{k}}\, \Vert \dpsiT\Vert_{n_{k}}
\end{eqnarray}
\setlength{\mathindent}{\mathindentorig}$\!\!$
holds. Since the norms were ordered, the above identity implies that once it 
holds, it holds for all $m\geq m_{k}$ and $n\geq n_{k}$ as well with some 
constants $C_{k,m,n}\in\R^{+}$. That is, the map $\MDS_{\psi_{0}}$ 
is a continuous linear map $(\HH{\otimes}\FT)\cap(H_{m}{\otimesHS}F_{n})\rightarrow \HH\cap H_{k}$ for 
large enough indices $m,n\in\N_{0}$, given the index $k\in\N_{0}$. The continuous 
extension of the map $\MDS_{\psi_{0}}$ will be denoted by the same symbol for brevity, 
and it is then a continuous linear map $\MDS_{\psi_{0}}:\,H_{m}{\otimesHS}F_{n}\rightarrow H_{k}$, for 
such indices.
 \item By means of (\ref{remM2nucl}), between distant enough 
indices, the inclusion maps $H_{m}\supset{\dots}\supset H_{m'}$ ($m<m'$) and $F_{n}\supset{\dots}\supset F_{n'}$ ($n<n'$) 
become Hilbert--Schmidt, and eventually become nuclear. Therefore, 
given $k\in\N_{0}$, for large enough indices $m,n\in\N_{0}$ the map 
$\MDS_{\psi_{0}}:\,H_{m}{\otimesHS}F_{n}\rightarrow H_{k}$ becomes 
Hilbert--Schmidt, and eventually becomes nuclear. (In concrete spaces, Maurin 
embedding theorem gives the concrete index bounds, see \cite{Laszlo2022s}-Remark\ref{S-remSobolev}(\ref{S-remSobolevII}).)
 \item As a particular case of the above statement, for all large enough indices 
$m,n\in\N_{0}$ one has that the linear map 
$\MDS_{\psi_{0}}:\,H_{m}{\otimesHS}F_{n}\rightarrow H_{0}$ 
is Hilbert--Schmidt. The adjoint of this map $\MDS_{\psi_{0}}^{\dagger}:\,H_{0}\rightarrow H_{m}{\otimesHS}F_{n}$ 
is then also Hilbert--Schmidt. Therefore, the operator 
$\MDS_{\psi_{0}}^{\dagger}\MDS_{\psi_{0}}:\,H_{m}{\otimesHS}F_{n}\rightarrow H_{m}{\otimesHS}F_{n}$, 
becomes a positive nuclear (trace class) operator.
 \item Fix a complete orthonormal basis $(e_{i})_{i\in I}$ in $F_{n}$ (since $F_{n}$ 
is separable, one may set $I\equiv\N$). Then, for all 
$G\in H_{m}$ the estimate
\begin{eqnarray}
 B(G,G) := \sum_{i\in I}\left<(G{\otimesHS}e_{i}),\,\MDS_{\psi_{0}}^{\dagger}\MDS_{\psi_{0}}(G{\otimesHS}e_{i})\right>_{H_{m}{\otimesHS}F_{n}} & < & \infty
\label{eqHSfiniteness}
\end{eqnarray}
is valid. That is because of the Hilbert--Schmidt property of the map 
$\MDS_{\psi_{0}}:\,H_{m}{\otimesHS}F_{n}\rightarrow H_{0}$. Namely, for some 
(and therefore: for any) complete orthonormal basis $(g_{j})_{j\in J}$ in $H_{m}$, 
one has that 
$\sum_{j\in J}\sum_{i\in I} \Vert\MDS_{\psi_{0}}(g_{j}{\otimesHS}e_{i})\Vert_{H_{m}{\otimesHS}F_{n}}^{2}<\infty$
holds (one may set $J\equiv\N$ as well, due to the separability of $H_{m}$). 
Taking specially an orthonormal basis $(g_{j})_{j\in J}$ in $H_{m}$, such that one 
of its elements is $G/\Vert G\Vert_{H_{m}}$, one infers that indeed the 
estimate Eq.(\ref{eqHSfiniteness}) holds.
 \item Given $G\in H_{m}$, the corresponding 
expression Eq.(\ref{eqHSfiniteness}) is independent of the chosen complete orthonormal 
basis $(e_{i})_{i\in I}$ in $F_{n}$. That is because for a Hilbert--Schmidt 
operator $A$ and an unitary operator $U$ in a Hilbert space, one has that 
the Hilbert--Schmidt norm of $A$ and $U^{\dagger}AU$ is the same.
 \item Due to the Hilbert--Schmidt property of $\MDS_{\psi_{0}}:\,H_{m}{\otimesHS}F_{n}\rightarrow H_{0}$, 
the quadratic form $H_{m}\rightarrow\C,\,G\mapsto B(G,G)$ is continuous, and 
therefore by the polarization formula it gives rise to a corresponding 
continuous sesquilinear form $H_{m}\times H_{m}\rightarrow\C,\,(G_{1},G_{2})\mapsto B(G_{1},G_{2})$. 
Therefore, by Riesz representation theorem, there is a corresponding 
unique continuous linear map $\MDShat^{2}_{\psi_{0}}:\,H_{m}\rightarrow H_{m}$, 
such that for all $G_{1},G_{2}\in H_{m}$, the identity $B(G_{1},G_{2})=\big<G_{1},\MDShat^{2}_{\psi_{0}}G_{2}\big>_{H_{m}}$ holds.
 \item Due to the positive semidefiniteness of $B$, the map 
$\MDShat^{2}_{\psi_{0}}$ is a positive operator. Moreover, due to the Hilbert--Schmidt 
property of $\MDS_{\psi_{0}}:\,H_{m}{\otimesHS}F_{n}\rightarrow H_{0}$, the 
map $\MDShat^{2}_{\psi_{0}}$ is nuclear (trace class). The nuclear (trace) norm 
of $\MDShat^{2}_{\psi_{0}}:\,H_{m}\rightarrow H_{m}$, by construction, equals to the Hilbert--Schmidt 
norm of $\MDS_{\psi_{0}}:\,H_{m}{\otimesHS}F_{n}\rightarrow H_{0}$. 
One can see that the operator $\MDShat^{2}_{\psi_{0}}$ is simply the absolute 
value squared version of the MDS operator, with its $\FT$ variable traced out. 
 \item \label{remM2Ker} It is obvious from the construction that 
$\mathop{\bigcap}\limits_{\dpsiT\in\FT}\Ker(\MDS_{\psi_{0},\dpsiT})=\Ker(\MDShat^{2}_{\psi_{0}})$.
\end{enumerate}
\label{remM2}
\end{Rem}

If the reference field 
$\psi_{0}\in F$ was chosen to be such that it satisfies the EL 
equation Eq.(\ref{eqELc}), then one has that 
$\MDS_{\psi_{0},\dpsiT}\,\1=-\I\,L_{\dpsiT}\,\1$ ($\forall\dpsiT\in\FT$). 
Because of that, in this situation, 
$b\,\MDShat_{\psi_{0}}^{2}\,\1=\big<\1,\MDShat_{\psi_{0}}^{2}\1\big>_{m} >0$, and therefore $\1\not\in\Ker(\MDShat_{\psi_{0}}^{2})$. 
Thus, generally, the trivial correlator $\1$ cannot be a solution of the regularized MDS equation. 
One could still aim to find a projection of $\1$ which (up to normalization) 
satisfies the regularized MDS equation. 
Let us denote the orthogonal projection onto $\Ker(\MDShat_{\psi_{0}}^{2})$ in $H_{m}$ by $P$. 
Then, $\Ker(\MDShat_{\psi_{0}}^{2})=\Ran(P)$. One can state the following theorem on $P\1$.

\begin{Thm}
Let $P$ denote the orthoprojection in $H_{m}$ onto $\Ker(\MDShat_{\psi_{0}}^{2})$. 
Then, the following statements are equivalent.
\begin{enumerate}[(i)]
 \item \label{thmP1i}   The solution space of the regularized MDS equation in $H_{m}$ is not empty.
 \item \label{thmP1ii}  One has that $P\1 \neq 0$.
 \item \label{thmP1iii} One has that $b\,P\1 \neq 0$.
\end{enumerate}
\label{thmP1}
\end{Thm}
\begin{Prf}
By construction, the MDS equation has solutions in $H_{m}$ if and only if 
$b\,\Ker(\MDShat_{\psi_{0}}^{2})\neq\{0\}$, i.e.\ if and only if there exists some 
$G\in H_{m}$, such that $b\,P\,G\neq 0$. (That is because $\Ker(\MDShat_{\psi_{0}}^{2})=\Ran(P)$ 
and because Remark\ref{remM2}(\ref{remM2Ker}).) However, the identity 
$b\,P\,G \;=\; \left<\1,P\,G\right>_{m} \;=\; \left<P\1,G\right>_{m}$ holds, because 
of Remark\ref{remTDubin}(\ref{remTDubinb}), and because $P$ was orthoprojection in $H_{m}$. 
Therefore, (\ref{thmP1i}) $\Leftrightarrow$ (\ref{thmP1ii}).

Moreover, one has that $\left<P\1,P\1\right>_{m}\;=\;\left<\1,P\1\right>_{m}\;=\;b\,P\1$, 
since $P$ was an orthoprojection in $H_{m}$, and because of Remark\ref{remTDubin}(\ref{remTDubinb}). 
Therefore, (\ref{thmP1ii}) $\Leftrightarrow$ (\ref{thmP1iii}).
\end{Prf}

It is seen that the orthoprojection $P$ in $H_{m}$ onto $\Ker(\MDShat_{\psi_{0}}^{2})$ 
plays an important role in the problematics of existence of MDS solutions. 
One can approximate $P$ as below.

\begin{Thm}
For all $T>0$ parameter, which is not smaller than the operator 
norm of $\MDShat_{\psi_{0}}^{2}$, and with the 
notation $\mathcal{P}:=I-T^{-1}\,\MDShat_{\psi_{0}}^{2}$, the 
operator sequence $k\mapsto\mathcal{P}^{k}$ converges strongly (pointwise) 
to $P$ in $H_{m}$.
\label{thmPapprox}
\end{Thm}
\begin{Prf}
The operator $\mathcal{P}=I-T^{-1}\,\MDShat_{\psi_{0}}^{2}$ is a positive continuous operator with 
spectrum in $[0,1]$. Therefore, $k\mapsto\mathcal{P}^{k}$ is a monotonically 
decreasing sequence of such operators, bounded from below by the zero 
operator. Therefore the sequence $k\mapsto\mathcal{P}^{k}$ 
converges strongly (pointwise). Since it converges strongly, it converges 
also weakly (i.e.\ matrix element-wise), and its weak limit equals to the strong limit. 
We evaluate its strong limit via evaluating its weak limit, below.

Take any $f,g\in H_{m}$, then there exists a unique complex valued bounded variation 
Radon measure $\mu_{\mathcal{P},f,g}$ over $\C$ with $\mathrm{supp}(\mu_{\mathcal{P},f,g})\subset\mathrm{Sp}(\mathcal{P})\subset[0,1]$, 
and $\left<f,\mathcal{P}g\right>_{m}=\int\limits_{\lambda\in[0,1]}\lambda\,\mathrm{d}\mu_{\mathcal{P},f,g}(\lambda)$ holds. 
Moreover, for any non-negative integer $k$, one has that
$\left<f,\mathcal{P}^{k}g\right>_{m}=\int\limits_{\lambda\in[0,1]}\lambda^{k}\,\mathrm{d}\mu_{\mathcal{P},f,g}(\lambda)$ 
holds. One has that 
$\int\limits_{\lambda\in[0,1]}\lambda^{k}\,\mathrm{d}\mu_{\mathcal{P},f,g}(\lambda)=\int\limits_{\lambda\in[0,1[}\lambda^{k}\,\mathrm{d}\mu_{\mathcal{P},f,g}(\lambda)+\int\limits_{\lambda\in\{1\}}\lambda^{k}\,\mathrm{d}\mu_{\mathcal{P},f,g}(\lambda)$, 
where the second term equals to $\left<f,P\,g\right>_{m}$ by construction. 
The function $\lambda\mapsto\lambda^{k}$ converges to zero pointwise on 
$[0,1[$, and is bounded by the constant $1$ function which is $\mu_{\mathcal{P},f,g}$ 
absolute integrable on $[0,1[$. Therefore, by Lebesgue's theorem of dominated convergence, 
the integral $\int\limits_{\lambda\in[0,1[}\lambda^{k}\,\mathrm{d}\mu_{\mathcal{P},f,g}(\lambda)$ 
tends to zero as a function of $k$. Therefore, $\left<f,\mathcal{P}^{k}g\right>_{m}$ 
converges to $\left<f,P\,g\right>_{m}$ in $k$, i.e.\ $\mathcal{P}^{k}$ converges 
weakly to $P$.
\end{Prf}

By combining Theorem\ref{thmP1} and Theorem\ref{thmPapprox}, one can draw the following conclusion.

\begin{Col}
For all $T>0$ parameter, which is not smaller than the operator norm of $\MDShat_{\psi_{0}}^{2}$, 
one has that the iteration
\begin{eqnarray}
G_{0}   :=  \1, \qquad
G_{k+1} :=  G_{k}-T^{-1}\,\MDShat_{\psi_{0}}^{2}\,G_{k}
\end{eqnarray}
converges in $H_{m}$. Therefore, there exists the finite real number $\nu:=\lim\limits_{k\rightarrow\infty}b\,G_{k}\in\R$.\newline
The MDS equation has solutions in $H_{m}$ if and only if $\nu\neq 0$.\newline
Moreover, if $\nu\neq 0$, then $\frac{1}{\nu}\,\lim\limits_{k\rightarrow\infty}G_{k}$ 
is a solution of the MDS equation in $H_{m}$.\newline
\label{colGk}
\end{Col}
\begin{Prf}
Clearly, by construction we have that $G_{k}=\mathcal{P}^{k}\,\1$ for all $k\in\N_{0}$, 
and we have just shown in Theorem\ref{thmPapprox}, that $k\mapsto\mathcal{P}^{k}$ converges strongly to $P$. 
Therefore, $\lim\limits_{k\rightarrow\infty}G_{k}=P\,\1$, and $\nu=b\,P\,\1$. 
Applying then Theorem\ref{thmP1}, we get the stated result.
\end{Prf}

\begin{Rem}
The following identities are useful for technical evaluation.
\begin{enumerate}[(i)]
 \item The minimal factor $T>0$, which can be used in the above existence test 
is the operator norm of $\MDShat^{2}_{\psi_{0}}$. The operator norms are generally 
hard to estimate. However, it can be estimated from above by the trace norm 
of $\MDShat^{2}_{\psi_{0}}$, or equivalently, by the Hilbert--Schmidt norm of 
$\MDS_{\psi_{0}}$, which are technically easier to evaluate.
 \item It is a useful fact that the indicator sequence $k\mapsto\nu_{k}:=b\,G_{k}\in\R$ 
consists of non-negative numbers, and is monotonically decreasing. 
That is because $\nu_{k}=b\,G_{k}=\left<\1,G_{k}\right>_{m}=\left<\1,\mathcal{P}^{k}\1\right>_{m}$. 
The operator sequence $k\mapsto\mathcal{P}^{k}$ consists of a sequence of positive 
operators, which are monotonically decreasing. Therefore $k\mapsto\nu_{k}$ 
inherits this property. Thus, it is enough to test whether the indicator sequence 
$k\mapsto\nu_{k}$ is bounded away from zero. 
Moreover, the scalar component $b\,G_{k}$ of the approximants $G_{k}$ start from 
$1$, they do stay real, and they do not flip sign from positive to negative, 
and they monotonically decrease.
 \item By means of Corollary\ref{colGk}, for concrete models, evaluating 
whether the indicator $\nu$ is bounded away from zero, is expected to involve 
elaborate Sobolev estimates. The corollary, however, pinpoints a well defined 
point where one has to invoke these estimates, and therefore this can be 
considered as a useful existence test condition.
 \item Since Corollary\ref{colGk} is a necessary and sufficient condition, and not 
merely a sufficient condition, one may also use it in the reverse direction. 
Namely, if for a concrete model the regularized MDS equation 
had any solutions, then the iteration scheme of Corollary\ref{colGk} is guaranteed 
to be good enough to be convergent, and to produce one particular MDS solution. 
This is a useful piece of information, even without actually performing the 
above Sobolev estimates.
\end{enumerate}
\end{Rem}

\section{Concluding remarks}
\label{secConclusion}

In the QFT literature, the master Dyson--Schwinger (MDS) equation on the field 
correlators is known to be a differential reformulation of the Feynman integral formalism. 
In this paper it is shown that the MDS equation can be cast into a particular 
presentation, in which the involved function spaces and operators are 
perfectly well defined, regardless of a fixed background spacetime metric, 
or causal structure, or signature. Moreover, the Wilsonian regularized version 
of the construction is also shown to be well defined in such a generally 
covariant setting. A necessary and sufficient condition is proved for the 
solution space of the regularized MDS equation to be nonempty, for conformally 
invariant Lagrangians. The pertinent theorem is constructive in the sense that 
it provides an iterative algorithm to obtain an MDS solution. The algorithm is 
guaranteed to converge whenever the solution space is nonempty, and could be 
eventually used for a lattice QFT-like nonperturbative numerical solution scheme, 
capable of working in the original metric signature.

\section*{Acknowledgments}

We would like to thank to the organizers of the Simplicity III workshop 
at the Perimeter Institute, and especially to Neil Turok and Job Feldbrugge 
for the inspiring exposition on the problematics of 
Feynman integral formulation in Lorentz signature, which inspired this work. 
We would also like to thank to Antal Jakov\'ac for enlightening discussion 
on Feyman integral formulation and ERGE from the physical point of view. 
We would also like to thank \'Aron Szab\'o, Bence Racsk\'o and Igor Khavkine for valuable 
feedback on the mathematical content of the manuscript. 
Special thanks to J\'anos Krist\'of for the enlightening mathematical 
inputs concerning the theory of topological vector spaces (TVS) and measure 
theory on them, moreover to Zsigmond Tarcsay for double-checking the 
mathematical content of the paper concerning the theory of TVS, 
especially regarding the questions of operator closability.

This work was supported in part by the Hungarian Scientific Research fund 
(NKFIH K-138152).

\appendix

\section{Continuity properties of the Euler--Lagrange functional}
\label{secEL}

Our presentation of the master Dyson--Schwinger (MDS) operator heavily relies 
on the precise definition of the Euler--Lagrange functional of a classical field 
theory. In order to pin down the topological properties of the involved spaces 
and precise continuity property of their operators, we need to briefly recall 
the standard variational formulation of a classical field theory. 
For the sake of simplified treatment, we will use neither the jet formalism, 
nor the theory of general connections over fiber bundles \cite{Sardanashvily2009,Cohen2017}. 
We will rather concentrate on the TVS theory side \cite{Laszlo2022s}, i.e.\ we 
keep the differential geometric treatment to a reasonable appropriate minimum. 
Let $\M$ denote throughout the paper a finite dimensional real smooth orientable and oriented manifold, and let $m:=\dim(\M)$. It may be compact or noncompact, and may be with or 
without boundary (if with boundary, we assume the cone condition for it, 
so that locally the Sobolev and Maurin embedding theorems hold, 
see also \cite{Laszlo2022s}-Section\ref{S-subsecSobolev}). 
The manifold $\M$ is meant to model the spacetime manifold, 
or eventually, the compact manifold with boundary underlying the conformal 
compactification (Penrose diagram) of a spacetime. 
Let $V(\M)$ be some real vector bundle over $\M$ 
with finite dimensional fibers. Denote, as usual, by $\Gamma\big(\cdot\big)$ 
the space of smooth sections. In particular, the space $\Gamma\big(V(\M)\big)$ 
denotes the real vector space of smooth sections of $V(\M)$ 
(these are meant to model the matter fields). 
The covariant derivation operators over $\Gamma\big(V(\M)\big)$ (which are meant to model 
the mediator fields) form an affine space with subordinate vector space $\Gamma\big(T^{*}(\M){\otimes}V(\M){\otimes}V^{*}(\M)\big)$, 
as it is common knowledge. 
More particularly, covariant derivation operators can be considered as sections of an affine bundle over $\M$, 
which we will denote by $\DV(\M)$, and then a covariant derivation is an element 
of $\Gamma\big(\DV(\M)\big)$. 
The bundle $\DV(\M)$ is an affine bundle with subordinate vector bundle 
$T^{*}(\M){\otimes}V(\M){\otimes}V^{*}(\M)$, the sections of which are 
the difference tensors (or gauge potentials, in field theory). 
As usual, the symbols $\otimes$ shall denote pointwise tensor product of vector 
bundles over the same base, whereas $\oplus$ will denote pointwise direct sum 
of vector bundles over the same base. 
The fibered product (pointwise cartesian product, or Whitney sum) of two generic fiber bundle 
over the same base shall be denoted by $V(\M)\timesW U(\M)$, 
and their elements by $(v,u)_{\!\!{}_{W}}$ in order to indicate that 
these are pairs of fields, over the same base points. The subscript ${}_{{}_{W}}$ 
is used in order to distinguish the above from sections 
$(v,u)$ of the cartesian product bundle $V(\M)\times U(\M')$, which would be 
a bundle over the product manifold $\M{\times}\M'$ (eventually $\M'=\M$).

\begin{Def}
Let $\M$, $V(\M)$ and $D\!V(\M)$ as above.

A \defin{Lagrange form} is a base point preserving, smooth fiber bundle homomorphism
\setlength{\mathindent}{0.0\mathindentorig}
\begin{eqnarray*}
 \LL:\; V(\M) \;\oplus\; T^{*}(\M){\otimes}V(\M) \;\oplus\; T^{*}(\M){\wedge}T^{*}(\M){\otimes}V(\M){\otimes}V^{*}(\M) \;\longrightarrow\; \mathop{\wedge}\limits^{m}T^{*}(\M).
\end{eqnarray*}
\setlength{\mathindent}{\mathindentorig}$\!\!$
By construction, a Lagrange form takes some sections
\setlength{\mathindent}{0.0\mathindentorig}
\begin{eqnarray*}
v\in\Gamma\big(V(\M)\big), \quad
D\!v\in\Gamma\big(T^{*}(\M){\otimes}V(\M)\big), \quad
P\in\Gamma\big(T^{*}(\M){\wedge}T^{*}(\M){\otimes}V(\M){\otimes}V^{*}(\M)\big)
\end{eqnarray*}
\setlength{\mathindent}{\mathindentorig}$\!\!$
into a maximal form field $\LL(v,\Dv,P)\in\Gamma\big(\mathop{\wedge}\limits^{m}T^{*}(\M)\big)$.

An element $(v,\nabla)_{\!\!{}_{W}}\in \Gamma\big(V(\M)\timesW \DV(\M)\big)$ 
is called a \defin{field configuration}. The field configurations 
form an affine space over the real vector space 
${\Gamma\big(V(\M)\oplus T^{*}(\M){\otimes}V(\M){\otimes}V^{*}(\M)\big)}$. 
An element $(\dv,\dC)_{\!\!{}_{W}}$ from that space is called a \defin{field variation}.

The map
\begin{eqnarray*}
 \Gamma\big(V(\M) \timesW \DV(\M)\big) \;\longrightarrow\; \Gamma\big(\mathop{\wedge}\limits^{m}T^{*}(\M)\big), \quad (v,\nabla)_{\!\!{}_{W}} \;\longmapsto\; \LL(v,\nabla{v},P(\nabla))
\end{eqnarray*}
is called the \defin{Lagrangian expression}, where $\nabla{v}$ is the covariant 
derivative of the section $v$, 
and $P(\nabla)$ is the curvature tensor of $\nabla$. 
(Note that the expression $(v,\nabla{v},P(\nabla))_{\!\!{}_{W}}$ ecodes the same 
information as the first jet of a field configuration $(v,\nabla)_{\!\!{}_{W}}$, but 
we do not intend to use the jet formalism in the present paper.)

Given a Lagrange 
form $\LL$, its \defin{action functional} is the real Radon measure valued map
\begin{eqnarray*}
 S^{\LL}:\quad \Gamma\big(V(\M) \timesW \DV(\M)\big) \;\longrightarrow\; \Rad(\M,\R), \quad (v,\nabla)_{\!\!{}_{W}} \;\longmapsto\; S^{\LL}(v,\nabla)_{\!\!{}_{W}}
\end{eqnarray*}
where on compact subsets $\K\subset\M$ the definition is $S^{\LL}_{\K}(v,\nabla)_{\!\!{}_{W}}:=\int\limits_{\K}\LL(v,\nabla{v},P(\nabla))$, 
i.e.\ the action functional is the Radon measure defined by local integrals 
of the Lagrangian expression, as usual.
\label{defMVM}
\end{Def}

We use the shorthand notation 
$F:=\Gamma\big(V(\M) \timesW \DV(\M)\big)$ for the space of field 
configurations, moreover $\F:=\Gamma\big(V(\M) \,\oplus\, T^{*}(\M){\otimes}V(\M){\otimes}V^{*}(\M)\big)$ 
for the space of field variations. The space $F$ is an affine space over the 
real vector space $\F$. 
The real vector space $\F$ may be naturally endowed with the standard $\EE$ smooth function topology. 
(The $\EE$ topology is defined by the family of arbitrary order Sobolev norms of over compact patches of $\M$.) 
With this topology $\F$ and thus $F$ become Hausdorff locally convex 
topological vector and affine spaces, respectively. It is also common knowledge 
\cite{Laszlo2022s}-Remark\ref{S-remNuclear}), that $\F$ with the $\EE$ topology 
becomes a \emph{nuclear Fr\'echet space}, which fact will be an important detail in the QFT construction.

The real vector space of real valued Radon measures $\Rad(\M,\R)$ can be 
also naturally endowed with a topology, defined by compact setwise 
total variations as family of seminorms, or equivalently, by the convergence of 
measure sequences over compact sets. With this, $\Rad(\M,\R)$ becomes a 
Hausdorff locally convex topological vector space.

\begin{Rem}
It is not true in general that the continuity of a map between topological spaces 
is equivalent to its sequential continuity. 
It is common knowledge, however, that metrizable topological spaces 
are \emph{sequential} (\cite{Laszlo2022s}-Remark\ref{S-remSequential}), i.e.\ their topology is 
completely characterized by the convergence of sequences. 
Since $\F$ is Fr\'echet space, by construction its topology is metrizable 
in a translationally invariant way, and therefore also is the topology of $F$. 
In particular, a map $S:\,F\rightarrow Y$ to any 
topological space $Y$ is continuous if and only if $S$ is sequentially continuous, 
i.e.\ it maps convergent sequences in $F$ to convergent sequences in $Y$.
\label{remMetrizable}
\end{Rem}

\begin{Rem} The following can be observed.
\begin{enumerate}[(i)]
 \item The action functional was defined to be a Radon measure valued map. That 
was motivated by the fact that no asymptotics was prescribed on the field 
configurations $F$, nor it was assumed that $\M$ is compact. 
Because of that, one cannot guarantee that the 
smooth maximal form field $\LL(v,\nabla{v},P(\nabla))$ is integrable throughout the full $\M$ for 
sufficiently many field configurations $(v,\nabla)_{\!\!{}_{W}}\in F$. It is, however, 
always locally integrable, hence the action functional 
as a Radon measure valued map is meaningful, and everywhere defined.
 \item Due to Lebesgue's theorem of dominated convergence, the action functional 
is sequentially continuous, and therefore by means of Remark\ref{remMetrizable}, 
it is continuous.
\end{enumerate}
\end{Rem}

The action functional is everywhere differentiable in 
the Fr\'echet--Hadamard sense (see also \cite{Laszlo2022s}-Section\ref{S-subsecFH}), as it is 
common knowledge in Lagrangian field theory. In order to show its explicit form, 
we recall some differential geometric identities. 
We will use Penrose abstract indices for the tangent tensors throughout the section.

\begin{Rem}
If $\nabla$ is a covariant derivation over $T(\M)$, then there is a unique 
covariant derivation $\tilde{\nabla}$ over $T(\M)$ associated to it, having 
vanishing torsion tensor and having the same affine parametrized geodesics as $\nabla$. The covariant 
derivation $\tilde{\nabla}$ is called the \emph{torsion-free part} of $\nabla$. 
In explicit formulae: whenever $v^{b}$ is a smooth section of $T(\M)$, then one has
$\tilde{\nabla}_{a}v^{b}=\nabla_{a}v^{b}+\frac{1}{2}T(\nabla)_{ac}^{b}v^{c}$, where 
$T(\nabla)_{ac}^{b}$ denotes the torsion tensor of $\nabla$.
\end{Rem}

\begin{Thm}
The action functional $S^{\LL}$ is everywhere differentiable, and its derivative 
at some fixed $(v,\nabla)_{\!\!{}_{W}}\in F$ is a continuous linear map 
$DS^{\LL}(v,\nabla)_{\!\!{}_{W}}:\,\F\rightarrow\Rad(\M,\R)$, given by the formula
\setlength{\mathindent}{0.0\mathindentorig}
\begin{eqnarray}
 \Big. (\dv,\dC)_{\!\!{}_{W}} \;\mapsto\; \big(DS^{\LL}_{\K}(v,\nabla)_{\!\!{}_{W}}\,\big\vert\,(\dv,\dC)_{\!\!{}_{W}}\big) = \cr
 \int\limits_{\K}\Bigl(D_{1}\LL(v,\nabla{v},P(\nabla))\,\dv+D_{2}^{a}\LL(v,\nabla{v},P(\nabla))\,(\nabla_{a}{\dv}+\dC_{a}v)+2\,D_{3}^{[ab]}\LL(v,\nabla v,P(\nabla))\,\tilde{\nabla}_{[a}\dC_{b]}\Bigr),\cr
\label{eqSderiv}
\end{eqnarray}
\setlength{\mathindent}{\mathindentorig}$\!\!$
when evaluated on some compact subset $\K\subset\M$. 
Here, $D_{1}\LL$, $D_{2}\LL$, $D_{3}\LL$ denote the spacetime pointwise 
partial derivative of $\LL$ against its first, second and third field variable, respectively.
It also follows that the derivative map $DS^{\LL}:\,F\times\F\rightarrow\Rad(\M,\R)$ is jointly continuous in its two variables.
\end{Thm}
\begin{Prf}
This is a simple consequence of the below elementary facts.
\begin{itemize}
 \item The Lagrange form evaluation as a map $(v,D\!v,P)_{\!\!{}_{W}}\mapsto\LL(v,D\!v,P)$ acting on the 
space of sections is continuously differentiable in the $\EE$ topology, 
and the map $(v,\nabla)_{\!\!{}_{W}}\mapsto(v,\nabla{v},P(\nabla))_{\!\!{}_{W}}$ is also continuously 
differentiable in the $\EE$ topology. Therefore, their composition, being 
the Lagrangian expression $(v,\nabla)_{\!\!{}_{W}}\mapsto\LL(v,\nabla{v},P(\nabla))$, 
is also differentiable in the $\EE$ topology, and its derivative is given by the 
integrand of Eq.(\ref{eqSderiv}).
 \item The local integral evaluation of a smooth 
maximal form over a compact subset $\K\subset\M$ is sequentially continuous 
map in the $\EE$ topology due to Lebesgue theorem of 
dominated convergence, and therefore by means of Remark\ref{remMetrizable} 
it is continuous in the $\EE\rightarrow\Rad(\M,\R)$ topologies. 
Due to its linearity then it is differentiable, and its derivative is itself.
 \item Chain rule for the differentiation of composite functions made out of the above two maps implies the first part of the theorem.
 \item Lebesgue's theorem of dominated convergence implies joint sequential 
continuity of $DS^{\LL}$. Therefore, by means of Remark\ref{remMetrizable}, the derivative 
functional is jointly continuous as a $DS^{\LL}:F\times\F\rightarrow\Rad(\M,\R)$ map, 
since $F$, $\F$ and thus their product $F\times\F$ is metrizable. This proves the second statement of the theorem.
\end{itemize}
\end{Prf}

\begin{Rem} Let us also recall the following differential geometric identities.
\begin{enumerate}[(i)]
 \item Let $J^{a}_{[c_{1}{\dots}c_{m}]}$ be a smooth section of 
$T(\M)\otimes \mathop{\wedge}\limits^{m}T^{*}(\M)$, i.e.\ a maximal form valued tangent vector 
field (the symbol ${}_{[\,]}$ denotes index antisymmetrization). 
Then, given any covariant derivation $\nabla$ on $T(\M)$, 
one has that the expression 
$\tilde{\nabla}_{a}J^{a}_{[c_{1}{\dots}c_{m}]}$ is independent of the choice of 
the covariant derivation $\nabla$, where $\tilde{\nabla}$ denotes the torsion-free part 
of $\nabla$. That is, the divergence of a maximal form valued vector field is 
naturally defined without further assumptions. Similarly, for a smooth section 
$K^{[ab]}_{[c_{1}{\dots}c_{m}]}$ of $\big(T(\M){\wedge}T(\M)\big)\otimes \mathop{\wedge}\limits^{m}T^{*}(\M)$ 
one has that $\tilde{\nabla}_{a}K^{[ab]}_{[c_{1}{\dots}c_{m}]}$ is independent 
of the choice of the covariant derivation $\nabla$, and thus the divergence of such field 
is naturally defined without further assumptions.
 \item Let $J^{a}_{[c_{1}{\dots}c_{m}]}$ be a smooth section of 
$T(\M)\otimes \mathop{\wedge}\limits^{m}T^{*}(\M)$, i.e.\ a maximal form valued tangent vector 
field. Then, given any covariant derivation $\nabla$ on $T(\M)$, 
one has that $\tilde{\nabla}_{a}J^{a}_{[c_{1}{\dots}c_{m}]}=m\;\de_{[c_{1}}(J^{a}_{\;a\,c_{2}{\dots}c_{m}]})$, 
where $\de$ denotes exterior differentiation (see \cite{Hawking1973}).
\end{enumerate}
\label{remDiv}
\end{Rem}

\begin{Thm}
The derivative $DS^{\LL}(v,\nabla)_{\!\!{}_{W}}$ of the action functional $S^{\LL}$ at a fixed 
$(v,\nabla)_{\!\!{}_{W}}\in F$ can be re-expressed as
\setlength{\mathindent}{0.0\mathindentorig}
\begin{eqnarray}
 \Bigg. (\dv,\dC)_{\!\!{}_{W}} \mapsto \big(DS^{\LL}_{\K}(v,\nabla)_{\!\!{}_{W}}\,\big\vert\,(\dv,\dC)_{\!\!{}_{W}}\big) = \cr
 \Bigg. \qquad \int\limits_{\K}\Bigl(D_{1}\LL(v,\nabla{v},P(\nabla))_{[c_{1}{\dots}c_{m}]}\,\dv-\big(\tilde{\nabla}_{a}D_{2}^{a}\LL(v,\nabla{v},P(\nabla))_{[c_{1}{\dots}c_{m}]}\big)\,\dv\Bigr)+ \cr
 \Bigg. \qquad\quad\; \Bigl(D_{2}^{a}\LL(v,\nabla{v},P(\nabla))_{[c_{1}{\dots}c_{m}]}\,\dC_{a}v-2\,\big(\tilde{\nabla}_{a}D_{3}^{[ab]}\LL(v,\nabla{v},P(\nabla))_{[c_{1}{\dots}c_{m}]}\big)\,\dC_{b}\Bigr) \cr
 \Bigg. \qquad +\,m\,\int\limits_{\partial{\K}}\Bigl(D_{2}^{a}\LL(v,\nabla{v},P(\nabla))_{[ac_{1}{\dots}c_{m-1}]}\,\dv+2\,D_{3}^{[ab]}\LL(v,\nabla{v},P(\nabla))_{[ac_{1}{\dots}c_{m-1}]}\,\dC_{b}\Bigr),
\label{eqEL}
\end{eqnarray}
\setlength{\mathindent}{\mathindentorig}$\!\!$
when evaluated over some compact subset $\K\subset\M$ with cone property boundary $\partial{\K}$.
\end{Thm}
\begin{Prf}
This can be proved as usual in Lagrangian field theory. Namely, we start out 
from the expression in Eq.(\ref{eqSderiv}), use Leibniz rule, apply 
the differential geometric identities of Remark\ref{remDiv}, 
and then apply Stokes theorem for the boundary term.
\end{Prf}

Let us introduce $\FT\subset\F$ to be either the vector space of compactly supported 
sections from $\F$, or if $\partial\M\neq\emptyset$ optionally they may be even 
required to vanish on $\partial\M$ together 
with all of their derivatives. Elements of $\FT$ will be called the \defin{test field variations}. 
The space $\FT$ can be endowed with the standard $\DD$ test function topology, 
being stronger that the $\EE$ topology, defined by the restricted $\EE$ 
topology for sections with their supports within each fixed compact set of $\M$. 
It is common knowledge (\cite{Laszlo2022s}-Remark\ref{S-remNuclear}), that $\FT$ 
with its natural $\DD$ test function topology is a \emph{strict inductive limit of a countable 
system of nuclear Fr\'echet spaces with closed adjacent images (LNF space)} whenever 
$\M$ is noncompact, and it is \emph{nuclear Fr\'echet (NF space)} if $\M$ is compact. 
These are important details in the QFT construction. 
It is seen that due to Lebesgue's theorem of dominated convergence the 
integrand within the expression $\big(DS^{\LL}_{\M}(v,\nabla)_{\!\!{}_{W}}\,\big\vert\,(\dv_{{}_{T}},\dC_{{}_{T}})_{\!\!{}_{W}}\big)$, 
see again Eq.(\ref{eqSderiv}) and Eq.(\ref{eqEL}), is absolutely integrable for all 
fields $(v,\nabla)_{\!\!{}_{W}}\in F$ and all test field 
variations $(\dv_{{}_{T}},\dC_{{}_{T}})_{\!\!{}_{W}}\in\FT$. 
In other words: the measure $\K\mapsto\big(DS^{\LL}_{\K}(v,\nabla)_{\!\!{}_{W}}\,\big\vert\,(\dv_{{}_{T}},\dC_{{}_{T}})_{\!\!{}_{W}}\big)$ 
has bounded total variation, and thus 
$\big(DS^{\LL}_{\M}(v,\nabla)_{\!\!{}_{W}}\,\big\vert\,(\dv_{{}_{T}},\dC_{{}_{T}})_{\!\!{}_{W}}\big)\in\R$ is finite. 
Consequently, the following definition is meaningful.

\begin{Def}
Let $\M$, $V(\M)$, $\LL$, $S^{\LL}$ as before. 
The map
\begin{eqnarray}
 E^{\LL}:\; F\times\FT\rightarrow\R,\; \big(\psi,\,\dpsiT\big)\mapsto \big(E^{\LL}(\psi)\,\big\vert\,\dpsiT\big) := \big(DS^{\LL}_{\M}(\psi)\,\big\vert\,\dpsiT\big)
\end{eqnarray}
is called the \defin{Euler--Lagrange (EL) functional}.
(Here, we used a shorthand notation $\psi:=(v,\nabla)_{\!\!{}_{W}}\in F$ for a field, 
and $\dpsiT:=(\dv_{{}_{T}},\dC_{{}_{T}})_{\!\!{}_{W}}\in\FT$ for a test field variation.)
\label{defELcl}
\end{Def}

Note that it was possible to define the EL functional as real valued 
at the price of restricting its second argument to compactly supported field variations. 
This setting also explains why one can automatically discard the EL boundary 
terms in classical variational problems over noncompact manifolds without boundary. 
It is clear that for all $\psi\in F$ the map $\big(E^{\LL}(\psi)\,\big\vert\,\cdot\big):\,\FT\rightarrow\R$ 
is well defined. Moreover, it is linear, and continuous in the $\DD$ topology due to Lebesgue's theorem 
of dominated convergence. 
Therefore the EL functional may be viewed either as map $E^{\LL}:\,F\times\FT\rightarrow\R$, 
or alternatively as a distribution valued map $E^{\LL}:\,F\rightarrow\FT^{*}$, 
where ${}^{*}$ denotes the strong dual. 
About their continuity properties, one can state the following.

\begin{Thm}
The EL functional $E^{\LL}:\,F\times\FT\rightarrow\R$, with $F$ and $\FT$ carrying the standard $\EE$ and $\DD$ topologies, respectively, 
has the following continuity properties.
\begin{enumerate}[(i)]
 \item It is jointly sequentially continuous.
 \item It is separately continuous in each variable.
 \item It is continuous as a $E^{\LL}:\,F\rightarrow\FT^{*}$ map.
\end{enumerate}
\label{thmEcont}
\end{Thm}
\begin{Prf}
Property (i) is obviously seen via applying Lebesgue theorem of dominated convergence in the 
joint variables.

To see (ii), take first a fixed $\dpsiT\in\FT$. Then, the 
map $E^{\LL}(\cdot,\dpsiT):\,F\rightarrow\R$ is sequentially continuous by means of (i), 
and due to the metrizability of $F$, by means of 
Remark\ref{remMetrizable}, then it is continuous. Take than a fixed 
$\psi\in F$. The linear map $E^{\LL}(\psi,\cdot):\FT\rightarrow\R$ is sequentially 
continuous by means of (i). Due to the facts in 
\cite{Laszlo2022s}-Remark\ref{S-remSequential}, the space $\FT$ carries the bornological 
property, by means of which the sequentially continuous linear map $E^{\LL}(\psi,\cdot):\FT\rightarrow\R$ is continuous.

To see (iii), observe that due to (i) the map $E^{\LL}:\,F\rightarrow\FT^{*}$ 
is sequentially continuous, whenever $\FT^{*}$ is endowed 
with the weak (pointwise) topology. Due to the facts in 
\cite{Laszlo2022s}-Remark\ref{S-remSequential}, the space $\FT^{*}$ carries the Montel 
property, therefore weakly convergent sequences are also strongly convergent 
in $\FT^{*}$. Thus, the pertinent map is also sequentially continuous when the 
target space $\FT^{*}$ is endowed with its standard strong dual topology ($\DD^{*}$ topology). 
Due to the metrizability of $F$, by means of Remark\ref{remMetrizable}, then it is $\EE\rightarrow\DD^{*}$ continuous.
\end{Prf}

\begin{Def}
A tuple $\left(\M,\,V(\M),\,F,\,\F,\,\FT,\,E,\,\mathcal{C}\right)$ is called a \defin{classical field theory}, 
where $\M$ and $V(\M)$ is as in Definition\ref{defMVM}, 
$F$ is the space of smooth sections of the affine bundle 
$V(\M)\timesW\DV(\M)$, the space $\F$ consists of the smooth sections of the 
subordinate vector bundle $V(\M)\oplus T^{*}(\M){\otimes}V(\M){\otimes}V^{*}(\M)$, 
the space $\FT$ consists of compactly supported sections 
from $\F$ (if $\partial\M\neq\emptyset$, optionally, elements of $\FT$ may be required to vanish on $\partial\M$ together with all of their derivatives --- variation with boundary included/excluded). 
Furthermore, the object $E$ is a map $F\times\FT\rightarrow\R$, such that 
there exists a Lagrange form $\LL$ as in Definition\ref{defMVM}, such that 
$E=E^{\LL}$. Finally, $\mathcal{C}:=\left\{\psi\in F\,\big\vert\,\forall\dpsiT\in\FT:\,(E(\psi)\vert\dpsiT)=0\right\}$. 
The set $\mathcal{C}$ is called the \defin{solution space} of the classical field theory.
\label{defClassf}
\end{Def}

\begin{Def}
Let $\left(\M',\,V'(\M'),\,F',\,\F',\,\FT',\,E',\,\mathcal{C}'\right)$ and $\left(\M,\,V(\M),\,F,\,\F,\,\FT,\,E,\,\mathcal{C}\right)$ 
be two classical field theories. These are called \defin{isomorphic}, if and only 
if there exists a vector bundle isomorphism $V'(\M')\rightarrow V(\M)$ with 
an underlying diffeomorphism $\M'\rightarrow\M$ of the base manifold, such that 
$\LL$ subordinate to $E$ is pulled back to $\LL'$ subordinate to $E'$. 
(Isomorphic classical field theories are postulated 
to describe the same physics.) Quite naturally, isomorphisms of a classical 
field theory with itself are called \defin{automorphisms}, or \defin{symmetries}.

A classical field theory $\left(\M,\,V(\M),\,F,\,\F,\,\FT,\,E,\,\mathcal{C}\right)$ is called 
\defin{generally covariant}, if and only if all the vector bundle automorphisms 
$V(\M)\rightarrow V(\M)$ are automorphisms of the classical field theory.

A classical field theory $\left(\M,\,V(\M),\,F,\,\F,\,\FT,\,E,\,\mathcal{C}\right)$ is called 
\defin{diffeomorphism invariant}, if and only if for all the diffeomorphisms 
$\M\rightarrow\M$ of the base manifold there exists a vector bundle automorphism 
$V(\M)\rightarrow V(\M)$, such that it is an automorphism of the classical field theory.

Those automorphisms of a classical field theory $\left(\M,\,V(\M),\,F,\,\F,\,\FT,\,E,\,\mathcal{C}\right)$, 
for which the underlying $\M\rightarrow\M$ diffeomorphism is the identity of $\M$, 
are called \defin{internal symmetries} or \defin{gauge transformations}.
\end{Def}

\begin{Def}
The \defin{observables} of a classical field theory $\left(\M,\,V(\M),\,F,\,\F,\,\FT,\,E,\,\mathcal{C}\right)$ 
are the continuous maps $O:\,F\rightarrow\R$.
\end{Def}

\begin{Rem}
The presented formulation of a classical Lagrangian field theory formalizes 
the Palatini type variational principle, when applied to a setting 
eventually containing general relativity. That is: the spacetime 
metric field or its ingredients, if present in the theory, is treated just like 
any other of the fields. In particular, it is not assumed a priori that on $T(\M)$ a 
Levi--Civita covariant derivation is present associated to some spacetime 
metric. If a metric and a covariant derivation on $T(\M)$ is present, they 
are varied independently in the presented formulation. We also remark, 
that in this formulation, the Lagrange form of general relativity 
can be chosen to be polynomial in the field variables: one variable can 
be chosen to be the inverse spacetime metric densitised with the metric 
volume form, i.e.\ a field $\mathbf{g}^{ab}_{[cdef]}$ (this is in one-to-one 
correspondence with the ordinary spacetime metric field $g_{ab}$), 
and the other variable can simply be the $T(\M)$ covariant derivation $\nabla_{a}$. 
The Einstein--Hilbert Lagrangian expression is then
$(\mathbf{g}^{ab}_{[cdef]},\nabla_{h})\mapsto \mathbf{g}^{ab}_{[cdef]}\,R(\nabla)_{ahb}{}^{h}$ 
which is a third degree polynomial of its field variables, where $R(\nabla)_{abc}{}^{d}$ denotes the Riemann tensor of $\nabla$.
\end{Rem}

\section{The Wilsonian renormalization}
\label{secWR}

\begin{Rem}
It is rather straightforward to see that the space of mollifying kernels 
form a real vector space, naturally carrying a Hausdorff sequential 
convergence (CVS) structure which is Cauchy complete. A sequence $(\kappa_{n})_{n\in\N}$ 
of mollifying kernels is said to converge to zero iff for all compact sets 
$\K\subset\M$ there exists some compact set 
$\K'\subset\M$, such that for all $n\in\N$ the closure of the sets 
${\{(x,y)\in\M{\times}\M\,\vert\,x\in\K,\;\kappa_{n}(x,y)\neq 0\}}$ and 
${\{(x,y)\in\M{\times}\M\,\vert\,y\in\K,\;\kappa_{n}(x,y)\neq 0\}}$ 
are contained in $\K{\times}\K'$ and $\K'{\times}\K$, respectively, moreover 
the sections $(\kappa_{n})_{n\in\N}$ along with all their polynomial 
derivatives converge uniformly to zero over the compact sets 
$\K{\times}\K'\subset\M{\times}\M$ and $\K'{\times}\K\subset\M{\times}\M$. 
We do not address in the present note whether this convergence structure 
originates from a TVS structure or not, since we do not need it. 
It is also rather easy to see, that whenever the base manifold $\M$ is affine, 
the convolution kernels form a sequentially closed vector subspace within 
the space of all mollifying kernels.
\end{Rem}

\begin{Def}
On the set of mollifying kernels, one may introduce 
a natural, vector bundle automorphism invariant pre-ordering relation. 
Namely, for mollifying kernels 
$\kappa''$ and $\kappa$ we say that $\kappa''\precsim\kappa$ 
(in words: \defin{$\kappa''$ is less ultraviolet than $\kappa$}) iff either 
$C_{\kappa''}=C_{\kappa}$ or there exists some mollifying kernel $\kappa'$ such 
that $C_{\kappa''}=C_{\kappa'}\,C_{\kappa}$ holds. It is evidently seen 
from the construction, that indeed this defines a pre-order, i.e.\ a relation which is transitive 
and reflexive. It is also seen that such relation may be 
also formulated on the set of convolution kernels, whenever convolution 
is meaningful, i.e.\ whenever the base manifold $\M$ is affine 
(and in that case, the pertinent relation is invariant to affine transformations of $\M$).
\label{defOrdering}
\end{Def}

\begin{Thm}
For a real valued smooth compactly supported test function $\varphiT$ over $\M$, 
denote by $M_{\varphiT}$ the multiplication operator by $\varphiT$. 
The pre-order relation $\precsim$, introduced in Definition\ref{defOrdering}, 
when restricted to the set of mollifying kernels $\kappa$ which admit some 
$\varphiT$ such that $C_{\kappa}M_{\varphiT}$ is not finite rank, 
becomes a partial order, i.e.\ it is antisymmetric.
\end{Thm}
\begin{Prf}
Let $\kappa$ and $\kappa'$ be any two mollifying kernels. We need to show that 
$\kappa'\precsim\kappa$ and $\kappa\precsim\kappa'$ implies $\kappa'=\kappa$ under the conditions of the theorem.

Writing out the condition $\kappa'\precsim\kappa$ and $\kappa\precsim\kappa'$ explicitely, 
there exist continuous linear operators $A,B:\,\F\rightarrow\F$, 
such that $C_{\kappa'}=A\,C_{\kappa}$ and $C_{\kappa}=B\,C_{\kappa'}$, where 
$A=I$ or $A=C_{\alpha}$ with some mollifying kernel $\alpha$, and 
$B=I$ or $B=C_{\beta}$ with some mollifying kernel $\beta$. Putting these together, 
they imply $C_{\kappa'}=A\,B\,C_{\kappa'}$ and $C_{\kappa}=B\,A\,C_{\kappa}$. 
Taking any real valued compactly supported smooth test function $\varphiT$ over $\M$, these imply 
$C_{\kappa'}M_{\varphiT}=A\,B\,C_{\kappa'}M_{\varphiT}$ and $C_{\kappa}M_{\varphiT}=B\,A\,C_{\kappa}M_{\varphiT}$. 
Since $\kappa$ and $\kappa'$ was properly supported, then there exists 
some large enough compact region $\K\subset\M$ containing $\supp(\varphiT)$, 
such that the supports of the images of $C_{\kappa}M_{\varphiT}$ and 
$C_{\kappa'}M_{\varphiT}$ are also contained within $\K$. Let $\eta_{{}_{T}}$ 
be a real valued smooth compactly supported test function, which takes the value 
$1$ within this set $\K$. Then, one has
\begin{eqnarray}
C_{\kappa'}M_{\varphiT}=A\,B\,M_{\eta_{{}_{T}}}\,C_{\kappa'}M_{\varphiT}  \;\mathrm{and}\; C_{\kappa}M_{\varphiT}=B\,A\,M_{\eta_{{}_{T}}}\,C_{\kappa}M_{\varphiT}.
\label{eqOrd}
\end{eqnarray}
One can choose an even larger compact region $\K'\subset\M$, which contains $\supp(\eta_{{}_{T}})$ and 
also contains the supports of the images of $A\,B\,M_{\eta_{{}_{T}}}$ and 
$B\,A\,M_{\eta_{{}_{T}}}$. 
Under such conditions, the kernel function of $C_{\kappa'}M_{\varphiT}$ and of 
$C_{\kappa}M_{\varphiT}$ are square integrable, and therefore are Hilbert--Schmidt 
on the space of $L^{2}$ sections over $\K'$, so they are compact operators. 
If any of $A$ or $B$ is not the unit operator, then it is a mollifying operator by 
our assumptions, and then both $A\,B\,M_{\eta_{{}_{T}}}$ and $B\,A\,M_{\eta_{{}_{T}}}$ 
are also Hilbert--Schmidt on the above $L^{2}$ function space over $\K'$, 
for the same above reason, so they are also compact. Eq.(\ref{eqOrd}) implies that 
$\Ran(C_{\kappa'}M_{\varphiT})$ is contained in the eigenspace of 
$A\,B\,M_{\eta_{{}_{T}}}$ with eigenvalue one, and 
$\Ran(C_{\kappa}M_{\varphiT})$ is contained in the eigenspace of 
$B\,A\,M_{\eta_{{}_{T}}}$ with eigenvalue one. But since nonzero eigenvalue 
eigenspaces of compact operators are finite dimensional, 
$\Ran(C_{\kappa'}M_{\varphiT})$ and $\Ran(C_{\kappa}M_{\varphiT})$ must be 
finite dimensional if any of $A$ or $B$ are not the unity operator. 
But it was assumed that $\kappa$ admitted some $\varphiT$ such that 
$\Ran(C_{\kappa}M_{\varphiT})$ is not finite dimensional (and for $\kappa'$ 
the same was assumed with some $\varphiT'$). Therefore, both $A$ and $B$ must be the unity operator, i.e.\ $\kappa'=\kappa$.
\end{Prf}

\begin{Rem}
The pre-ordering $\precsim$ becomes a partial order, under mild conditions.
\begin{enumerate}[(i)]
 \item If for some test function $\varphiT$ the mollifying kernel $\kappa$ is 
such that $C_{\kappa}M_{\varphiT}$ is injective on an infinite dimensional 
linear subspace of the $L^{2}$ sections, then $C_{\kappa}M_{\varphiT}$ is not finite rank. 
That is because in the pertinent case, the $L^{2}$ adjoint of the continuous operator 
$C_{\kappa}M_{\varphiT}$ is evidently non-finite rank, due to which the operator 
itself cannot be finite rank.
 \item If the mollifying kernel $\kappa$ is such that the operator $C_{\kappa}$ 
is injective over the space of test field variations $\FT$, then it satisfies 
the above condition, and thus $C_{\kappa}M_{\varphiT}$ is not finite rank, for any test function $\varphiT$.
 \item If the base manifold $\M$ is affine, then the convolutions are meaningful, 
and the convolution kernels by test functions are such that their $C_{\kappa}$ 
operators are injective over the space of test field variations $\FT$. 
That claim can be verified in Fourier space, using a consequence of the 
Paley--Wiener--Schwartz theorem (\cite{Hormander1990}~Theorem7.3.1), namely the fact that the Fourier 
transform of a compactly supported distribution (and hence, of a function) 
is an analytic function. (Alternatively, it also follows from \cite{Andersson2015}~Theorem4.4.) 
Therefore, if $\kappa$ were a convolution kernel, by means of the above observation, 
$C_{\kappa}M_{\varphiT}$ is not finite rank.
\end{enumerate}
\end{Rem}

The above leads us to the following conclusion.

\begin{Col}
On the set of mollifying kernels which are injective on the space of test 
field variations, the pre-ordering $\precsim$ is antisymmetric, i.e.\ it is 
a partial order.

In particular, when the base manifold is affine, over the set of nonvanishing convolution 
kernels the pre-ordering $\precsim$ is a partial order.

In such cases, we may use the symbol $\preceq$ instead of $\precsim$ for clarity.
\end{Col}

\begin{Rem}
In Section~\ref{secFeyn} it was argued that the Wilsonian regularization 
justifies our regularized MDS equation Eq.(\ref{eqMDSr}). 
Applying the heuristic integral substitution (measure pushforward) formula 
for composite maps in the Wilsonian Feynman integral Eq.(\ref{eqExpHeurW}), it would follow that 
if $(\psi_{0},G_{\psi_{0},\kappa})\in F\times A(\F_{(\C)})$ 
were a solution of the $\kappa$-regularized MDS equation, and $\kappa''\precsim\kappa$, 
then there should exist a solution $(\psi_{0},G_{\psi_{0},\kappa''})\in F\times A(\F_{(\C)})$ of 
the $\kappa''$-regularized MDS equation, such that the identity
\begin{eqnarray}
 H_{C_{\kappa'}} \, G_{\psi_{0},\kappa} & = & G_{\psi_{0},\kappa''}
\label{eqERGE}
\end{eqnarray}
holds, where $\kappa'$ is the corresponding mollifying kernel satisfying 
$C_{\kappa''}=C_{\kappa'}C_{\kappa}$ (because of $\kappa''\precsim\kappa$), 
and $H_{C_{\kappa'}}$ is defined as $\mathop{\otimes}\limits^{n}C_{\kappa'}$ 
on the $n$-vectors of $A(\F_{(\C)})$, and is thus a unital algebra 
homomorphism of $A(\F_{(\C)})$ generated by the 
continuous linear operator $C_{\kappa'}:\,\F\rightarrow\F$. 
This equation is called the exact renormalization equation (ERGE) in the QFT literature, 
and $H_{C_{\kappa'}}$ is called a blocking transformation. It is seen that if 
Feynman integrals existed as a proper finite measure, the ERGE equation would be just 
the consequence of the fundamental formula for integral substitution, for the pushforward measures. 
In our rigorous formalism, defined on the field correlators, one needs to 
impose that by hand, as stated below.
\label{remWilsonRen}
\end{Rem}

\begin{Def}
Let the index set $\mathcal{I}$ be the set of mollifying kernels, and denote by 
$A(\F_{(\C)})^{\mathcal{I}}$ the set of all maps $\mathcal{I}\rightarrow A(\F_{(\C)})$. 
Then, the \defin{solution space of the Wilsonian renormalized MDS equation} is
\setlength{\mathindent}{0.0\mathindentorig}
\begin{eqnarray}
 Q_{r} := \Big\{ (\psi_{0},G_{\psi_{0},\cdot})\in F\times A(\F_{(\C)})^{\mathcal{I}} \,\Big\vert\, \forall\kappa,\kappa''\in\mathcal{I}:\; \kappa''\precsim\kappa \;\mathrm{(with}\; \kappa'\mathrm{)} \;\Rightarrow\; H_{C_{\kappa'}}\,G_{\psi_{0},\kappa}=G_{\psi_{0},\kappa''} \cr
 \quad \;\mathrm{and}\; \forall\kappa\in\mathcal{I}:\; \forall\dpsiT\in\FT:\; b\,G_{\psi_{0},\kappa}=1 ,\; \MDS_{\hbar,\psi_{0},\kappa,\dpsiT}\,G_{\psi_{0},\kappa}=0 \Big\},
\end{eqnarray}
\setlength{\mathindent}{\mathindentorig}$\!\!$
i.e.\ they are the solution families of the regularized MDS equation, 
satisfying the ERGE relation. 
We say that a model is \defin{Wilsonian renormalizable}, if $Q_{r}$ is not empty.
\end{Def}

One may recognize that the solution families satisfying the ERGE relation are 
so-called projective families, and therefore, the solution space of the Wilsonian 
renormalized MDS equation is the corresponding projective limit (\cite{DeJong2021}~Chapter4.21). 
The theory is Wilsonian renormalizable whenever the corresponding projective limit exists as 
a nonempty set.

\begin{Rem}
It is not uncommon in QFT that running coupling factors need to be introduced. 
In that case, it is assumed that the EL function can be specified as a finite sum 
$E=g_{1}\,E_{1}+{\dots}+g_{n}\,E_{n}$, with each $E_{i}:\,F{\times}\FT\rightarrow\R$ being 
jointly sequentially continuous, called the Euler--Lagrange terms, and $g_{i}$ 
being nonzero real numbers, called to be the coupling factors ($i=1,{\dots},n$). 
Recall that the space of mollifying kernels $\mathcal{I}$ was a Hausdorff complete 
sequential convergence vector space, due to which one can define (sequentially) 
continuous functions from $\mathcal{I}$ to other convergence vector spaces. 
Given some (sequentially) continuous functionals 
$\gamma_{i}:\,\mathcal{I}\rightarrow\R$ ($i=1,{\dots},n$), one may define the running regularized 
MDS operator as
\setlength{\mathindent}{0.0\mathindentorig}
\begin{eqnarray}
 \MDS_{\hbar,\psi_{0},\kappa,(\gamma_{1},{\dots},\gamma_{n}),\dpsiT}:\quad A(\F_{(\C)})\rightarrow A(\F_{(\C)}) ,\cr
  G\mapsto \MDS_{\hbar,\psi_{0},\kappa,(\gamma_{1},{\dots},\gamma_{n}),\dpsiT}\,G \,:=\,
  \Big(\,\iotabig_{\gamma_{1}(\kappa)\,(\E_{1,\psi_{0}}\vert\dpsiT)+{\dots}+\gamma_{n}(\kappa)\,(\E_{n,\psi_{0}}\vert\dpsiT)} \;-\; \I\,\hbar\,L_{C_{\kappa}\dpsiT}\,\Big)\,G,
\end{eqnarray}
\setlength{\mathindent}{\mathindentorig}$\!\!$
for fixed $\hbar\in\R$, reference field $\psi_{0}\in F$, test field variation $\dpsiT\in\FT$, 
mollifying kernel $\kappa\in\mathcal{I}$ and running couplings 
$\gamma_{i}$ ($i=1,{\dots},n$). The solution space of the Wilsonian renormalized 
MDS equation with running couplings is then
\setlength{\mathindent}{0.0\mathindentorig}
\begin{eqnarray}
\Big\{ (\psi_{0},(\gamma_{1},{\dots},\gamma_{n}),G_{\psi_{0},\cdot})\in F\times C(\mathcal{I},\R)^{n}\times A(\F_{(\C)})^{\mathcal{I}} \,\Big\vert\, \cr
 \quad \Big. \forall\kappa,\kappa''\in\mathcal{I}:\; \kappa''\precsim\kappa \mathrm{(with}\;\kappa'\mathrm{)} \;\Rightarrow\; H_{C_{\kappa'}}\,G_{\psi_{0},\kappa}=G_{\psi_{0},\kappa''}  \cr
  \qquad \;\mathrm{and}\; \forall\kappa\in\mathcal{I}:\; \forall\dpsiT\in\FT:\; b\,G_{\psi_{0},\kappa}=1 ,\; \MDS_{\hbar,\psi_{0},\kappa,(\gamma_{1},{\dots},\gamma_{n}),\dpsiT}\,G_{\psi_{0},\kappa}=0 \Big\}.
\end{eqnarray}
\setlength{\mathindent}{\mathindentorig}
\end{Rem}

\section*{References}

\bibliographystyle{JHEP}
\bibliography{mds}

\end{document}


\title[Some recalled facts on topological vector spaces]{Some recalled facts on topological vector spaces}

\author{Andr\'as L\'aszl\'o}
\address{Wigner Research Centre for Physics, Budapest}
\ead{laszlo.andras@wigner.hu}

\begin{abstract}
The paper ``On generally covariant mathematical formulation of Feynman integral in Lorentz signature'' 
heavily relies on the theory of non-normable topological vector spaces (TVS). 
Therefore, a supplementary material is dedicated to a recollection of important 
and sometimes counterintuitive theorems on TVS.
\end{abstract}

\noindent{\it Keywords}: Topological vector spaces (TVS)

\tableofcontents

\title[Some recalled facts on topological vector spaces]{(Supplementary material to: On generally covariant mathematical formulation of Feynman integral in Lorentz signature)}

\maketitle


We intend to review here some notions and fundamental results on the theory 
of topological vector spaces. For a concise introduction see \cite{Pietsch1972}~Chapter0,4,5,7 
and also \cite{Rudin1991} as well as \cite{Taylor1995}.

\section{Fr\'echet--Hadamard derivative}
\label{subsecFH}

Let $F$ and $G$ be real topological affine spaces, both with some 
Hausdorff locally convex topology. Let their underlying vector spaces 
denoted by $\F$, $\mathbb{G}$, respectively. A mapping $S:\,F\rightarrow G$ is said 
to be \defin{Fr\'echet--Hadamard differentiable at a point $\psi\in F$} 
\cite{Schechter1984,Montaldi2017}, 
whenever there exists a continuous linear map 
$DS(\psi):\,\F\rightarrow\mathbb{G}$, called the derivative of $S$ at $\psi$, 
such that for any convergent sequence $n\mapsto h_{n}$ in $\F$ and 
any nowhere zero sequence $n\mapsto t_{n}$ in $\R$ which converges to zero, the 
continuous linear map $DS(\psi)$ satisfies 
\begin{eqnarray}
 \lim_{n\rightarrow\infty} \left(\frac{S(\psi+t_{n}\,h_{n})-S(\psi)}{t_{n}} - DS(\psi)\,h_{n}\right) & = & 0.
\end{eqnarray}
It is called Fr\'echet--Hadamard differentiable, if it is so in every point of its domain. 
It is called continuously Fr\'echet--Hadamard differentiable, if it is differentiable, 
and the derivative map 
$F\rightarrow\mathcal{L}(\F,\mathbb{G}),\,\psi\mapsto DS(\psi)$ is 
continuous. Note that in general, in order to make sense of this latter 
definition, one needs to specify a topology on 
the space $\mathcal{L}(\F,\mathbb{G})$ of $\F\rightarrow\mathbb{G}$ 
continuous linear maps (there can be many natural topologies on $\mathcal{L}(\F,\mathbb{G})$, see also the followings).

\section{Fundamentals on topological vector spaces}
\label{subsecTVS}

\begin{Rem}
We recall some basic definitions and facts on topological vector spaces.
\begin{enumerate}[(i)]
 \item A Hausdorff topological vector space is called locally convex whenever 
it has a topological basis consisting of convex neighborhoods. 
One of the many equivalent definitions of a Hausdorff locally convex 
topological vector space (HLCTVS) is that it is a real or complex 
vector space, with a topology defined by a family of seminorms which 
separate points 
(\cite{Rudin1991}~ChapterI.1~Theorem1.36,1.37 and 
\cite{Treves1970}~ChapterI.3,I.4,I.7~PropositionI.7.7 and 
\cite{Horvath1966}~Chapter2.1-4).
 \item \label{remTVSReorg} Without loss of generality, the above HLCTVS topology can always 
be defined with an increasing family of seminorms 
(\cite{Horvath1966}~Chapter2.4 and \cite{Bourles2018}~Remark3.19). 
Moreover, if the family of seminorms is countable, it can be recombined such 
that they are indexed by $\N_{0}$ in increasing order (using the above cited algorithm).
 \item Similarly to normed spaces, the notion of Cauchy completeness and 
(up to natural isomorphism) a unique Cauchy completion of a HLCTVS can 
always be defined
(\cite{Taylor1995}~Chapter1 and \cite{Treves1970}~ChapterI.5 and \cite{Horvath1966}~Chapter2.9).
 \item A HLCTVS is metrizable if and only if its topology can be specified 
by an at most countable system of seminorms, and if complete it is called Fr\'echet. 
The TVS-sense Cauchy completion and the metric-sense Cauchy completion coincides.
(See: \cite{Rudin1991}~ChapterI.1 and \cite{Treves1970}~ChapterI.8,I.10 and \cite{Horvath1966}~Chapter2.6.)
 \item A set $B$ of a HLCTVS is called von Neumann bounded or shortly as bounded, 
whenever for all open neighborhoods $U$ of the origin there exists some $t_{U}\in\R^{+}$, 
such that for all numbers $s$ with $|s|\geq t_{U}$ the relation $B\subset s\,U$ holds. If the space is 
metrizable, the von Neumann boundedness is usually not the same as the metric boundedness. 
On Fr\'echet spaces, the former is stronger or equal to the latter. The two notions coincide on normed spaces. 
By default, by boundedness we mean von Neumann boundedness (\cite{Rudin1991}~ChapterI.1.29).
 \item A HLCTVS is normable if and only if its origin has a bounded neighborhood 
(Kolmogorov's normability criterion, \cite{Rudin1991}~ChapterI.1~Theorem1.39 and 
\cite{Treves1970}~ChapterI.14~Proposition14.4 and 
\cite{Horvath1966}~Chapter2.6 and 
\cite{Schaefer1999}~ChapterII.2~Theorem2.1 and 
\cite{Pietsch1972}~Theorem0.8.2).
 \item Let $X$ be a HLCTVS, then $X'$ denotes the vector space of the continuous 
linear functionals from $X$ to the numbers, i.e.\ the continuous dual of $X$. 
The vector space $X'$ may be endowed with many natural topologies. 
The most important two are the strong and weak dual topologies. 
The weak dual topology is generated by the open zero neighborhoods 
$G^{X'}_{A,\varepsilon}(0):={\big\{y\in X'\,\big\vert\,\sup\limits_{x\in A}\vert(y\vert x)\vert<\varepsilon\big\}}$ 
where $\varepsilon>0$ and $A$ runs through the finite sets of $X$. 
This topology is the weakest one such that the evaluation map $y\mapsto (y\vert x)$ 
is continuous for all $x\in X$. The convergence in this topology is equivalent 
to pointwise convergence. 
The strong dual topology is generated by the open zero neighborhoods 
$G^{X'}_{A,\varepsilon}(0):={\big\{y\in X'\,\big\vert\,\sup\limits_{x\in A}\vert(y\vert x)\vert<\varepsilon\big\}}$ 
where $\varepsilon>0$ and $A$ runs through the von Neumann bounded sets of $X$. 
The convergence in that topology is the uniform convergence on von Neumann bounded sets. 
We denote by $X^{*}$ the space $X'$ endowed with the strong dual topology. 
If $X$ were normed, the strong dual topology on $X'$ would coincide with the 
natural norm topology on the continuous linear functionals. 
(See: \cite{Taylor1995}~Chapter4~Definition4.1 and 
\cite{Treves1970}~ChapterII.19 and 
\cite{Bourles2018}~Chapter3.6.2-4 and 
\cite{Schaefer1999}~ChapterIV.1-5 and 
\cite{Pietsch1972}~Chapter0.6 and 
\cite{Becnel2016}~Chapter2.)
 \item It is common knowledge that the space of smooth sections of a vector 
bundle over a smooth manifold with its natural $\EE$ topology forms a 
Fr\'echet space. So does its closed subspace $\DD_{\K}$, the sections 
with support within a fixed compact subset $\K$ of the manifold. 
(These facts can be also deduced by following arguments of \cite{Taylor1995}~Chapter6.)
\end{enumerate}
\label{remTVS}
\end{Rem}

\section{Tensor product and nuclear spaces}
\label{subsecNuclear}

\begin{Rem}
We recall some basic definitions and facts related to topological tensor products.
\begin{enumerate}[(i)]
 \item Let $X$ and $Y$ be HLCTVS. As usual in linear algebra, their algebraic 
tensor product is a space $X\otimes_{a}Y$ such that all bilinear maps 
$X\times Y\rightarrow Z$ to a third space $Z$ factors through a unique 
linear map $X\otimes_{a}Y\rightarrow Z$. From the definition it follows that the 
algebraic tensor product is unique up to linear isomorphism, so one can speak 
about ``the'' algebraic tensor product space. 
The algebraic tensor product $X\otimes_{a}Y$ may be endowed 
with a natural HLCTVS topology by the projective topology. 
This is the finest locally convex topology such that the map 
$X\times Y\rightarrow X\otimes_{a} Y$ is jointly continuous. 
An other important topology, weaker than the projective, is the 
equicontinuous topology. That is based on viewing 
$X\otimes_{a}Y$ as a space of linear forms on $X'\otimes_{a}Y'$, 
and the equicontinuous topology is the one defined by uniform convergence 
on sets $S\otimes_{a}T\subset X'\otimes_{a}Y'$ where $S\subset X'$ and 
$T\subset Y'$ are equicontinuous sets. 
Since generally we will be dealing with Cauchy complete spaces, 
by default $X\otimes Y$ shall denote the Cauchy completion of $X\otimes_{a}Y$ 
in the projective tensor product topology. 
(See: \cite{Taylor1995}~Chapter3 and 
\cite{Treves1970}~ChapterIII and 
\cite{Schaefer1999}~ChapterIII.6.1,5 and 
\cite{Pietsch1972}~Chapter7.1.)
 \item It is a well known phenomenon that in infinite dimensions 
the completed tensor product of Hilbert spaces leads out from the cathegory 
of Hilbert spaces (\cite{Garrett2020}~Appendix19). It is possible, however, to 
introduce the Hilbert--Schmidt (HS) tensor product within the Hilbert space 
cathegory. Care should be taken, however, that the HS tensor product fails to obey 
the universality property, i.e.\ it is not truly a tensor product operation 
(\cite{Garrett2020}~Appendix19). The HS tensor product is the operation used 
e.g.\ when constructing a Fock space.
 \item Let $X$ and $Y$ be Fr\'echet spaces and let $X\otimes Y$ denote their 
completed tensor product in the projective tensor product topology. 
Then, for all $u\in X$ there exists a sequence $(x_{n})_{n\in\N_{0}}$ and 
$(y_{n})_{n\in\N_{0}}$ both converging to zero in $X$ and $Y$, respectively, 
and an absolute summable sequence of numbers $(\lambda_{n})_{n\in\N_{0}}$, such 
that $u=\sum_{n=0}^{\infty}\lambda_{n}\,x_{n}\otimes y_{n}$ 
(\cite{Taylor1995}~Chapter3.9~Theorem3.9 and 
\cite{Treves1970}~ChapterIII.45~Theorem45.1
\cite{Schaefer1999}~ChapterIII.6~Theorem6.4).
 \item One of the equivalent definitions of a HLCTVS to be nuclear is the following. 
A HLCTVS $X$ is nuclear if and only if for all HLCTVS $Y$ 
the completed projective and equicontinuous tensor product of $X$ and $Y$ 
are isomorphic. The completed tensor product of two nuclear spaces is nuclear. 
That is, nuclear spaces are distinguished by the fact 
that they behave well against the completed tensor product with any HLCTVS.
(See: \cite{Treves1970}~ChapterIII.50.3~Theorem50.1 and 
\cite{Pietsch1972}~Chapter5.)
 \item \label{remNuclearProj} An important equivalent defining property of nuclear spaces is the 
following.
A complete nuclear HLCTVS space is the projective limit of a family of Hilbert spaces, with nuclear linking maps. 
A Fr\'echet space is nuclear if and only if it is the projective limit of 
an at most countable such family. 
(See e.g.:\ \cite{Taylor1995}~Chapter5~Corollary5.14 and \cite{Schaefer1999}~ChapterIII.7.3~Corollary3 and \cite{Pietsch1972}~Chapter7.3.)
 \item In infinite dimensions, nuclear spaces cannot be normed: 
a normable HLCTVS is nuclear if and only if it is finite dimensional 
(\cite{Treves1970}~ChapterIII.50.11~Corollary2).
 \item Cauchy completion of a nuclear space is nuclear 
(\cite{Taylor1995}~Chapter5~Corollary5.10 and \cite{Treves1970}~ChapterIII.50.5~Proposition50.1).
 \item A Fr\'echet space is nuclear if and only if its strong dual is nuclear 
(\cite{Treves1970}~ChapterIII.50.15~Proposition50.6). 
A nuclear F\'echet space is reflexive to the strong duality 
(\cite{Taylor1995}~Chapter5~Proposition5.12).
 \item The projective limit of a family of nuclear spaces is nuclear, 
and the strict inductive limit of a countable family of nuclear spaces is nuclear 
(\cite{Taylor1995}~Chapter5~Corollary5.17 and \cite{Treves1970}~ChapterIII.50.5~Proposition50.1).
 \item Every linear subspace of a nuclear space is nuclear, and the quotient with 
a closed linear subspace is nuclear 
(\cite{Taylor1995}~Chapter5~Proposition5.16 and \cite{Treves1970}~ChapterIII.50.5~Proposition50.1
and \cite{Pietsch1972}~Chapter5.1).
 \item \label{remNuclearMult} The cathegory of nuclear Fr\'echet (NF) spaces and the cathegory 
of strong dual of nuclear Fr\'echet (DNF) spaces is closed under the 
completed tensor product, and they are dual to each-other in terms of the 
strong duality. Moreover, if $X$ and $Y$ are both NF or both DNF, then 
\begin{equation*}
 \mathcal{L}(X,Y^{*}) \;\equiv\; X^{*}\otimes Y^{*} \;\equiv\; (X\otimes Y)^{*} \;\equiv\; \mathcal{B}(X,Y)
\end{equation*}
holds, where the space $\mathcal{L}(X,Y^{*})$ of continuous linear maps 
is understood with the topology of uniform convergence on von Neumann bounded 
sets, and $\mathcal{B}(X,Y)$ is the space of jointly continuous bilinear 
functionals of $X\times Y$ into the numbers 
(Schwartz kernel theorem, see e.g.:\ \cite{Taylor1995}~Chapter5~Theorem5.25). 
That is, for NF or for DNF spaces, tensor product can be implemented via 
the multiplicative realization, in the analogy to finite dimensional vector spaces. 
The closed linear subspace of an NF or DNF space is NF or DNF, respectively, 
and the quotient by closed linear subspace also preserves the NF or DNF 
property (\cite{Pietsch1972}~Chapter5.1.5,6,7,8). 
All NF or DNF spaces are separable (\cite{Pietsch1972}~Chapter4.4.10).
 \item It is common knowledge that the space of smooth sections of a vector 
bundle over a smooth manifold with its natural $\EE$ topology forms a 
nuclear Fr\'echet space. So does its closed subspace $\DD_{\K}$, the sections 
with support within a fixed compact subset $\K$ of the manifold. 
(These facts can be also deduced by following arguments of \cite{Taylor1995}~Chapter6.)
 \item A topological vector space is called LF space, if it is a countable 
strict inductive limit of Fr\'echet spaces (\cite{Treves1970}~ChapterI.13). 
An LF space is Cauchy complete (\cite{Treves1970}~ChapterI.13~Theorem13.1). 
An LF space is metrizable iff the inductive system of subspaces are finite dimensional. 
An LF space is called LNF if it is countable strict inductive limit of nuclear Fr\'echet spaces, 
and as such, an LNF space is nuclear. 
Let us require from this point on that an LNF space is such a strict inductive 
limit of a countable system of NF spaces, that each member of the family has closed 
image within its adjacent member. 
It can be shown that such an LNF space is reflexive against strong duality 
and its dual, a DLNF space, is also nuclear 
(\cite{Taylor1995}~Chapter6~Proposition6.8,9,10).
 \item It is common knowledge that the space of compactly supported smooth sections of a vector 
bundle over a smooth non-compact manifold with its natural $\DD$ topology forms an 
LNF space, and its strong dual $\DD^{*}$ a corresponding DLNF space. 
(These facts can be also deduced by following arguments of \cite{Taylor1995}~Chapter6.)
\end{enumerate}
\label{remNuclear}
\end{Rem}

\section{Joint and separate continuity of bilinear maps}
\label{subsecJoint}

\begin{Rem}
In a generic HLCTVS, certain continuity properties of bilinear forms are 
rather counterintuitive. E.g., separately continuous bilinear maps need not be jointly continuous.
\begin{enumerate}[(i)]
 \item The duality pairing form on a HLCTVS and its strong dual is jointly continuous 
if and only if the space is normable 
(\cite{Horvath1966}~Chapter.4.7~p.359). 
Therefore, in an infinite dimensional nuclear space, the duality pairing cannot 
be jointly continuous. There are many examples in such spaces which are separately 
continuous, but not jointly.
 \item Let $X$, $Y$, $Z$ be HLCTVS and consider a separately continuous bilinear 
map $X\times Y\rightarrow Z$. If $X$ is Fr\'echet, then separate continuity 
implies joint sequential continuity. 
Moreover, if in addition $Y$ is metrizable, then separate continuity implies joint continuity 
(\cite{Rudin1991}~ChapterI.2~Theorem2.17).
 \item Let $X$, $Y$, $Z$ be HLCTVS and consider a separately continuous bilinear 
map $X\times Y\rightarrow Z$. Assume that $X$ and $Y$ are strong duals of 
nuclear Fr\'echet spaces. Then, the pertinent bilinear map is jointly continuous
(\cite{Bourles2018}~Chapter3.9.1~Theorem3.137(b)).
 \item \label{remJointDD} For instance the bilinear map defined by ${\EE(\R^{n})\times\DD(\R^{n})\rightarrow\DD(\R^{n})},\;{(\dpsi,\dpsiT)\mapsto\dpsi\dpsiT}$ 
(pointwise product) is separately continuous, but not jointly continuous. It is merely so-called hypocontinuous. 
(See: \cite{Treves1970}~ChapterIII.41.4~p.423.)
On the other hand, the pointwise product ${\DD(\R^{n})\times\DD(\R^{n})\rightarrow\DD(\R^{n})},\;{(\dpsiT{}_{1},\dpsiT{}_{2})\mapsto\dpsiT{}_{1}\dpsiT{}_{2}}$ 
is jointly continuous (\cite{Hirai2001}~Proposition2.2). 
But 
${\DD(\R^{n})\times\DD(\R^{m})\rightarrow\DD(\R^{n+m})},$ ${(\dpsiT{}_{1},\dpsiT{}_{2})\mapsto\dpsiT{}_{1}\otimes\dpsiT{}_{2}}$ 
is not jointly continuous because although the completed tensor product 
${\DD(\R^{n})\otimes\DD(\R^{m})}\equiv\DD(\R^{n+m})$ as vector spaces, but the 
tensor product topology is weaker than the original (inductive limit) topology of $\DD(\R^{n+m})$
(\cite{Hirai2001}~Theorem2.4). 
The above phenomena are explained by the fact that even forming a finite cartesian product topology 
cannot be exchanged with taking the (strict) inductive limit topology.
\end{enumerate}
\label{remJoint}
\end{Rem}

\section{Cartesian products, locally convex direct sums}
\label{subsecCartesian}

\begin{Rem}
We collect some fundamental identities on the cartesian products and 
locally convex direct sums of HLCTVS spaces.
\begin{enumerate}[(i)]
 \item Let $\big(X_{i}\big)_{i\in I}$ a family of LCTVS. 
The cartesian product set $\mathop{\bigtimes}\limits_{i\in I}X_{i}$ 
endowed with the entrywise linear operations is denoted by 
$\bigoplus\limits_{i\in I}X_{i}$, which can be naturally endowed with 
the product (Tychonoff or initial) topology, being the coarsest topology such that 
the canonical projection maps are continuous. The Tychonoff topology is, 
quite evidently, a LCTVS topology. Therefore, 
$\bigoplus\limits_{i\in I}X_{i}$ is called the Tychonoff direct sum. 
The box topology on 
$\bigoplus\limits_{i\in I}X_{i}$ is defined by the zero neighborhoods being 
the convex hulls of cartesian boxes composed of zero neighborhoods. 
The linear subspace of $\bigoplus\limits_{i\in I}X_{i}$ which consists 
of tuples with only finitely many nonzero entries is denoted by 
$\mathop{\oplus}\limits_{i\in I}X_{i}$, and is called the algebraic direct sum. 
It may be endowed with the locally convex direct sum topology (final topology), 
which is the finest locally convex topology on which the canonical injections 
are continuous. Clearly, the Tychonoff and the box topologies may be as well 
restricted to the algebraic direct sum space. These topologies are Hausdorff 
whenever each element of the family is Hausdorff.
(See: \cite{Horvath1966}~Chapter2.7 and \cite{Bourles2018}~Chapter3.3.7.)
 \item For a general family of LCTVS, the relation: product (Tychonoff) topology $\subset$ 
box topology $\subset$ locally convex direct sum topology holds. 
For a countable family of LCTVS, the relation: product (Tychonoff) topology $\subset$ 
box topology $=$ locally convex direct sum topology holds. 
For a finite family of LCTVS, the relation: product (Tychonoff) topology $=$ 
box topology $=$ locally convex direct sum topology holds. 
(See e.g.: \cite{Chasco2003}~Proposition11,Theorem21,Corollary22.)
 \item The Cauchy completion of a Tychonoff direct sum is the 
Tychonoff direct sum of Cauchy completed spaces, the Cauchy completion of 
a locally convex direct sum is the locally convex direct sum of Cauchy completed 
spaces 
(\cite{Horvath1966}~Chapter2.7 and \cite{Schaefer1999}~ChapterII.6.1,2).
 \item The Tychonoff direct sum of countable family of metrizable HLCTVS is 
metrizable, the locally convex direct sum of a countable family of metrizable 
HLCTVS is metrizable 
(\cite{Horvath1966}~Chapter2.7).
 \item The Tychonoff direct sum of arbitrarily many nuclear spaces is nuclear, 
the locally convex direct sum of at most countable many nuclear spaces is nuclear 
(\cite{Taylor1995}~Proposition5.15 and \cite{Pietsch1972}~Chapter5.2.1,2).
 \item Let $\big(X_{i}\big)_{i\in I}$ be a family of HLCTVS. Then, 
$\big(\mathop{\bigoplus}\limits_{i\in I}X_{i}\big)^{*}=\mathop{\oplus}\limits_{i\in I}X_{i}^{*}$ and 
$\big(\mathop{\oplus}\limits_{i\in I}X_{i}\big)^{*}=\mathop{\bigoplus}\limits_{i\in I}X_{i}^{*}$, where 
$(\cdot)^{*}$ denotes strong duality 
(see e.g.\ \cite{Bourles2018}~Theorem3.105).
\end{enumerate}
\label{remCartesian}
\end{Rem}

\section{Continuity and sequential continuity}
\label{subsecSequential}

\begin{Rem}
Since not all the HLCTVS have sequential topology, it is useful to recall some 
relations of continuity and sequential continuity of maps.
\begin{enumerate}[(i)]
 \item Since an NF space is metrizable, its topology is sequential 
       (\cite{Rudin1991}~Appendix~A6).
 \item A DNF space is sequential (\cite{Webb1968}~Proposition5.7).
 \item A Hausdorff topological vector space is sequential if and only if there exists no strictly finer topology with the same convergent sequences 
       (\cite{Dudley1964}~Theorem7.4).
 \item A LNF or a DLNF space which are not metrizable or dual metrizable, respectively, is not sequential 
       (\cite{Gabriyelyan2017}~Theorem1.1). An LNF is metrizable iff it is a 
       strict inductive limit of a countable sequence of finite dimensional vector spaces. 
       In particular, the test functions on non-compact manifolds and their strong duals are not sequential.
 \item \label{remSequentialMb} An NF or a DNF space is Montel (\cite{Taylor1995}~Theorem5.25).\newline
       An NF or a DNF space is bornological (\cite{Taylor1995}~Theorem5.25).\newline
       Strict inductive limit preserves the Montel property (\cite{Bourles2018}~Theorem3.68).\newline
       Strict inductive limit preserves the bornological property (\cite{Taylor1995}~p.11).\newline
       Strong duality preserves the Montel property (\cite{Horvath1966}~Chapter3.9~Proposition9, \cite{Bourles2018}~Theorem3.123, 
       {\cite{Schaefer1999}~ChapterIV.5.9 -- assumes reflexivity, permanence properties automatic}).\newline
       Strong duality does not preserve the bornological property.\newline
       But an DLNF space is still bornological, since LNF is the strict inductive limit of 
       Montel and bornological Fr\'echet spaces and these are bornological (\cite{Horvath1966}~Chapter3.16~Theorem2).\newline
       In summary: NF, DNF, LNF, DLNF spaces are Montel and bornological.
 \item On Montel spaces the weakly bounded sets are strongly bounded.
       On the strong dual of a Montel space the weak and the strong topology 
       coincides over bounded sets (\cite{Treves1970}~ChapterII.34.7~Proposition34.6).\newline
       Therefore, in Montel spaces sequences are weakly convergent if 
       and only if they are strongly convergent (\cite{Treves1970}~ChapterII.34.7~Corollary1).
 \item Arbitrary cartesian product of Montel spaces is Montel (\cite{Horvath1966}~Chapter3.9~Proposition4 and Remark1).\newline
       At most countable cartesian product of bornological spaces are bornological (\cite{Bourles2018}~p.147).\newline
       Therefore at most countable cartesian product of NF, DNF, LNF, DLNF spaces are Montel and bornological.
 \item \label{remSequentialMB} If a Hausdorff locally convex topological vector space $X$ is 
       bornological, and $Y$ is locally convex topological vector space, 
       then a \underline{linear} map $F:X\rightarrow Y$ is continuous if and only if it is 
       sequentially continuous (\cite{Schaefer1999}~ChapterII.8~Theorem.8.3).
       If $X$ is Montel and bornological, then weak sequential continuity (in $X$) also 
       implies continuity, for linear maps.
 \item Sequential topology is not preserved by cartesian product, but 
       metrizability is preserved by at most countable cartesian product, which 
       is then sequential. Therefore, at most countable cartesian product of an NF 
       space is sequential.\newline
       A DNF space is sequential, but since it is not metrizable, its cartesian 
       products with itself or other spaces are generally not sequential.
 \item LNF or DLNF spaces are not sequential. But they are still Montel and 
       bornological, similarly to NF and DNF. Therefore, if $X$ consists of at most 
       countable cartesian product of NF, DNF, LNF, DLNF spaces, and $Y$ is a locally 
       convex topological vector space, then a \underline{linear} map 
       $F:X\rightarrow Y$ is continuous if and only if it is sequentially continuous 
       (weak sequential continuity in $X$ is enough).
 \item There is a theory of convergence vector spaces (CVS), capturing only 
       the convergence aspects instead of topological aspects over some vector 
       space \cite{beattie2002}. In the analogy of TVS, is possible to define 
       such a space to be Hausdorff as well as being locally convex (HLCVS), 
       and to be complete. A CVS may or may not 
       be originating from a corresponding topological vector space (TVS). 
       In addition, a CVS may eventually be detemined by convergence of 
       countable sequences \cite{beattie1987,beattie1996,Dudley1964}, called 
       to be sequential CVS. An 
       example for a CVS which does not originate from a TVS is the space of 
       Lebesgue measurable functions over $\R^{N}$ with the pointwise almost 
       everywhere convergence. Most of the CVS, however admit some underlying 
       TVS. The CVS cathegory, however, harmonizes better with the notion 
       of joint sequential continuity, see e.g.\ \cite{beattie1996}. 
       In this paper, as is most common, we will use an approach through 
       TVS, and only refer to CVS properties when unavoidable.
 \item Let each of $X$, $Y$ and $Z$ be one of these spaces: NF, DNF, LNF or DLNF. 
       Then, any separately sequentially continuous bilinear map 
       $F:\,X{\times}Y\rightarrow Z$ is also jointly sequentially continuous 
       (\cite{Beattie1991}~Proposition4.1 and Theorem4.8). That is, in such 
       topological spaces, when considered as CVS, separate continuity 
       implies joint continuity in the sense of CVS continuity 
       (but usually not as TVS continuity).
\end{enumerate}
\label{remSequential}
\end{Rem}

\section{Closure and sequential closure}
\label{subsecClosure}

\begin{Rem}
We recall some results from the literature on the closure of linear operators.
\begin{enumerate}[(i)]
 \item Let $X$ and $Y$ be TVS, and $F:X\rightarrowtail Y$ a linear map defined on 
       a subspace $\Dom(F)\subset X$, which may be smaller than $X$. The operator 
       $F$ is called closed iff its graph is closed in $X\times Y$. 
       It is called sequentially closed iff its graph is sequentially closed 
       in $X\times Y$. It is called closable or sequentially closable, 
       iff it has a closed or sequentially closed extension as a linear map, 
       and the smallest such extension is called its closure or its sequential closure, respectively. 
       Clearly, closability and closure are stronger than sequential closability and sequential closure, 
       even for maps between complete spaces. These are equivalent, however, if 
       both $X$ and $Y$ are complete metrizable (e.g.\ for both being NF spaces). 
       These are also equivalent whenever both $X$ and $Y$ are DNF spaces.
 \item Let $X$ and $Y$ be TVS, and $F:X\rightarrowtail Y$ 
       a linear operator, defined on $\Dom(F)\subset X$. Then, $F$ is closable 
       iff $Y$ can be endowed with a Hausdorff vector topology which is not stronger 
       than the original topology on $Y$, and with respect to which $F$ becomes continuous 
       (\cite{Falkner1983}~Theorem1). Moreover, if $X$ and $Y$ was LCTVS, then $F$ 
       is closable iff $Y$ can be edowed with a respective HLCTVS topology, 
       not stronger than the original, making $F$ continuous 
       (see \cite{Falkner1983}~p.108 or in proof of Theorem1 there).
 \item Let $X$ and $Y$ be TVS, and $F:X\rightarrowtail Y$ a densely defined 
       closable linear operator, with its closure $\tilde{F}$. Then, 
       the subspace $\Ker(\tilde{F})$ is closed. (Consequence of \cite{Falkner1983}~Theorem2.)
 \item It is elementary to check that a linear map $F:X\rightarrowtail Y$ is 
       sequentially closable if and only if: for all $x\in X$, for any 
       sequences $(x_{n})_{n\in\N}$ and $(x'_{n})_{n\in\N}$ with $\lim\limits_{n\rightarrow\infty}x_{n}=\lim\limits_{n\rightarrow\infty}x'_{n}=x$, 
       one has that whenever both $(F\,x_{n})_{n\in\N}$ and $(F\,x'_{n})_{n\in\N}$ are convergent, 
       then $\lim\limits_{n\rightarrow\infty}F\,x'_{n}=\lim\limits_{n\rightarrow\infty}F\,x_{n}$ holds. 
       An other reformulation is that for all sequences $(x_{n})_{n\in\N}$ with 
       $\lim\limits_{n\rightarrow\infty}x_{n}=0$ one has that whenever 
       $(F\,x_{n})_{n\in\N}$ is convergent, then $\lim\limits_{n\rightarrow\infty}F\,x_{n}=0$ holds. 
       A further equivalent reformulation of sequential closability is the following:
       for all $x\in X$, for any sequences $(x_{n})_{n\in\N}$ and $(x'_{n})_{n\in\N}$ with 
       $\lim\limits_{n\rightarrow\infty}x_{n}=\lim\limits_{n\rightarrow\infty}x'_{n}=x$ 
       and with $\lim\limits_{n\rightarrow\infty}F\,x_{n}=0$, one has that 
       whenever $(F\,x'_{n})_{n\in\N}$ is convergent, then $\lim\limits_{n\rightarrow\infty}F\,x'_{n}=0$ holds. 
       The latter in words means that: the operator $F$ is sequentially closable iff it is sequentially closable at its 
       approximate kernel points (which will then be the actual kernel points of its sequential closure).
 \item If $X$ and $Y$ are not both metrizable (e.g.\ NF) or DNF, then generally the sequential closability does 
       not imply closability of a densely defined linear operator 
       $F:X\rightarrowtail Y$, even if $X$ and $Y$ were complete Montel and bornological 
       HLCTVS. That is because the dense subspace $\Dom(F)\subset X$ may not 
       inherit the bornological property of $X$, and therefore one cannot conclude 
       from sequential behavior, despite of the linearity of $F$.
 \item For a densely defined linear operator $F:X\rightarrowtail Y$ between two 
       complete HLCTVS, the notion of multivalued set can be introduced:
       $\mathrm{Mul}(F):=\big\{{y\in Y}\,\big\vert\,{\exists(x_{n})_{n\in\N}\text{ in }\Dom(F):\;{\lim\limits_{n\rightarrow\infty}x_{n}=0}\text{ and }{\lim\limits_{n\rightarrow\infty}F\,x_{n}=y}}\big\}$. 
       By construction, $\mathrm{Mul}(F)$ is a linear subspace (moreover, in 
       Banach spaces it is also closed, see e.g.\ \cite{Antonevich1997}). 
       The operator $F$ is sequentially closable iff $\mathrm{Mul}(F)=\{0\}$. 
       That is, $\mathrm{Mul}(F)$ measures the non-closability of $F$. 
       The operator $F$ is called maximally non-closable or maximally singular, 
       whenever $\mathrm{Mul}(F)$ is dense in the closure of $\Ran(F)$. In 
       particular, whenever $\mathrm{Mul}(F)$ is dense in $Y$, the operator 
       $F$ is necessarily maximally non-closable.
 \item Let $\F$ denote the space of smooth sections of some vector undle over the base manifold $\M$, as previously, with the 
       standard $\EE$ smooth function topology. Recall that 
       $\F{\otimes}\F$ may be identified with the corresponding space of smooth 
       sections over $\M{\times}\M$ with their $\EE$ topology. Introduce the diagonal evaluation 
       map $\mathrm{ev}:\,\F\otimes\F\longrightarrow\F,\,\big((x,y)\mapsto G(x,y)\big)\longmapsto\big(z\mapsto G(z,z)\big)$. 
       Such map appears e.g.\ in the interaction term of the MDS operator 
       generated by an interacting Euler--Lagrange functional. The above map 
       may be considered as well as a densely defined linear map 
       $\widehat{\mathrm{ev}}:\,{(\FxT)^{*}{\otimes}(\FxT)^{*}}\rightarrowtail(\FxT)^{*}$ on the space of 
       distributions, since there is the natural dense inclusion $\F\subset(\FxT)^{*}$. 
       If this operator were (sequentially) closable, it would provide a straightforward solution 
       for the problematics of renormalization. It can be shown, however, that 
       this map is maximally non-closable. In order to demonstrate that, 
       take any function $g\in\F$ and a corresponding sequence 
       $(G_{n})_{n\in\N}$ in $\F{\otimes}\F$ such that for all $n\in\N$ 
       the equality $G_{n}(z,z)=g(z)$ ($\forall z\in\M$) is satisfied, and 
       that $(G_{n})_{n\in\N}$ converges locally uniformly to zero around 
       any point of $\M{\times}\M\setminus{\{(x,y)\in\M{\times}\M\vert x=y\}}$. 
       Clearly, one can construct such a sequence for all $g\in\F$. 
       (One can construct this explicitely on $\R^{N}$, and then one can bring this 
        to a manifold $\M$ via partition of unity arguments.)
       The above finding means, however, that all $g\in\F\subset(\FxT)^{*}$ is in $\mathrm{Mul}(\widehat{\mathrm{ev}})$. 
       Therefore, $\widehat{\mathrm{ev}}$ is maximally non-closable.
\end{enumerate}
\label{remClosure}
\end{Rem}

\section{Nuclear Fr\'echet spaces with countable Hilbertian structure}
\label{subsecCH}

\begin{Rem}
We recall some fundamental facts on a special type of nuclear Fr\'echet 
spaces, which will plays an important role in QFT constructions.
\begin{enumerate}[(i)]
 \item \label{remCHc} Whenever the topology of a HLCTVS is defined by a countable family of 
seminorms, and there exists a continuous norm on that HLCTVS, then by means 
of Remark\ref{remTVS}(\ref{remTVSReorg}): without loss of generality, the defining family of 
seminorms of the topology of the space can be taken to be an increasing 
family of \emph{norms} instead of seminorms, indexed by $\N_{0}$. 
Specially, the pertinent continuous norm may be chosen as one of the elements 
of the topology defining norm family. Moreover, it may be chosen also to be the weakest one. 
Let $\HH$ be such a HLCTVS, and let 
$\left\Vert\cdot\right\Vert_{n}$ ($n\in\N_{0}$) be a countable increasing family of 
topology defining norms on it. Denote the completion of $\HH$ in the 
$\left\Vert\cdot\right\Vert_{n}$ norm by $H_{n}$, then the natural inclusion 
$i_{\infty,n}:\,\HH\rightarrow H_{n}$ is a continuous injective linear map. 
One may write $H_{n}\supset \HH$ is dense (for all $n\in\N_{0}$).
 \item If the topology of a HLCTVS is defined by a countable system of Hilbertian 
seminorms, and there exists a continuous Hilbertian norm, then the analogy of 
the statements in (\ref{remCHc}) will hold, respectively, with Hilbertian norms 
and Hilbert spaces as completed spaces.
 \item \label{remCHcomp} Let $\HH$ be a HLCTVS as in (\ref{remCHc}), and let 
$\left\Vert\cdot\right\Vert_{n}$ ($n\in\N_{0}$) be a countable increasing family of 
topology defining norms on it, the completed Banach spaces denoted by $H_{n}$. 
The $\HH\rightarrow\HH$ identity 
map is continuous when the norm topology of $\left\Vert\cdot\right\Vert_{n+1}$ is taken on 
the starting space and the norm topology of $\left\Vert\cdot\right\Vert_{n}$ is 
taken on the image space. Therefore, its continuous extension in these norms is a continuous linear map 
$i_{n+1,n}:\,H_{n+1}\rightarrow H_{n}$. The 
map $i_{n+1,n}$ is  the identity over $\HH$, and therefore it is injective on that 
subspace, but its continuous extension $i_{n+1,n}:\,H_{n+1}\rightarrow H_{n}$ may or may not be injective 
over the full completed space $H_{n+1}$. Whenever 
$i_{n+1,n}$ is injective, the adjacent norms are called Gel'fand compatible 
(see \cite{Gelfand1968}~Chapter2.2 and \cite{Bogachev2017}~Chapter2.2.8-2.2.11 
and \cite{Merkle1998} and \cite{Merkle1990}~AppendixB.10-13). 
An other way to formulate the Gel'fand compatibility of the adjacent norms is that 
any sequence in $\HH$ which is Cauchy in both of the adjacent norms, and 
which is convergent to zero in either of the norms, then it is convergent to zero 
also in the other norm. If Gel'fand compatibility of adjacent norms hold, then one may regard the 
completed spaces as nested in each-other:
$H_{0}\supset{\dots}\supset H_{n}\supset H_{n+1}\supset{\dots}\supset\HH$.
 \item \label{remCHdef} Combining the above observations and Remark\ref{remNuclear}(\ref{remNuclearProj}), the 
topology of a nuclear Fr\'echet space $\HH$ admitting a continuous Hilbertian norm 
can be described by a countable system of increasing Hilbertian norms, indexed 
by $\N_{0}$. Whenever these adjacent topology defining Hilbertian norms are 
Gel'fand compatible, then (\ref{remCHcomp}) applies, and in addition 
$\mathop{\bigcap}\limits_{n\in\N_{0}}H_{n}=\HH$ holds. Moreover, for all 
$n\in\N_{0}$ there exists an integer $m\geq 1$, such that the inclusion map 
$i_{n+m,n}:\,H_{n+m}\rightarrow H_{n}$ is nuclear. 
(Specially, one may choose the system such, that all the adjacent inclusion maps $i_{n+1,n}$ are nuclear.) 
Nuclear Fr\'echet spaces of this kind are called countably Hilbert (CH) type NF spaces.
(See \cite{Bogachev2017}~Chapter2.2.8, and e.g.\ the review paper \cite{Becnel2016}. Note 
that the Gel'fand compatibility condition is often overlooked in the literature.) 
The CH type NF spaces give realization to a special form of a projective limit: 
the projection maps $i_{\infty,n}$ as well as the linking maps $i_{n+1,n}$ are all injective.
 \item On a noncompact manifold, the space of smooth sections of a vector bundle 
with the standard $\EE$ topology is an NF space, but not of CH type. It is not difficult 
to see that on a compact manifold, they are of CH type NF spaces. 
Also, on a finite dimensional 
real affine space, one can define the space of rapidly decreasing 
(Schwartz) sections, which are well known to be NF spaces of CH type 
(see a review paper: \cite{Becnel2015}).
 \item \label{remCHdual} The strong topological dual (or complex conjugate dual) space of CH type NF 
spaces have a corresponding inductive limit structure. 
Namely, let $\HH$ be a CH type NF space, with the notations as in (\ref{remCHdef}). Taking the strong 
topological dual (or complex conjugate dual) of the inclusion chain 
${H_{0}\supset{\dots}\supset H_{n}\supset H_{n+1}\supset{\dots}\supset\HH}$, one infers that 
${\HH^{*}\supset{\dots}\supset H_{n+1}^{*}\supset H_{n}^{*}\supset{\dots}\supset H_{0}^{*}}$ and 
${\bar{\HH}^{*}\supset{\dots}\supset \bar{H}_{n+1}^{*}\supset \bar{H}_{n}^{*}\supset{\dots}\supset \bar{H}_{0}^{*}}$
holds. More precisely, for all $n\in\N_{0}$ one has the continuous linear injective maps 
$i_{\infty,n}:\,\HH\rightarrow H_{n}$ and $i_{n+1,n}:\,H_{n+1}\rightarrow H_{n}$, 
whose topological transpose (or complex conjugate topological transpose) 
give rise to the continuous linear maps 
$i_{\infty,n}^{*}:\,H_{n}^{*}\rightarrow \HH^{*}$ and $i_{n+1,n}^{*}:\,H_{n}^{*}\rightarrow H_{n+1}^{*}$, 
as well as 
$\bar{i}_{\infty,n}^{*}:\,\bar{H}_{n}^{*}\rightarrow \bar{\HH}^{*}$ and $\bar{i}_{n+1,n}^{*}:\,\bar{H}_{n}^{*}\rightarrow \bar{H}_{n+1}^{*}$, 
respectively. Since for all $n\in\N_{0}$ the image of $i_{n+1,n}$ was dense (it 
contained $i_{\infty,n}[\HH]$ whose completion was $H_{n}$), the transpose 
(or complex conjugate transpose) linking maps cannot have a kernel. Thus, 
$i_{\infty,n}^{*}$, $i_{n+1,n}^{*}$, $\bar{i}_{\infty,n}^{*}$, $\bar{i}_{n+1,n}^{*}$ 
are injective, which justifies the above inclusion chain of the dual (or complex conjugate dual) spaces. 
Moreover, one has $\HH^{*}\equiv \mathop{\bigcup}\limits_{n\in\N_{0}}H_{n}^{*}$ and 
$\bar{\HH}^{*}\equiv \mathop{\bigcup}\limits_{n\in\N_{0}}\bar{H}_{n}^{*}$. 
Since the inclusion $H_{n}\supset H_{n+m}$ was eventually Hilbert--Schmidt (and eventually nuclear), 
the above corresponding dual (or complex conjugate dual) inclusions also eventually become Hilbert--Schmidt (and eventually become nuclear).
 \item \label{remCHGt} 
Whenever a preferred Hilbertian norm $\Vert{\cdot}\Vert_{0}$ is fixed on a 
countable Hilbert type NF space, it give rise to the well known rigged Hilbert spaces 
or Gel'fand triples (see \cite{Gelfand1968,Bogachev2017}). 
More concretely, using the notations of 
(\ref{remCHdual}), for all $n\in\N_{0}$ the Riesz representation theorem guarantees 
the natural linear unitary isomorphism $r_{n}:\,H_{n}\rightarrow \bar{H}_{n}^{*}$, and 
correspondingly, the natural antilinear antiunitary antiisomorphism 
$\bar{r}_{n}:\,H_{n}\rightarrow H_{n}^{*}$, via the relations 
$r_{n}(x):=\overline{\left<x,\cdot\right>}_{H_{n}}$ 
and 
$\bar{r}_{n}(x):=\left<x,\cdot\right>_{H_{n}}$ for all $x\in H_{n}$. Using this, 
in particular for $H_{0}$, one has the continuous linear inclusions with dense image 
${\bar{\HH}^{*}\supset{\dots}\supset H_{-n-1}\supset H_{-n}\supset{\dots}\supset H_{0}\supset{\dots}\supset H_{n}\supset H_{n+1}\supset{\dots}\supset\HH}$ 
where the inclusions are eventually Hilbert--Schmidt (and eventually nuclear). 
Here, we used the notation $H_{-n}:=\bar{H}_{n}^{*}$ for all $n\in\N_{0}$, and 
the identification $\bar{H}_{0}^{*}\equiv H_{0}$ via $r_{0}$. One should note that the 
injective continuous linear maps (which are eventually Hilbert--Schmidt and 
eventually nuclear) 
$\bar{i}^{*}_{n,n+k}\circ r_{n}\circ i_{n+m,n}:\,H_{n+m}\rightarrow \bar{H}^{*}_{n+k}$ 
and 
$\bar{i}^{*}_{n',n'+k'}\circ r_{n'}\circ i_{n'+m',n'}:\,H_{n'+m'}\rightarrow \bar{H}^{*}_{n'+k'}$ 
($n,m,k,n',m',k'\in\N_{0}$) are in general different maps for $n'\neq n$ even 
if $n'+m'=n+m$ and $n'+k'=n+k$ holds. That is because the Riesz identification 
maps $r_{n}$ and $r_{n'}$, even if restricted to the common set $\HH$, 
are different for $n'\neq n$. That is, the continuous linear injection 
$\HH\rightarrow H_{0}\rightarrow \bar{\HH}^{*}$ is not natural, it explicitely 
depends on the choice of the Hilbertian norm $\left\Vert\cdot\right\Vert_{0}$ on $\HH$. 
In Section~\ref{subsecSobolev} we shall show an example when a similar embedding 
chain construction is natural.
\end{enumerate}
\label{remCH}
\end{Rem}

\section{Sobolev and Maurin embedding theorems}
\label{subsecSobolev}

\begin{Rem}
For some of the proofs, recalling the Sobolev and Maurin embedding theorems will be helpful. 
Let $\M$ be $\R^{m}$, or an open set of $\R^{m}$ with a boundary 
having the cone property, or a compact smooth $m$-dimensional 
real manifold with such boundary. 
Let $V(\M)$ be a real vector bundle over $\M$. 
Fix some smooth Riemannian metric on $V(\M)$. The symbol $C_{b}^{k}(V(\M))$ will denote 
the Banach space of $k$-times continuously differentiable sections of $V(\M)$ with bounded derivatives, 
with the $C^{k}$ supremum norm. The symbol $H_{l}(V(\M))$ will denote the Hilbert space 
of $l$-times weakly differentiable section Lebesgue equivalence classes of $V(\M)$ 
which are square integrable together with all of their derivatives, equipped 
with the corresponding $L^{2}$ norm including all the derivatives 
(i.e.\ $H_{l}$ is the $l$-th order $L^{2}$ type Sobolev space).
\begin{enumerate}[(i)]
 \item \label{remSobolevI} For $\M$ as above, and $l>k+\frac{\dim(\M)}{2}$ one has the inclusion 
$H_{l}(V(\M))\subset C_{b}^{k}(V(\M))$ and the embedding is continuous, 
i.e.\ the $H_{l}$ norm is stronger than the $C_{b}^{k}$ norm. 
(Sobolev embedding theorem, see e.g.: 
\cite{Adams2003}~Chapter4~Theorem4.12(I)(A)
and 
\cite{ChoquetBruhat2000}~ChapterVI.2.II.3, 
\cite{ChoquetBruhat2000}~ChapterVI.16.3.)
 \item \label{remSobolevII} For $\M$ being as above, but assumed to be bounded for the case $\M\subset\R^{m}$, 
then for $l>k+\frac{\dim(\M)}{2}$ one has that the inclusion 
$H_{l}(V(\M))\subset H_{k}(V(\M))$ is Hilbert--Schmidt. 
(By construction of the $H$-type Sobolev spaces, one has that for all $l>k$ the 
inclusion $H_{l}(V(\M))\subset H_{k}(V(\M))$ is trivially valid, and is continuous, 
i.e.\ the $H_{l}$ norm is stronger than the $H_{k}$ norm.) 
The analogous statement holds for the 
$H$-type Sobolev spaces obtained from the completion of compactly supported 
sections. 
(Maurin embedding theorem, see e.g.: 
\cite{Adams2003}~Chapter6~Theorem6.61 
or 
\cite{ChoquetBruhat2000}~ChapterVI.2.II.3, 
\cite{ChoquetBruhat2000}~ChapterVI.2.II.7, 
\cite{ChoquetBruhat2000}~ChapterVI.16.3 etc.)
 \item Knowing result (\ref{remSobolevI}) and H\"older's inequality, 
it is easy to understand why on the $\EE$ or $\DD$ type spaces the family 
of supremum type $C^{k}$ local norms and the $L^{2}$ type $C^{k}$ local norms ($k\in\N_{0}$) 
are equivalent.
 \item Knowing result (\ref{remSobolevII}), it is easy to understand the 
nuclearity of the $\EE$ or $\DD$ spaces. The theorem implies that 
the $L^{2}$ type $C^{k}$ local norms over each fixed compact region 
are eventually getting gradually stronger with growing $k\in\N_{0}$, and 
eventually the embedding becomes Hilbert--Schmidt with large enough $k$. 
Consequently, the embedding eventually becomes nuclear (trace class) with 
large enough $k$ (since composition of two Hilbert--Schmidt maps are nuclear).
 \item Knowing that for all $l>k$ the inclusion $H_{l}(V(\M))\rightarrow H_{k}(V(\M))$ 
is injective, it is easy to understand that over compact manifolds the 
$\EE$ (or $\DD$) spaces are not only NF spaces, but are also CH type NF spaces. 
It is obvious that they admit continuous Hilbertian norms, and because of the 
above injective inclusion, the topology defining countable family of increasing 
Hilbertian norms are also automatically Gel'fand compatible (see Remark\ref{remCH}).
 \item Denote by $\bar{V}^{\times}(\M)$ the densitised complex conjugate dual vector bundle of $V(M)$, 
and their spaces of smooth sections by $\bar{\EE}^{\times}$ and $\EE$, respectively. 
Whenever the base manifold $\M$ is compact, there is the natural jointly continuous 
sesquilinear form $\bar{\EE}^{\times}\times\EE\rightarrow\C,\;(p,\dpsi)\mapsto\int\limits_{\M}\bar{p}\,\dpsi$. 
This gives rise to a natural continuous complex-linear injection $r:\,\EE\rightarrow(\bar{\EE}^{\times})^{*}$. 
If a smooth complex (sesquilinear) Riemann metric on $V(\M)$ is chosen which is densitized by a positive 
volume form on $\M$, it gives rise to corresponding Hilbertian norms 
$\left\Vert\cdot\right\Vert^{\times}_{0}$ and $\left\Vert\cdot\right\Vert_{0}$ on 
$\bar{\EE}^{\times}$ and $\EE$, respectively. The corresponding norm equivalence class, and therefore 
the norm topology does not depend on the particular choice of the densitized Riemann metric 
(this is elementary, but see also \cite{Laszlo2013}~AppendixA). 
The above sesquilinear form is continuous in these norms. On the other hand, both 
$\bar{\EE}^{\times}$ and $\EE$ are CH type NF spaces, giving rise to the 
continuous complex-linear injections $\EE\rightarrow E_{0}$ and $(\bar{E}^{\times}_{0})^{*}\rightarrow(\bar{\EE}^{\times})^{*}$, 
referring to the notations of Remark\ref{remCH}(\ref{remCHGt}). 
With these choices, the above injection $r$ induces a continuous 
complex-linear bijection $r:\,E_{0}\rightarrow(\bar{E}_{0}^{\times})^{*}$, which is 
isometric, and therefore is a unitary isomorphism. 
Consequently, one has the natural inclusions 
${(\bar{\EE}^{\times})^{*}\supset{\dots}\supset E_{-n-1}\supset E_{-n}\supset{\dots}\supset E_{0}\supset{\dots}\supset E_{n}\supset E_{n+1}\supset{\dots}\supset\EE}$, 
where for $n\in\N_{0}$ one has $E_{-n}:=(\bar{E}^{\times}_{n})^{*}$, furthermore 
on $E_{0}$ and $\bar{E}^{\times}_{0}$ the above particular Hilbertian norms 
were used, and therefore one has the natural unitary isomorphism 
$E_{0}\equiv\bar{E}^{\times}_{0}$ induced by the map $r$, regardless of the 
particular choice of the densitized Riemann metric.
 \item If $l>\frac{\dim(\M)}{2}$, and the base base manifold $\M$ is compact, the Sobolev 
space $H_{l}(V(\M))$ becomes a reproducing kernel Hilbert space 
(see also \cite{Aronszajn1950} and e.g.\ \cite{Laszlo2013}~AppendixD), due to the Sobolev inequality (\ref{remSobolevI}). 
That is, for every $x\in \M$ the point evaluation $H_{l}(V(\M))\rightarrow\C,\;f\mapsto (p_{x}|f(x))$ 
becomes a well defined continuous complex-linear map, where $p_{x}$ is any element of 
$V_{x}^{*}(\M)$. The Riesz representation theorem ensures that for each $x\in \M$ and 
$p_{x}\in V_{x}^{*}(\M)$, there exists a unique $K_{x}^{\bar{p}_{x}}\in H_{l}(V(\M))$ such 
that $\left<K_{x}^{\bar{p}_{x}},f\right>=(p_{x}|f(x))$ holds for all $f\in H_{l}(V(\M))$. 
If Penrose abstract indices $\mathcal{A},\mathcal{B},{\dots}$ are used on $V(\M)$, this reads as 
$\left<\bar{p}_{x}{}_{\mathcal{A}'}K_{x}^{\mathcal{A}'},f\right>=p_{x}{}_{\mathcal{A}}\,f^{\mathcal{A}}(x)$. 
As $K_{x}^{\bar{p}_{x}}$ itself is a section of $V(\M)$, it may also be evaluated at any 
point. The reproducing kernel function $K:\bar{V}^{*}(\M)\times V^{*}(\M)\rightarrow\C$ is defined 
as $K(x,y)^{\bar{p}_{x},q_{y}}:=\big(q_{y}\big\vert K_{x}^{\bar{p}_{x}}(y)\big)$, or in Penrose abstract 
indices, $q_{y}{}_{\mathcal{B}}\,\bar{p}_{x}{}_{\mathcal{A}'}K(x,y)^{\mathcal{A}'\mathcal{B}}:=q_{y}{}_{\mathcal{B}}\,\bar{p}_{x}{}_{\mathcal{A}'}K_{x}^{\mathcal{A}'}(y)^{\mathcal{B}}$, 
and thus it can be regarded as a section of the vector bundle $(\bar{V}^{*}(\M){\times}V^{*}(\M))^{*}$ over the manifold 
$\M{\times}\M$. It may be verified that for any $x,y\in \M$:
\begin{itemize}
\item[(A)] $\left<K(x,\cdot)^{\mathcal{A}'\cdot},K(y,\cdot)^{\mathcal{B}'\cdot}\right>=K(y,x)^{\mathcal{B}'\mathcal{A}}$, 
\item[(B)] $\overline{K(y,x)^{\mathcal{B}'\mathcal{A}}}=K(x,y)^{\mathcal{A}'\mathcal{B}}$,
\item[(C)] if $\left(\Phi_{i}\right)_{i\in \mathcal{I}}$ comprises a
  complete  orthonormal system in $H_{l}(V(\M))$, then the identity 
  $K(x,\cdot)^{\mathcal{A}'\mathcal{B}}=\sum_{i\in \mathcal{I}} \bar{\Phi}_{i}^{\mathcal{A}'}(x)\Phi_{i}^{\mathcal{B}}(\cdot)$ 
  holds, where the infinite summation is understood in the $H_{l}$ norm topology.
\end{itemize}
\end{enumerate}
\label{remSobolev}
\end{Rem}

\section*{References}
\addcontentsline{toc}{section}{\refname}

\bibliographystyle{JHEP}
\bibliography{mdssuppl}